\documentclass[aps,amsmath,amssymb,prd,floatfix,preprint,superscriptaddress,nofootinbib,12pt]{JHEP3}
\usepackage{epsfig}
\usepackage{amsmath}
\usepackage{amssymb,amsfonts}
\usepackage{comment}
\usepackage{graphicx}
\usepackage{epsf,colordvi,shadow,pifont}
\usepackage{latexsym}
\usepackage{slashed}
\usepackage{xcolor}
\usepackage{subfigure}
\usepackage{subfig}
\let\normalcolor\relax

\relax
\renewcommand{\theequation}{\arabic{section}.\arabic{equation}}
\def\be{\begin{equation}}
\def\ee{\end{equation}}
\def\bs{\begin{subequations}}
\def\es{\end{subequations}}
\newcommand{\een}{\end{subequations}}
\newcommand{\ben}{\begin{subequations}}
\newcommand{\beq}{\begin{eqalignno}}
\newcommand{\eeq}{\end{eqalignno}}

\DeclareMathOperator\Tr{tr}

\DeclareMathOperator\sech{sech}
\DeclareMathOperator\sn{sn}
\DeclareMathOperator\cn{cn}
\DeclareMathOperator\dn{dn}
\def\square{\large\hbox{{$\sqcup$}\llap{$\sqcap$}}}

\def\l{\lambda}
\def\m{\mu}
\def\n{\nu}

\def\p{\partial}

\def\sp{\;\;\;,\;\;\;}

\newcommand\fverb{\setbox\pippobox=\hbox\bgroup\verb}
\newcommand\fverbdo{\egroup\medskip\noindent%
                        \fbox{\unhbox\pippobox}\ }
\newcommand\fverbit{\egroup\item[\fbox{\unhbox\pippobox}]}
\newbox\pippobox
\def\tr{\tilde\rho}
\def\tu{\tilde{u}}

\def\hri#1#2{\href{http://arxiv.org/abs/#1}{[ArXiv:#1]#2}}
\def\hre#1#2{\href{http://arxiv.org/abs/#1/#2}{[ArXiv:#1/#2]}}

\def\beq{\begin{equation}}
\def\eeq{\end{equation}}

\def\4R{{{}^{(4)}R}}

\def\K5{{\kappa}}
\def\K52{{\kappa^2}}

\newcommand{\half}{\frac{1}{2}}

\newcommand{\A}{A}

\def\bea{\begin{eqnarray}}
\def\eea{\end{eqnarray}}
\def\nn{\nonumber}
\newcommand{\ba}{\begin{aligned}}
\newcommand{\ea}{\end{aligned}}

\newcommand{\rd}{{\rm d}}
\newcommand{\re}{{\rm e}}
\newcommand{\ri}{{\rm i}}

\def\IC{{\mathbb C}}

\def\eqp{\;\;\;.}
\def\eqc{\;\;\;,}

\title{Euclidean Wormholes and Holography}
\author{P. Betzios$^{\flat}$, E. Kiritsis$^{\flat,\natural}$, O. Papadoulaki$^*$\\
~\\
$^\flat$ \href{http://hep.physics.uoc.gr}{Crete Center for Theoretical Physics}, Institute for Theoretical and Computational Physics,
Department of Physics,  P.O. Box 2208,\\
University of Crete, 70013, Heraklion, Greece
~\\
~\\
$^\natural$ \href{http://www.apc.univ-paris7.fr}{APC, AstroParticule et Cosmologie}, Universit\'e Paris Diderot, CNRS/IN2P3, CEA/IRFU,
Observatoire de Paris, Sorbonne Paris Cit\'e,\\
 10, rue Alice Domon et L\'eonie Duquet, 75205 Paris
Cedex 13, France\\
~\\
$^*$ \href{https://www.ictp.it/}{International Centre for Theoretical Physics} \\
Strada Costiera 11, Trieste 34151 Italy.
}

\preprint{CCTP-2019-3\\
ITCP-IPP-2019/3}

\abstract{Asymptotically AdS wormhole solutions are considered in the context of holography.
Correlation functions of local operators on distinct boundaries are studied.
It is found that such correlators are finite at short distances. Correlation
functions of non-local operators (Wilson loops) on distinct boundaries are also studied,
with similar conclusions. Deformations of the theory with multi-trace operators on distinct boundaries
are considered and studied. As a consequence of these results, the dual theory is expected to factorize in the UV, and the two sectors to be coupled by a soft non-local interaction. A simple field theory model with such behavior is presented.}


\begin{document}

\section{Introduction}

In the past two decades, developments revolving around the holographic correspondence, have provided a rich dictionary and elucidated deep connections between Quantum Field Theory (QFT) and Gravity/String Theory. Nevertheless, there exist gravitational backgrounds for which the inner workings of the correspondence are still shrouded in mystery. An example of this kind are Euclidean wormholes that describe positive signature spaces having more than a single asymptotic boundary. An indicative list of different types of Euclidean wormhole solutions is~\cite{Giddings:1987cg,Coule:1989xu,Halliwell:1989ky,Hosoya:1989zn,Rey:1998yx,Maldacena:2004rf,ArkaniHamed:2007js,Hertog:2017owm}, the more recent works being geared towards embedding them in string theory. A general feature of such solutions is the presence of (at least locally) negative Euclidean ``energy" that supports the throat/s connecting the different asymptotic regions. In Euclidean signature there is no obvious issue with the presence of negative Euclidean ``energy" (for example a Euclidean magnetic field has a negative ``energy" contribution to the Euclidean stress tensor) and therefore a no-go theorem for such configurations does not exist.

Turning to holography, even though their two boundary Lorentzian analogues such as the eternal AdS-Schwarzschild black hole have been known to correspond to a pair of entangled QFT's~\cite{Maldacena:2001kr}, once they are analytically continued to Euclidean signature the space smoothly caps off disconnecting the different asymptotic boundaries, \cite{Skenderis:2009ju}.
The reason is that the two boundaries are always separated by a horizon, once the averaged null energy condition (ANEC) is assumed for the gravitational theory. Therefore upon Euclidean continuation, the presence of the horizon gives rise to a disconnected Euclidean geometry. Said in other words, entanglement cannot provide the ``glue'' to hold the space together in a Euclidean setup and therefore one might expect some more direct form of interaction between different Euclidean QFT's residing on each boundary.

A construction where the two boundary QFT's are directly coupled has appeared in the Lorentzian context~\cite{Gao}~\footnote{A different type of ``bridges" between distinct asymptotic regions using quivers is given in~\cite{Bachas:2017rch}.}. This involves a double trace deformation on the pair of entagled QFT's that holographically produces  negative null energy in the bulk. For such a theory at the quantum level the bulk null energy condition can fail, and this leads to protocols where horizons can become traversable~\cite{Gao,MSY,vP}. One could also then argue that the counterpart of the presence of negative Euclidean ``energy" in Lorentzian signature is a violation of the null energy condition. What is then left is to devise the appropriate interaction term between different Euclidean QFT's that can lead to a wormhole geometry.

The question of a holographic interpretation and construction of such solutions gains an even greater importance if we ask questions related to quantum gravity in the bulk, since their role in the gravitational Euclidean path integral is still a subject of debate. The main issues here are those of stability i.e. if such solutions are local minima of the path integral, and whether summing over multiple wormholes imposes an inherent randomness in the coupling constants~\cite{Coleman:1988cy,Lavrelashvili:1987jg,Giddings:1988cx} (Coleman's $\alpha$-parameters) that define the theory in question.

In addition, there have been many works in the past related to further explorations of their putative physical properties. We briefly mention here the Baby-Universe~\cite{BabyUniverses} and Wheeler de-Witt interpretations~\cite{Hawking:1988ae}, violation of global conservation laws~\cite{Gupta:1989bs} in the bulk of the space at question - which also resonates with modern holography where global symmetries exist only for the dual boundary theory - and a different interpretation of the wormhole gas as one that contains ``holes of nothing'' by imposing appropriate boundary conditions (antipodal identification)~\cite{Betzios:2017krj}. A recent review of these and several other developments can be found in~\cite{Hebecker:2018ofv}.

On the other hand, holographic considerations seem to distinguish between different kinds/setups of studying Euclidean Wormholes at least at large-N in a limit where a semi-classical geometry is trustworthy.
A first thorough study of the problem of Euclidean semiclassical wormhole geometries and their holographic interpretation has been undertaken in \cite{Maldacena:2004rf} where a study of the issues has been performed, including various solutions and a preliminary stability analysis.

Our main focus will be an extension of these results, mainly by studying different observables such as correlation functions of local and non-local observables (Wilson lines), trying to elucidate universal properties that any putative holographic dual should satisfy.

\subsection{Results and outlook}

In this paper we analyse asymptotically AdS wormhole solutions in two, three and four dimensions. Our goal is, by studying various dynamical observables, to obtain hints for the interpretation of such solutions in terms of dual Euclidean CFTs.
Although we shall not have a clear and crisp proposal for this in the end, we will be able to ascertain several properties and to propose tentative toy dual theories.  The solutions we shall examine are presented in section~\ref{sol_asym_reg}.

Our setup and goals in this paper are orthogonal to that in ~\cite{Rey:1998yx, ArkaniHamed:2007js} which investigated in a holographic context the fate of the fluctuation wormhole physics and the fate of  $\alpha$-parameters {\`a} la Coleman.

After first discussing how the usual holographic rules get extended in the presence of multiple boundaries in section~\ref{multipleboundariescorrelators}, in section~\ref{HomogeneousODES} we compute explicitly the non-trivial two-point correlators between operator insertions across the two boundaries. . This cross-correlation diminishes in the UV and grows in the IR while remaining bounded. {\it It does not exhibit any short distance singularity}. This fact suggests that in the UV, the two boundaries correspond to distinct QFTs. This is a common property of this cross-correlator among our holographic examples and we expect this to be universal.

To further corroborate those results, we also extend the analysis to non-local operators in the form of Wilson lines in section~\ref{Wilsonlines}. The expectation value of a single Wilson loop approaches an area law in the IR (that is visible when the size of the boundary spatial sphere becomes large) signalling a confining behaviour for the dual theory.

We continue with the computation of a correlator of two Wilson loops, one on the first and the other on the second boundary. The dual bulk string world-sheet has then the possibility to stretch across the two boundaries. There is also the possibility that the string forms two disconnected surfaces that interact only via the exchange of bulk perturbative modes.
The connected semiclassical contribution exists for all loop sizes.
The contribution mediated by a supergraviton exchange is, as usual, difficult  to compute but is expected to be non-zero. Therefore, the total connected contribution to the cross Wilson loop is generically non-trivial.

We then proceed to study the effects that multi-trace deformations have on the wormhole state in section~\ref{Doubletracedeformations}.
The result is that the IR structure of the correlators is qualitatively not affected, but in some cases, the cross-correlator acquires a short distance singularity, because of the local (UV) couplings of the multitrace operators.
This is an independent argument, that local UV interactions between the two boundaries are incompatible with the finiteness of the cross-correlator at short distances.

Our results are then  consistent with the holographic interpretation in terms of a system of coupled QFT's with a softer than usual UV coupling that grows in the IR (non-local interaction). In section~\ref{field theory analogue} we provide a simple field theory example that realizes this idea, with a soft non-local UV coupling that is IR relevant, and which is regular and meaningful  only in Euclidean space: there is no sensible Minkowski rotation of this construction. Such an obstruction also seems to hold for Euclidean wormholes as we describe below and in appendices~\ref{LorentzianHopf} and~\ref{Bang-Crunch}.

There are several open questions that are worth addressing.
A first question  is related to  the stability of such wormhole solutions. In Appendix~\ref{fluctuationstability} we provide a preliminary check of perturbative stability - whether there exist normalisable modes of the fluctuation operator with negative Euclidean energy. The operators of scalar perturbations that we study in our examples are found to have a positive definite spectrum and this has an immediate translation through Holography, to the fact that the dual scalar operators have real scaling dimensions. An issue might arise from modes such as those of the gauge field in the meron-solution as discussed in~\cite{Maldacena:2004rf}. The relevant analysis however has not been completed since the non-abelian gauge field fluctuations are also coupled with metric perturbations. Therefore the fluctuation problem is complex as shown in \cite{Betzios:2017krj}
 and has not been solved.
 Nevertheless, we expect on general grounds that the following condition will still hold: If the dual theory has physical operators with only positive scaling dimensions, then the dual bulk theory will be stable.

Another issue is related to the Osterwalder-Schrader theorem, \cite{Osterwalder:1973dx}, which assumes, among others, positivity of the spectrum and the cluster decomposition property for widely separated operators, and proves the existence of a regular continuation to Minkowski space. In the holographic theories that we study, there is no signal of a violation of cluster decomposition although it is fair to say that this question is subtle because all such theories have a boundary with finite volume. we can however consider the limit of such a volume to become large and therefore ask the question. Similarly in our toy QFT model, cluster decomposition and reflection positivity are intact. We would have expected, from the  Osterwalder-Schrader theorem, that the Euclidean QFT has a sensible Minkowski continuation, but we do not find any.

Finally, we would like to conclude with some interesting comments that appear to have not been discussed in the literature. The first is related to the symmetries of the dual field theory. We consider global symmetries, by taking as a concrete example the simplest case of a $U(1)$ gauge field propagating on a wormhole background having two asymptotic boundaries. One can perform the usual asymptotic analysis along the lines of \cite{Bianchi:2001kw} and discover two locally conserved currents $\partial_\mu \langle J^\mu_{1,2} \rangle = 0$ originating in the subleading term in the expansion of the gauge field near each boundary. Even though this naively would indicate the presence of two independent global $U(1)$ symmetries, there is a single gauge field in the bulk and a single bulk Gauss-Law constraint to be satisfied. What happens in particular is that the two asymptotic fall-off's at the first boundary can be mapped to the other two on the second, using connection formulae that result from the second order bulk ODE. This is also clear since the cross-correlator of charged operators across the two boundaries with matching charges is non-zero. This indicates that there is one and not two independent global symmetries.

 Another issue is related to the gauge groups of the dual field theory system. For Lorentzian wormholes, (the boundaries in that case are separated by a horizon)  there exist two independent gauge groups for the QFT's residing on each asymptotic boundary and therefore a $U(N_1)\times U(N_2)$ gauge symmetry. Here one could envisage a microscopic construction where the strong cross interactions result into the two gauge groups merging in the IR and therefore  only a smaller part remaining intact. This is something that can appear in systems with confining behaviour in the IR such as those exhibiting a cascading behaviour (see the review~\cite{Strassler:2005qs}, and~\cite{Aharony:2009fc} for a system related to ABJM that could describe the meron wormhole of section~\ref{EinsteinYangMillswormhole}).

We therefore conclude that it is of great importance to find a microscopic model that can describe Euclidean wormhole solutions. A simple putative construction that seems in line with our analysis could involve a system of two gauge theories with a $U(N_1) \times U(N_2)$ gauge group, in the presence of a bi-fundamental instanton that connects the two gauge group factors~\footnote{We would like to thank Costas Bachas for suggesting to us this construction.}. We plan to analyse the properties of such models in the future.

 We end by mentioning a few more tangential questions emerging from our investigations.
It is natural to ponder about the possible relevance of such solutions to Cosmology, since one can think of two distinct analytic continuations into Lorentzian signature. The first one is along the radial direction that connects the two boundaries. This generically results in a Big-Bang Big-Crunch Cosmology~\cite{Maldacena:2004rf}. This is badly singular geometry, and does not make sense as a semiclassical solution. The resolution of such Cosmological spacelike singularities in string theory have already been discussed in the past (see~\cite{Elitzur:2002rt,Betzios:2016lne} and references within), but no conclusive picture has yet emerged.

The second possible analytic continuation is along one of the boundary directions (as an example for the case of a compact $S^3$ one obtains a wormhole whose slices would be $dS_3$).
This possibility has issues depending on the specific case at hand. For example in the four dimensional Einstein-Yang-Mills solution, one finds that the gauge field becomes complex, and  it is not clear whether one can put it again into a real section by some gauge transformation. We expect that this is not possible. On the other hand, in the three-dimensional Einstein-Dilaton example, it is not clear how one should analytically continue quotients of the two dimensional hyperbolic space $H_2$ into Lorentzian signature. Finally, the simplest case of $AdS_2$, seems to be in better shape since the analytic continuation exists and one can define both the Euclidean and Lorentzian theories. However, we know that Lorentzian AdS$_2$ cannot exist on its own as it is a sick geometry\cite{Maldacena:1998uz}~\footnote{For the Lorentzian theory one actually needs to consider the case of nearly-$AdS_2$, see the recent works~\cite{KM}.}, and any embedding in a higher dimensional geometry is expected to destroy the wormhole interpretation.
We conclude that it is highly probable that there is no sensible Minkowski continuation for such wormhole geometries in accord with the field theory examples in~\ref{field theory analogue}.

\section{Solutions with two asymptotic regions}\label{sol_asym_reg}

In this section we introduce wormhole solutions with two asymptotic AdS boundaries. They will serve as prototypical examples for our study.
We describe what are the common physical properties they share, as well as point out their differences. More precisely, in the subsection~\ref{EinsteiDilatonwormhole} we present solutions of Einstein - Dilaton theory with spherical and hyperbolic slicings and in~\ref{EinsteinYangMillswormhole} we present solutions of Einstein - Yang - Mills theory in four dimensions with spherical slices. For all the examples, we also introduce coordinates such that the solutions map to compact domains and in particular to metrics that are conformal to the Einstein static universe (ESU) metric, since in these coordinates the analysis of perturbations will take the most transparent form, see section~\ref{HomogeneousODES}.

\subsection{Einstein Dilaton theory}\label{EinsteiDilatonwormhole}

Our first example comes from a study of an Einstein-Dilaton gravitational action of the form
\be
S=\int d^{d+1}x\sqrt{g}\left[R-{1\over 2}(\p\phi)^2-V(\phi)\right]
\label{0}
\ee
with equations of motion
\be
G_{\m\n}-{1\over 2}\p_{\m}\phi\p_{\nu}\phi+{1\over
4}g_{\m\n}(\p\phi)^2+{V\over 2}g_{\m\n} = 0 \, ,
\label{1}\ee
$$
\square \phi - V' = 0 \, .
$$
We study separately the cases of metrics with spherical ($S_d$) or hyperbolic ($H_d$) slicing. We find regular wormhole solutions only in the case of $H_d$ slicing with the slices being compact. More details on Einstein-dilaton theory with examples of solutions with such AdS slicing, are given in~\cite{F}.

\subsubsection{$AdS_{d+1}$ deformations with $S_d$ slicing}

We use the following ansatz for the metric
\be
ds^2=dr^2+e^{2A(r)}~ d\Omega_d^2=dr^2+e^{2A(r)}g_{ij}dy^idy^j
\label{2}\ee
where
$d\Omega_d^2$ the round metric of a unit radius d-sphere, and $A(r)$ governs the size of the transverse space and depends on the holographic coordinate $r$. One can normalise units using the curvature on $S^d$
\be\label{3}
R^{(g)} = \frac{d(d-1)}{\alpha^2} \, ,
\ee
with $\alpha$ the curvature length scale.
Substituting the metric ansatz \eqref{0} in the equations of motion of \eqref{1} we obtain the following set of equations
\be
2(d-1)A''+\phi'^2+{2(d-1) \over \alpha^2 }e^{-2A}=0
\label{4}\ee
\be
d(d-1)A'^2-{\phi'^2\over 2}+V-{d(d-1) \over \alpha^2 }e^{-2A}=0
\label{5}\ee
\be
\phi''+d\A'\phi'- \frac{d V}{d \phi}=0
\label{6}\ee
where primes stand for derivatives with respect to the holographic coordinate $r$.

We observe from \eqref{4} that for a space that shrinks in the throat and grows asymptotically near the two boundaries, the warping factor $A(r)$ should be convex ($A''>0$). Since the last term of \eqref{4}  is positive, then one needs an imaginary $\phi'$ to satisfy this equation, or equivalently a kinetic term for the scalar with the opposite sign ---to provide for negative Euclidean energy---. Such a theory would violate reflection positivity and we therefore disregard such a possibility in the rest. Nevertheless, conformally coupled scalars can give rise to wormhole solutions as found in~\cite{Coule:1989xu,Halliwell:1989ky}. We provide also an explicit demonstration of the pathologies encountered when
\be
V=-{d(d-1)\over \alpha^2}
\ee
 in Appendix~\ref{scalarpathologies}.

\subsubsection{$AdS_{d+1}$ deformations with $H_d$ slicing}\label{Hyperbolicslicing}

To study the hyperbolically sliced case, we use the following ansatz for the metric
\be
ds^2=dr^2+e^{2A(r)}~ dH_d^2=dr^2+e^{2A(r)}g_{ij}dy^idy^j
\label{0a}\ee
where
$dH_d^2$ is the round metric of a d-dimensional hyperboloid,  and $A(r)$ contains the holographic direction dependence of the hyperboloid's radius. One can again normalise units using the curvature on $H_d$
\be\label{2a}
R^{(g)} = -\frac{d(d-1)}{\alpha^2} \, .
\ee
\\
\\
Substituting the metric ansatz in the equations of motion of \eqref{1},  we obtain the following set of equations
\be
2(d-1)A''+\phi'^2-{2(d-1) \over \alpha^2}e^{-2A}=0
\label{4a}\ee
\be
d(d-1)A'^2-{\phi'^2\over 2}+V+{d(d-1) \over \alpha^2}e^{-2A}=0
\label{5a}\ee
\be
\phi''+d\A'\phi'- \frac{dV}{d \phi}=0 \, .
\label{6a}\ee
We observe that the obstruction that arises for the spherical slicing is not present for the hyperbolic slicing, since the last term of \eqref{4a} has the opposite sign and provides a negative Euclidean energy contribution. In the following we search for the simplest solutions with a constant potential of the form
\be
V=-{d(d-1)\over \alpha^2}
\label{7a}\ee
such that the solutions are asymptotically $AdS$ and the size of the space is in units of $\alpha$. Solutions with non-constant potential were studied in~\cite{F}.\footnote{It should kept in mind that such solutions have singularities  at the boundary that need to be resolved, \cite{F}.}  We shall therefore focus on the constant potential case in the rest\footnote{In this case the scalar $\phi$ is essentially an axion field.}.
The solutions of \eqref{6a} and \eqref{5a} are (in a convenient parametrization)
\be \phi'= \sqrt{2d(d-1)}{C\over \alpha} ~e^{-d A}
\label{8a}\ee

\be
A'=\pm {1\over \alpha}\sqrt{1-{\alpha^2 }e^{-2A}+{C^2}e^{-2dA}}
\label{9a}\ee
From now on we restrict to the case of a three-dimensional solution foliated by hyperboloids of $d=2$ dimensions. We shall also use appropriate quotients of the transverse space hyperboloid $\mathcal{M}_T = \mathbb{H}/\Gamma$ with $\Gamma$ a Fuchsian discrete $PSL(2,\mathbb{R})$ subgroup, so as to render the global slice compact. For more details see section~\ref{ODEforEinsteinDilaton}.
\\
\\
In this case, we find the solution
\be
\left(-{\alpha^2 }+2 (e^{2A}+\sqrt{C^2-{\alpha^2 }e^{2A}+e^{4A}})\right)= \alpha^2 e^{2 {r \over \alpha}}
\label{10a}\ee
which gives the following polynomial equation for the scale factor
\be
4e^{2A}=Z+{\tilde C\over Z}+2{\alpha^2}\sp Z\equiv \alpha^2 e^{2 {r \over \alpha}}\sp \tilde C\equiv {\alpha^4 }-4C^2
\label{11aa}\ee
The scale factor vanishes at
\be
Z=-{\alpha^2 } + 2|C|
\label{11ab}\ee
When $r\to\infty$, the metric  to leading order is
\be
ds^2=dr^2+{\alpha^2 \over 4}e^{2 {r\over \alpha}} ~d H_2=dr^2+e^{2 {r \over \alpha}}R_{\rm uv}^2~d H^2_2\sp R_{\rm uv}={1\over 2} \alpha
\label{12a}\ee
Therefore  $R_{\rm uv}$  corresponds to the asymptotic ``size parameter'' of the transverse space.
In the $H_2$ slicing, however, the space shrinks and then turns around at $A'=0$. This happens when
\be
e^{2A_{\pm}}={\alpha^2\over 2}\pm \sqrt{{\alpha^4\over 4 }-C^2}=
{\alpha^2\over 2}\pm {1\over 2}\sqrt{\tilde C}
\label{12ab}
\ee
The first time it happens is at $e^{2A_+}$. The solution for the scalar $\phi$ is
\be
\phi=\phi_0-2 {\rm arctanh}{Z+{\alpha^2}\over 2C}
=\phi_0 + \log{ Z-2C+{\alpha^2 }\over Z+2C+{\alpha^2 }}
\label{13a}\ee
$$
\simeq \phi_0-{4C\over Z}+{\cal O}(Z^{-2})
$$
Therefore, we find that $C$ controls the vev of the operator dual to the scalar.
\\
\\
In the following, we look at the two different cases $\tilde{C} >0$ and $\tilde{C}<0$ as well as the special case $\tilde{C}=1$.

$\bullet$ $\tilde C>0$.
In this case the scale factor never vanishes, but it bounces at
\be
e^{2A_{+}} =
{\alpha^2\over 2 }+ {1\over 2}\sqrt{\tilde C}
\label{14a}
\ee
The second boundary is near $Z\to 0$ (or $r\to -\infty$) where the metric asymptotes to \eqref{12a} but with the slice curvature $ R_{\rm uv}'$ given by
\be\label{15a}
 R_{\rm uv}'=\sqrt{\tilde C} ~ R_{\rm uv}
\ee
The scalar interpolates between $\phi_0$ at the first boundary and
\be\label{16a}
\phi\simeq \phi_1+16{C\over \tilde C}Z+{\cal O}(Z^2)\sp \phi_1=\phi_0+ \log{-2C+{\alpha^2}\over 2C+{\alpha^2\over r^2}}
\ee
$\bullet$ $\tilde{C}=1$.
A special case that we shall examine in more detail in the next sections is when $\tilde{C}=1$. Then the slices acquire the same asymptotic curvature on both sides and we find that
\be\label{17a}
ds^2 = d r^2 + \frac{\alpha^2}{2 } \left(  \cosh \left(\frac{2 r}{\alpha} \right) + 1 \right) dH_2^2
\ee
This is a symmetric wormhole centered around $r=0$. The metric is similar to the ones that we shall encounter in the Einstein Yang-Mills system as well as in global $AdS_2$.
\\
\\
$\bullet$ $\tilde C<0$.
In this case, the scale factor vanishes at an intermediate point where $\phi\to -\infty$. This is a solution with a single boundary and the point where $e^A$ vanishes is a singularity. Moreover on can observe that at this point $(\p \phi)^2$ diverges.

\subsubsection{Einstein Static Universe in Elliptic coordinates}

In this paragraph we perform  a coordinate transformation such that the metric can be brought into a Euclidean Einstein Static Universe form (ESU). Such a transformation allows us to bring the fluctuation equation for the scalar into a Schr\"{o}dinger form where the physical properties of the system are encoded in the potential, (for more details see section~\ref{HomogeneousODES}).
This transformation can be realised parametrising the metric with elliptic functions. This feature is not unique to the hyperbolically sliced three-dimensional wormhole.
Thereafter, for simplicity, we treat in more detail only the special case of symmetric asymptotic radii, since the most general solution is complicated to write in terms of elliptic functions.
The interested reader can find more information about elliptic integrals and functions in Appendix~\ref{Elliptic functions} and in reference~\cite{Zwillinger}.
\\
\\
First, we define $\tilde{r}= r/ \alpha , \, B = 1/2\alpha$. We then perform the following transformation
\be\label{17ab}
y = \frac{\sqrt{2B} \sinh \tilde{r}}{\sqrt{2B \sinh^2 \tilde{r} + B + \half}}\, , \qquad y \in [-1,1]
\ee
in order to bring our space in a compact form. The metric becomes
\be\label{18a}
\frac{ds^2}{\alpha^2} = \frac{(2B+1)}{2} \left(\frac{2 d y^2}{(1-y^2)^2 (4B+ y^2(1-2B))} + \frac{d H_2^2}{(1-y^2)} \right)
\ee
with the two boundaries being at $y = \pm 1$.
One then introduces the elliptic parametrization $k^2 = (B-1/2)/2B < 1$ and $y = \sn (u, k)$ to write
\be\label{elliptichyperbolic}
\frac{ds^2}{\alpha^2} = \frac{B+ \half}{\cn^2 u} \left(  \frac{du^2}{2B} + d H_2^2 \right)\, , \qquad u \in [-K(k), K(k)]
\ee
where the boundaries are at $\pm K(k)$. This mapping is plotted in~\ref{fig:UHP} in Appendix~\ref{Elliptic functions}. Before further discussing properties of this solution, we first analyze the case of Einstein Yang-Mills, since we shall find that there exists a wormhole solution with an ESU metric \eqref{ellipticcoords2} that is very similar to the one found here, the essential difference being that the transverse space will now be an $S^3$.

\subsection{The Einstein Yang Mills system}\label{EinsteinYangMillswormhole}

In this subsection, we describe four-dimensional Euclidean wormholes with two asymptotic boundaries as solutions of the  Einstein Yang-Mills system with action
\begin{equation}\label{1b}
S=\int d^4 x\sqrt{g} \left(-\frac{1}{16{\pi}G}R + \Lambda +\frac{1}{4  g_{YM}^2 }\left(F^{a}_{\mu\nu}\right)^2\right)
\end{equation}
where $R$ is the Ricci scalar, $\Lambda$ is the cosmological constant and $F^{a}_{\mu\nu}$ is the field strength for an $SU(2)$ gauge field $A^{a}_{\mu}$. The equations of motion and our conventions can be found in Appendix~\ref{EYMconventions}.

This theory admits a wormhole solution in all the cases of zero, positive and negative cosmological constant~\cite{Hosoya:1989zn}. This solution was analyzed in the context of the holography of Euclidean wormholes in \cite{Maldacena:2004rf}. The metric of these wormhole solutions takes the form
\be\label{4b}
ds^2 = dr^2 + e^{2 w(r)} d \Omega_3^2 \, ,
\ee
with
\bea\label{5b}
\qquad e^{2 w(r)} &=& \half -  B \cos (2 r) \, , \qquad B=\sqrt{\frac{1}{4}- r_0^2 H^2}\, , \, \quad (\text{dS}) \, \nn \\
\qquad e^{2 w(r)} &=& B \cosh (2 r) - \half \, , \qquad B=\sqrt{\frac{1}{4} - r_0^2 H^2}\, , \, \quad (\text{AdS})
\eea
where $r_0^2 = 4 \pi G_N/g_{YM}^2$ and $H^2= 8 \pi G \Lambda /3\, $ with $\Lambda < 0 ,$ $\Lambda>0$ for the cases of AdS/dS.
We observe that the two cases are related by changing the sign of the cosmological constant $\Lambda$ and the size of the wormhole throat remains positive for $B \leq \half$ or $B \geq \half$ respectively. From here on we focus on the AdS case ($\Lambda<0$). The geometry is conformally flat and has two asymptotic boundaries, one at $r\to\infty$ with
\be\label{8b}
e^{2 w}\simeq {B \over 2}e^{2 r} \ee
and one at $r\to-\infty$ with
\be\label{8b1}
e^{2 w}\simeq {B \over 2}e^{- 2 r}\ee
Therefore, the space-time approaches AdS and the three-sphere slices have the same asymptotic radii.
The throat connecting them obtains a minimum size equal to
\be\label{9b}
r_{min}^2 = B - \half .
 \ee
For $B=\half$. the two sides pinch off in a smooth fashion and the metric becomes exactly that of two disconnected Euclidean $AdS_4$ ($EAdS_4$) copies
\be\label{10b}
ds^2_{EAdS} = dr^2 + \sinh^2 r d \Omega_3^2\, , \qquad r \in [0, \infty) \, .
\ee
We notice that in order to achieve this one needs to send $g_{YM} \rightarrow \infty$.
\\
\\
This wormhole background also involves a non-trivial non-abelian magnetic field configuration also known as the meron~\cite{deAlfaro:1976qet,Callan:1977gz} whose magnetic flux is responsible for supporting the wormhole throat.

In order to understand this configuration, it is useful to define the metric on the three sphere in terms of Euler angles as
\begin{equation}\label{6b}
 d\Omega^2_3 = {1 \over 4} \biggl( d t_1^2 + d t_2^2 + d t_3^2 + 2 \cos\, t_1 \, d t_2 d t_3   \biggr) =  \frac{1}{4} \omega^a \omega^a \, ,
\end{equation}
where $\omega^a$ are the Maurer-Cartan (left-invariant) forms on $S^3$ and $t_i$ the Euler angles, for more details see Appendix~\ref{MaurerHopf}.
The non-trivial gauge field then can be described as ($A^a = A^a_\mu dx^\mu$)
\be\label{7b}
A = \half g^{-1} d g \, , \qquad  \text{or} \quad A^a = \half \omega^a  \, , \qquad  \text{with} \quad F^a =  \frac{1}{8} \epsilon^{a b c} \omega^b \wedge \omega^c \, ,
\ee
with $g$ an $SU(2)$ group element. This is the \emph{meron} configuration, which provides the appropriate magnetic flux to support the throat from collapsing~\footnote{We should also note that there exists also an anti-meron wormhole solution for which $A =\half g d g^{-1}$, which is expressed in terms of right-invariant forms. It would be interesting if there exists a solution that asymptotes to a meron on the first boundary and to an antimeron on the second one.}. In order to understand better the holographic interpetation of this solution, we pick a radial gauge $A^a_r = 0$ and use the explicit expression for the Maurer-Cartan forms given in~\eqref{m12} to read off the components of the $SU(2)$ gauge field from the matrix
\be\label{gauge_f}
A^{a}_{\mu}=
\begin{pmatrix}
0\,\, & \frac{1}{2}\cos t_{3}\,\, & \frac{1}{2}\sin t_{3}\, \sin t_{1}\,\, & 0\\
0\,\, & \frac{1}{2}\sin t_{3}\,\, & -\frac{1}{2}\cos t_{3}\, \sin t_{1}\,\, & 0\\
0\,\, & 0\,\, & \frac{1}{2}\cos t_{1}\,\, & \frac{1}{2}
\end{pmatrix}
\ee
where the index $a = 1,2,3$ labels the rows of the matrix and is the gauge group index and $\mu$ is the bulk space index that labels the columns of the matrix. One can notice that the gauge field is independent of the radial direction. This corresponds to having a constant magnetic source turned on in the dual theory. This source breaks explicitly the global $SU(2)$ defined on each boundary. In addition if we compute the total topological charge on the wormhole manifold $\mathcal{M}_4$ it reduces to a sum of two integrals over the boundaries of the topological current $J_\mu^T$
\bea\label{gauge_g}
q_T &=& \int_{\mathcal{M}_4} \Tr ( \epsilon^{\mu \nu \rho \sigma} F_{\mu \nu}  F_{\rho \sigma}) = \oint_{\sum_{i=1}^2 S_i^3} d \Omega_\mu  J_\mu^T  \, \nn \\
&=& \oint_{\sum_{i=1}^2 S_i^3} d \Omega_\mu \Tr \epsilon^{\mu \nu \rho \sigma} \left( A_\nu \partial_\rho A_\sigma + \frac{2}{3} A_\nu  A_\rho  A_\sigma \right) = 0 \, .
\eea
This means that the total topological charge of our configuration is zero because the integral of the current on the first asymptotic boundary cancels the one from the other due to the opposite sign of the normal vector perpendicular to the two boundaries. If the topology was trivial (for a pure meron) we would instead have found half a unit of topological charge $q_T = 1/2$, due to the integral on a single three-sphere at infinity.
A further analysis can be performed for each fixed $S^3$ in the radial gauge, if we use the Hopf fibration $S^{1}\hookrightarrow S^{3}{\xrightarrow {\ p\,}} \, S^{2} $ that describes the three sphere as an $S^1$ bundle over $S^2$, see Appendix~\ref{MaurerHopf}. We can then compute the magnetic flux along various submanifolds of $S^3$ and in particular we find that the only non-zero result is
\be\label{gauge_i}
\Phi = \int_{S^2} \Tr F = \frac{1}{4}\int_0^{\pi} dt_1 \int_0^{2 \pi} dt_2 \left(\sin t_1 + (\cos t_3 - \sin t_3) \cos t_1 \right)  = \pi \, .
\ee
This means that there is a constant magnetic flux piercing the $S^2$ pointing along the fiber coordinate $t_3$. Since this coordinate is also periodic we have closed magnetic lines on each $S^3$ following the fibers. This constant magnetic flux is responsible for supporting the throat. Using this result we also find that the meron has first Chern-number $c_1= \Phi/2 \pi = 1/2$, unlike the magnetic monopoles for which it is an integer. For an anti-meron the topological numbers and the flux pick a minus sign (it reverses orientation along the fiber).


\subsubsection{ESU elliptic representation}\label{ESUellipticmetric}

To bring the metric in ESU form, first we perform the following coordinate transformation
\be\label{11b}
y = \frac{\sqrt{2B} \sinh r}{\sqrt{2B \sinh^2 r + B - \half}}\, , \qquad y \in [-1,1]
\ee
One then finds
\bea\label{compactcoords}
ds^2 &=& \left(B - \half \right) \left[ \frac{d y^2}{2B (1-y^2)^2 (1- \frac{B+\half}{2B}y^2)} + \frac{d \Omega_3^2}{1-y^2} \right] \, \nn \\
&=& \left(B - \half \right) \left[ \frac{d \tilde{y}^2}{(1- 2B \tilde{y}^2)^2 (1- (B+\half) \tilde y^2)} + \frac{d \Omega_3^2}{1-2B \tilde{y}^2} \right] \,
\eea

Defining the elliptic modulus $k^2 = (B+1/2)/2B < 1$ and the coordinate transformation $y= \sn (u, k)$ the metric can be written as
\be\label{ellipticcoords2}
ds^2 = \frac{(B-\half)}{\cn^2 (u,k)} \left(\frac{du^2}{2B} + d \Omega_3^2 \right),\quad u \in [-K , K]
\ee
with $K$ the elliptic period. We notice the similarity with the Einstein-dilaton wormhole~\eqref{elliptichyperbolic}. Their conformal factors are related with a simple change of elliptic modulus to the complementary one.

The periodic structure of the metric points to a possibility of describing fragmented disjoint manifolds~\footnote{One might imagine some deformation that can connect all these disconnected geometries similar to the deformations studied in~\ref{Doubletracedeformations}. This would be a multi-wormhole background.}.
Taking the limit $B\rightarrow \half, k\rightarrow 1$, one obtains the global ${EAdS_{4}}$ metric
\be\label{AdSESU}
ds^2_{GAdS} = \frac{1}{ \sinh^2 (u)} \left(du^2 + d \Omega^2_3 \right)\, , \qquad u \in [-\infty, 0]
\ee
which now covers another part of the Euclidean ESU. The UV boundary is now at $u=0$ and the IR at $u= - \infty$.
\\
In the limit where $k \rightarrow 0 \,,\; B \rightarrow -1/2$, then $\cn(u, k) \rightarrow \cos u$ and one obtains the ESU form of a flat space wormhole ($\Lambda = 0$).

As a final coordinate description, we can also bring the metric to a manifestly Fefferman-Graham (FG) form. To do that one transforms $\tilde{u}=u-K$ and then
\be\label{13b}
z^2 = \frac{1- \cn \tilde{u} }{1+ \cn \tilde u}\, , \qquad z \in [0, \infty)
\ee
to obtain
\be
ds^2_{FGworm} = \frac{dz^2}{z^2} + \frac{1}{4} \left(2- \frac{(\half + B)^2}{B^2} + \frac{1}{z^2}+ z^2 \right) d \Omega_3^2
\label{FGformWorm}
\ee
This is an explicit FG form with the two boundaries at $0, \infty$ (the metric is invariant under the inversion $z \rightarrow 1/z$ that exchanges the two boundaries). When we send $B\to 1/2$   we find the global AdS FG metric
\be
ds^2_{FGAdS} = \frac{dz^2}{z^2} + \frac{1}{4} \left( \frac{1}{z} - z \right)^2 d \Omega_3^2
\label{FGformAdS}
\ee
In this limit the UV is at $z=0$ and the IR is at $z=1$ ---which was the point of inversion symmetry for the wormhole metric~\eqref{FGformWorm}--- where the sphere shrinks to zero size.
In this limit, the solution factorizes in two AdS spaces.

\subsection{The special case of $AdS_2$}\label{AdS2_metr}

In this subsection we study the case of global $AdS_2$, which, unlike higher-dimensional AdS spaces,  has two boundaries both in Lorentzian and Euclidean signature. It  shares features with both the three-dimensional Einstein dilaton hyperbolic wormhole and the Einstein - Yang - Mills wormhole. The global $EAdS_2$ metric can be written as
\be\label{AdS2metric}
ds^2 = \frac{d u^2 + d \tau^2}{\cos^2 u}
\ee
with $u \in (- \pi/2 , \pi/2)$ and $\tau \in (- \infty, \infty)$. The metric can be analytically continued either in $u$  resulting in a Bang-Crunch universe or in $\tau$ resulting in the usual Lorentzian $AdS_2$. In the latter case there are still two asymptotic regions. We call this solution special since $AdS_2$ is a two-dimensional gravity solution and gravity in two dimensions is non-dynamical. In such case one needs a more elaborate model with extra fields (for example the Jackiw-Teitelboim gravity or an Einstein-Maxwell-Dilaton (EMD) model) to support the $AdS_2$ solution , \cite{Tei,J}. This is similar in spirit with our higher-dimensional wormholes that have non-trivial backgrounds of scalars or gauge fields and two boundaries. For the case at hand, there exist both possibilities. The first one is the case of a non-trivial dilaton, which has been analysed with detail in the old and recent works by~\cite{Maldacena:1998uz,KM,Almheiri:2014cka}. In that case one finds that while the pure $AdS_2$ Lorentzian background is unstable and does not support finite energy excitations, one can still make sense of nearly-$AdS_2$ backgrounds with the conformal symmetry broken both spontaneously in the IR as well as explicitly from UV-effects as exemplified by the studies of the SYK model, \cite{KM,JE}. In these studies the $AdS_2$-dilaton gravity captures the IR physics of the model and the running dilaton encodes the breaking of the conformal symmetry.

To make an analogy with our higher-dimensional example of the meron wormhole, we use the construction of Papadimitriou - Cveti\v{c}~\cite{Cvetic:2016eiv}, who concluded that there can be non-trivial solutions of the EMD theory with constant dilaton, but with a running gauge field, much in line to wormhole solutions encountered in higher dimensions (wormholes supported either by p-forms~\cite{Giddings:1987cg} or the meron solution studied here). The generic action of the two-dimensional EMD theory is
\be\label{1c}
S= \frac{1}{2 \kappa_2^2} \left( \int_{\mathcal{M}} d^2 x \sqrt{g} e^{- \phi_D} \left(R[g] + \frac{2}{L^2} - \frac{1}{4} e^{- 2 \phi_D} F_{\mu \nu} F^{\mu \nu} \right)  \, + \, \int_{\partial \mathcal{M}} \sqrt{\gamma} e^{-\phi_D} 2 K \right)
\ee
where the dilaton is denoted by $\phi_D$. The most general Euclidean solution with constant dilaton takes the form
\bea\label{2c}
e^{-\phi_D} & = & C \, , \qquad ds^2 = dr^2 + \left(a(\tau) e^{r} + \frac{b(\tau) e^{- r}}{C} \right)^2 d \tau^2 \nn \\
A_\tau & = & \mu(\tau) - \frac{1}{C}\left( a(\tau) e^{r} - \frac{b(\tau) e^{-r}}{C} \right)\, ,
\eea
with $a(\tau),b(\tau)$ arbitrary functions. This is an analytic continuation of the general Lorentzian solution presented in~\cite{Cvetic:2016eiv}, in Euclidean signature. The simplest solution in this class of solutions that still admits a Euclidean wormhole interpretation is the one for which the metric is just $ds^2= d r^2 + \cosh^2 r d \tau^2$ and the gauge field is $A_\tau = \mu - Q \sinh r $ with $Q$ having the interpretation of a conserved charge~\cite{Cvetic:2016eiv} $\partial_\tau Q = \partial_r Q = 0$~\footnote{The asymmetric solutions generalise the higher-dimensional case of hyperbolic slicing where the geometry can also be asymmetric near the two boundaries, but in the two-dimensional case there can also be non-trivial dependence in the transverse coordinate $\tau$.}. In this case one can simply transform the metric using $u = 2 \arctan \tanh r/2$ which is 1-1 for $u \in [-\pi/2,\pi/2]$ to obtain the global metric~\eqref{AdS2metric} along with a non-trivial gauge field $A_\tau = \mu - Q \tan u $. This is the two dimensional wormhole background on which we compute Holographic correlators in~\ref{AdS2correlators}.

Even if this solution is quite simple, there exist several aspects that make a holographic interpretation quite subtle. Notice then that asymptotically near the boundaries $u = \pm \pi/2$ the gauge field and the metric become
\be\label{3c}
A_\tau \sim \mu - \frac{Q}{\frac{\pi}{2} \mp u}\, , \qquad ds^2 \sim \frac{d \tau^2 + du^2}{\left(\frac{\pi}{2} \mp u \right)^2} \, .
\ee
One then finds that the leading mode is the one corresponding to the conserved charge (while both modes turned on are normalisable). Without going into the details that can be found in~\cite{Cvetic:2016eiv}, there are two options for interpreting such a solution. Either to consider $\mu$ as a source for the dual conserved local charge $Q$, or $Q$ as a source of the non-local Polyakov operator $\int d \tau \mu(\tau)$. In this later case $\mu(\tau)$ should be thought of as a dynamical gauge field for the boundary theory. In this work we choose the first option, from here on.

We have now finally introduced all the gravity solutions that we work with in the following sections of our paper. Namely, the three-dimensional Einstein - dilaton hyperbolic wormhole, the four-dimensional Einstein - Yang - Mills wormhole and the Einstein - Maxwell - dilaton global $AdS_{2}$ solution. We shall now turn to the study of correlation functions of scalar operators on such wormhole backgrounds.

\section{Correlators in a saddle-point  with multiple boundaries}\label{multipleboundariescorrelators}

A central  observable in holography is the set of correlation functions of boundary operators that are dual to bulk fields. In this paper we focus on scalar operators. In this section we provide an abstract and generic construction that allows to compute correlation functions in theories with multiple boundaries, and we supplement it with concrete examples in the case of two boundaries.
\\
\\
The action of a scalar field $\phi$ with mass $m$ in a space $\mathcal{M}$ with boundaries $\partial \mathcal{M} = \cup_i (\partial \mathcal{M})_i$ can be written as
\be\label{1d}
S[\phi] =\half \int_{\mathcal{M}} d^{d+1} x \sqrt{g} ~\phi \left(-\Box + m^2 \right) \phi -  \half  \int_{\partial \mathcal{M}} d^d x \sqrt{\gamma} ~\phi ~\vec{n} \cdot \partial \phi
\ee
where $\gamma_{ij}$ the induced metric on the boundaries and $\vec{n}$ the normal unit vector pointing outwards of the boundaries.
The bulk piece vanishes on-shell. We define the Bulk-to-Bulk propagator (BtB) or Green's function $G$
\be\label{2d}
(-\Box + m^2)G(u,\Omega ; u', \Omega') = \frac{1}{\sqrt{g}} \delta(u-u')\delta^d(\Omega- \Omega')
\ee
with $u$ a radial coordinate and $\Omega$ transverse coordinates. Here one has made an implicit choice of Dirichlet $G \rightarrow 0$ or Neumann $n \cdot \partial G \rightarrow 0$ boundary conditions on each of the respective boundaries.
The BtB propagator being the Euclidean Green's function can be constructed out of the normalisable bulk solutions of
\be\label{3d}
(-\Box + m^2) \phi_n = \lambda_n \phi_n \, , \qquad G(u,\Omega ; u', \Omega') = \sum_n \frac{\phi_n(u,\Omega) \phi_n(u',\Omega')}{\lambda_n} \, ,
\ee
assuming that the $\phi_n$ form a complete basis\footnote{For Neumann bc's one obtains an extra constant term in the definition of the Green's function arising from the zero mode.}. The summation could also be an integral depending on whether the spectrum is discrete or continuous.

One then defines the boundary to Bulk (btB) propagators as a limiting case of the BtB propagator sending one of its points towards one of the boundaries ---denoted by $u_i^*$. From now on we mainly focus on the Dirichlet case (for which $G, \, \phi_n \rightarrow 0$ at the boundaries)
\be\label{4d}
K^i(u,\Omega ; \Omega') = \lim_{u' \rightarrow u_i^* } \frac{\vec{n} \cdot  {\partial}' G(u,\Omega ; u', \Omega')}{\sqrt{\gamma_i}}
\ee
with $\sqrt{\gamma_i}$ the induced metric on the respective boundary and $\vec{n}$ the unit normal outwards pointing vector at the boundary.

If we now assume that the solution has the two asymptotic falloffs $\sim u^{\Delta^i_\pm}$ near each of the boundaries,
the btB propagators also satisfy
\bea\label{btBdfn}
&& (-\Box + m^2) K^i (u,\Omega ; \Omega') = 0\, , \nn \\
&& K^i (u,\Omega ; \Omega')|_{\partial \mathcal{M}_i} = \epsilon_i^{\Delta^i_-} \delta_{\epsilon_i} (\Omega - \Omega') \, , \nn\\
&& K^i (u,\Omega ; \Omega')|_{\partial \mathcal{M}_j} = 0 \, , \quad \text{for} \, i \neq j \nn \\
&& \phi(u, \Omega) = \sum_i \int d^d {\Omega}'  K^i (u,\Omega ; \Omega') \phi^s_i (\Omega')
\eea
where $\epsilon_i$ is a small parameter that represents the cutoff near each boundary. The last formula gives the reconstruction of a bulk field from given sources ($\phi^{s}_{i}$) at the boundaries and is based on the linearity of the fluctuation equation.

Notice that $\Delta_i$ can be different, in general, at different boundaries since the sizes of the asymptotic AdS regions can be different. Such a possibility has appeared for both the $3d$ Einstein Dilaton wormhole and the $AdS_2$ solution in eqns.~\eqref{15a} and~\eqref{2c}.

Two useful formulae that give away the scaling of the propagators near the boundaries are
\be\label{6d}
\epsilon_i \partial_{\epsilon_i} G(u,\Omega ; \epsilon_i, \Omega') = \Delta^i_+ G(u,\Omega ; \epsilon_i, \Omega')\, , \qquad \epsilon_i \partial_{\epsilon_i} K^i (\epsilon_i ,\Omega ; \Omega') = \Delta^i_- K^i (\epsilon_i ,\Omega ; \Omega')\;.
\ee
Hence the btB propagators have a leading asymptotic behaviour since they are capturing the excitations due to a $\delta$-function  source at any given boundary, while the BtB has a subleading behaviour since it is made out of normalisable solutions (as a Green's function) that vanish at the boundaries.
Another important equation is the second Green's identity
\be\label{7d}
\int_{\mathcal{M}} d^{d+1} x \sqrt{g}\left[ \psi  (\Box-m^2) \phi - \phi (\Box - m^2) \psi \right] = \int_{\partial \mathcal{M}} d^d x \sqrt{\gamma} \left[  \psi \vec{n} \cdot \partial \phi - \phi \vec{n} \cdot \partial \psi  \right]\, ,
\ee
because if $\phi$ is a solution of the homogeneous equation $(-\Box+m^2) \phi= 0$ and $\psi = G$ one finds
\be\label{8d}
\phi(u, \Omega) = \int_{\partial \mathcal{M}} d^d x \sqrt{\gamma} \left[  G n^u \partial_u \phi - \phi n^u  \partial_u G  \right] \, .
\ee
This indicates that one can use either Dirichlet $G=0$, Neumann $n^u \partial_u G = 0$ or mixed boundary conditions to reconstruct the bulk field. This choice is to be done independently for all the boundaries as we already mentioned.
Using the same identity one can relate the BtB and btB propagators by setting $\phi = K^i$ and $\psi = G$,
\be\label{9d}
K^i (u, \Omega ;\Omega' ) = \lim_{\epsilon_i \rightarrow 0} \frac{2 \Delta^i_+ - d}{\epsilon_i^{\Delta^i_+}} G(u,\Omega ; \epsilon_i , \Omega')
\ee
Comparing with~\eqref{4d} this relation is simpler and does not involve any derivatives. It also shows how to enhance the behaviour of the BtB propagator near one of the boundaries, so that to isolate a source part $K \sim \epsilon_i^{\Delta^i_-}$.
\\
\\
If the topology of the manifold allows it (for example in the case of $\mathcal{M} = \mathbb{R} \times \mathcal{M}_T$), one could expand the transverse coordinates into Fourier modes (or discrete harmonics in case of a compact manifold) and keep the radial coordinate in position space. Then, the normalised btB propagators simply become
\be\label{10d}
f^i(p, u) = \epsilon_i^{\Delta^i_-} \frac{\sqrt{\gamma_i(\epsilon_i)}}{\sqrt{\gamma_i(u)}} \frac{K^i(p, u)}{K^i(p, \epsilon_i)}
\ee
with $K^i(p, u)$ satisfying the homogeneous fluctuation ODE and $p$ is a Fourier space index for the transverse coordinates.

The transverse Fourier space modes are also useful for an alternative computation of the BtB propagator via the Wronskian method where only solutions to the homogeneous equation are needed.

In the case of two compact $S^3$ boundaries, one expands $\phi$ in terms of the $S^3$ harmonics $\phi(u, \Omega) =\sum_{j_1, k , p} \xi^{j_1}(u) Y_{j_1 k p}(\Omega) $ to find a second order ODE for $\xi^{j_1}(u)$ with two solutions $\xi^{j_1}_{1,2}(u)$. Then the BtB Green's function is given by the mode sum
\be\label{11d}
G(u,\Omega ; u', \Omega') =\sum_{j_1=0}^\infty \left[\Theta(u' - u) \frac{\xi^{j_1}_{1}(u) \xi^{j_1}_{2}(u')}{W(\xi_{1}, \xi_{2})} - (u \leftrightarrow u') \right] \sum_{k=0}^{j_1} \sum_{p=-k}^k Y_{j_1 k p}(\Omega)  Y_{j_1 k p}(\Omega')
\ee
with $W(\xi_{1}, \xi_{2})$ the Wronskian of the solutions of the radial ODE.
The simplest summation to perform is
\be\label{12d}
\sum_{k=0}^{j_1} \sum_{p=-k}^k Y_{j_1 k p}(\Omega)  Y_{j_1 k p}(\Omega') = \frac{j_1+1}{2 \pi^2} \frac{\sin (j_1+1)\Delta \Omega}{\sin \Delta \Omega} \, ,
\ee
with $\Delta\Omega = \Omega - \Omega'$ the distance between two points on the $S^3$.
In the end we are left with a single summation
\bea
& G(u,\Omega ; u', \Omega') = \sum_{j_1=0}^\infty \left[\Theta(u' - u) \frac{\xi^{j_1}_{1}(u) \xi^{j_1}_{2}(u')}{W(\xi_{1}, \xi_{2})} - (u \leftrightarrow u') \right] \frac{j_1+1}{2 \pi^2} \frac{\sin (j_1+1)\Delta \Omega}{\sin \Delta \Omega} \nn \\
& = -\frac{1}{2 \pi^2 \sin \Delta \Omega}  \frac{\partial}{\partial \Delta \Omega} \sum_{j_1=0}^\infty \left[\Theta(u' - u) \frac{\xi^{j_1}_{1}(u) \xi^{j_1}_{2}(u')}{W(\xi_{1}, \xi_{2})} - (u \leftrightarrow u') \right] \cos (j_1+1)\Delta \Omega \, , \nn \\
\label{BtBcorrelator}
\eea
in terms of the radial ODE modes $\xi^{j_1}_{1,2}(u)$.
\\
\\
We now turn to the boundary correlation functions. The correlation functions are given by boundary to boundary propagators (btb) as double limits of the BtB propagator (or as single limits of the corresponding btB propagator)
\be
\langle \mathcal{O}_i (\Omega) \mathcal{O}_j (\Omega') \rangle =(2 \Delta^{i}_+ - d)(2 \Delta^{j}_+ - d) \lim_{\epsilon_i , \epsilon_j \rightarrow 0} (\epsilon_i)^{- \Delta_+^i}  (\epsilon_j)^{- \Delta_+^j} G(\epsilon_i,\Omega ; \epsilon_j , \Omega')
\label{CorrelatorsBTB}
\ee
One then finds several correlation functions for a common or different boundaries (cross correlators).
In particular, in the case that the problem can be reduced to a radial ODE, the correlators are fully determined if one is given the connection formulae (global monodromy) of the bulk ODE.

Another approach to compute correlation functions involves the on-shell action. To properly define the on-shell action one needs a set of cutoffs for each boundary and the addition of the appropriate counterterms on all the boundaries. In the case of a free scalar the leading counterterm is a mass term for the scalar of the form
\be
S_c \, = \, {\Delta^i_+\over 2}  \int_{(\partial \mathcal{M})_i} d^d x \sqrt{\gamma} ~ \phi^2 \, ,
\ee
and one should include such terms on all the boundaries.
We leave a detailed analysis of holographic renormalisation in the case of multiple boundaries for the future.

In terms of the previously defined quantities, the on-shell action becomes
\bea
S[\phi]^{(on-shell)} &=& -  \half  \int_{\partial \mathcal{M}} d^d x \sqrt{\gamma} \phi \vec{n} \cdot \partial \phi = -  \half  \int_{\partial \mathcal{M}} d^d x \sqrt{\gamma} \phi n^u  \partial_u \phi \nn \\
&=&  -\half
\sum_i \sum_j \sum_k  \int_{\partial \mathcal{M}_j} d^d \Omega_1 \sqrt{\gamma_j} \int_{\partial \mathcal{M}_k} d^d \Omega_2  \sqrt{\gamma_k} \phi^s_j (\Omega_1)  \mathcal{F}^{j k}_{\epsilon_i}(\Omega_1, \Omega_2) \phi^s_k ({\Omega_2}) \nn \\
\mathcal{F}^{j k}_{\epsilon_i}(\Omega_1, \Omega_2) &=& \left[ \int_{\partial \mathcal{M}_i} d^d \Omega \sqrt{\gamma_i}  K^j (u, \Omega ;\Omega_1 ) n_i^u \partial_u K^k (u, \Omega ;\Omega_2 ) \right]_{u \rightarrow \partial \mathcal{M}_i } \, .
\label{Onshellaction}
\eea
One of the summations above comes from the integration over all the boundaries and the extra two from the expansions of the field $\phi$ in terms of btB propagators, see~\eqref{btBdfn}. Nevertheless, even though it naively seems that we need $n^3$ terms with $n$ the number of boundaries, the symmetry of the problem typically leads to less terms. In particular for the case of two symmetric boundaries one just has a correlator on a single boundary and a cross-correlator among the two boundaries as can be verified from the symmetry of~\eqref{Onshellaction}.
In this case the on-shell action is simply given by four terms (using $n_u = \frac{1}{\sqrt{g_{uu}}}$ and $\epsilon_{1,2}= \epsilon$ - the same cutoff on both sides)
\be
S[\phi] =  -\half \int_{\partial \mathcal{M}_1} \frac{d^d \Omega_1}{\epsilon^d}  \int_{\partial \mathcal{M}_1} \frac{d^d \Omega_2}{\epsilon^d} \phi^s_1 (\Omega_1)  \mathcal{F}^{1 1}_{1}(\Omega_1, \Omega_2) \phi^s_1 ({\Omega_2})\,  +  (1 \leftrightarrow 2) -
 \label{13d} \ee
 $$
 - \half \int_{\partial \mathcal{M}_1} \frac{d^d \Omega_1}{\epsilon^d}  \int_{\partial \mathcal{M}_2} \frac{d^d \Omega_2}{\epsilon^d} \phi^s_1 (\Omega_1)  \mathcal{F}^{1 2}_{1}(\Omega_1, \Omega_2) \phi^s_2 ({\Omega_2})\, -(1 \leftrightarrow 2)
$$
\be
\mathcal{F}^{1 1}_{1}(\Omega_1, \Omega_2) = \left[ \epsilon^{-\Delta_+} \epsilon  \partial_\epsilon K^1 (\epsilon, \Omega_1 ;\Omega_2 ) \right]_{\partial \mathcal{M}_1}\,,
\label{14d} \ee
 \be
 \quad \mathcal{F}^{1 2}_{1}(\Omega_1, \Omega_2) = \left[ \epsilon^{-\Delta_+} \epsilon  \partial_\epsilon K^2 (\epsilon, \Omega_1 ;\Omega_2 ) \right]_{\partial \mathcal{M}_1}\, ,
\label{Onshellaction2}
\ee
We notice that $\mathcal{F}^{1 1}_{1}$ will give the usual correlator for a single side, while $\mathcal{F}^{1 2}_{1}$ will give the cross correlator. This second term is less trivial to compute, since one needs the limit where the points of the BtB propagator stretch across two separated boundaries. For completeness we also present the on-shell action in momentum space variables,
\bea
S[\phi] = &-& \half \int_{\partial \mathcal{M}_1} \frac{d^d k}{(2 \pi)^d}  \phi^s_1 (k)  \mathcal{F}^{1 1}_{1}(k, \epsilon) \phi^s_1 (-k)\,  +  (1 \leftrightarrow 2) \nn \\ &-&  \half \int_{\partial \mathcal{M}_1} \frac{d^d k}{(2 \pi)^d}  \phi^s_1 (k)  \mathcal{F}^{1 2}_{1}(k, \epsilon) \phi^s_2 (-k)\, + (1 \leftrightarrow 2)
\label{Onshellactionmomentum}
\eea
\be
\mathcal{F}^{1 1}_{1}(k, \epsilon) = \left[ u^{-d} u  \partial_u f^1_k(u) \right]^{u=\epsilon}_{\partial \mathcal{M}_1}\, ,
\label{15d}\ee
 \be
\mathcal{F}^{1 2}_{1}(k, \epsilon) = \left[ u^{-d} u  \partial_u f^2_k(u) \right]^{u=\epsilon}_{\partial \mathcal{M}_1}\, .
\label{16d}\ee
To summarise,  we presented  the formalism on how to extend the usual holographic computation of BtB, btB and btb propagators in spaces with multiple boundaries. We shall now move to the study of the homogeneous fluctuation ODEs and in the explicit calculation of scalar correlators in our respective examples.

\section{Homogeneous ODEs and correlators}\label{HomogeneousODES}

In this section we use the formalism introduced in section \ref{multipleboundariescorrelators} to compute boundary correlators for the gravity solutions introduced in section \ref{sol_asym_reg}.

To study correlators for boundary operators we just need to study the homogeneous fluctuation equation. To bring the fluctuation equation into a Schr\"{o}dinger form the appropriate coordinates are the ones that bring the metric into conformal Einstein static universe form, since only in these coordinates the scale between the radial and transverse part is fixed.
In particular, for a metric of the form
\be\label{1e}
d s^2 = e^{2 \Omega} \left(du^2 + ds_{d}^2 \right)\, , \qquad ds_{d}^2 = g_{i j} dx^i dx^j
\ee
where $g_{i j}$ is independent of $u$, one finds the scalar fluctuation equation
\be\label{Genericfluctuationeqn}
\left( \partial_u^2 + \Box_g \right) \phi + (d-1) \Omega' \phi' - m^2 e^{2 \Omega} \phi = 0
\ee
or upon rescaling $\Psi = e^{(d-1)\Omega/2} \phi$,  the Schr\"{o}dinger form
\be\label{2e}
-\Psi'' + \left(m^2 e^{2 \Omega} + \frac{d-1}{2} \Omega'' + \left(\frac{d-1}{2} \Omega' \right)^2 \right) \Psi =  \Box_g \Psi = \lambda \Psi
\ee
To compare with the single boundary case, as well as a warm-up, the reader can first consult appendix~\ref{AdSODE}, where we treat the case of global AdS with a single boundary. To define the correlators, it is then important to discuss the possible choices of boundary conditions. A generic feature of our solutions in contrast with the global AdS case~\eqref{AdSGlobalE=0} where there is only one UV boundary, is that we now have two boundaries, where the solution can potentially diverge or become a constant.
In AdS, by imposing a regularity condition in the IR, we automatically fix a linear combination of the two solutions near the boundary, so there is only a one parameter freedom left.
The extra freedom in the two boundary case, is allowing the possibility to define two types of correlation functions, one on a single boundary which we label by $\langle \mathcal{O}_1 \mathcal{O}_1 \rangle$ or $\langle \mathcal{O}_2 \mathcal{O}_2 \rangle$,  and one cross-correlator across the two boundaries $\langle \mathcal{O}_1 \mathcal{O}_2 \rangle$ according to the notation of the generic discussion in~\ref{multipleboundariescorrelators}. We now proceed to study our respective examples in more detail.

\subsection{$AdS_2$}\label{AdS2correlators}

The simplest case with two boundaries is this of $AdS_2$, where one can obtain analytic formulae for the scalar correlators. This is a good model calculation that looks similar in spirit with the higher-dimensional cases to be treated subsequently. These correlators are also presented (up to normalisation) in position space in~\cite{Maldacena:2004rf} and~\cite{Maldacena:2018lmt}. In this section, we shall provide analytic results both in momentum and position space along with their normalisation. We use the Papadimitriou - Cveti\v{c} solution described at the end of~\ref{AdS2_metr}. Since the solution is supported by the flux of a gauge field, in the next subsection we shall also compute the correlator for a charged scalar that couples to this non-trivial flux of the background.

\begin{figure}[t]
\vskip 10pt
\centering
\includegraphics[width=80mm]{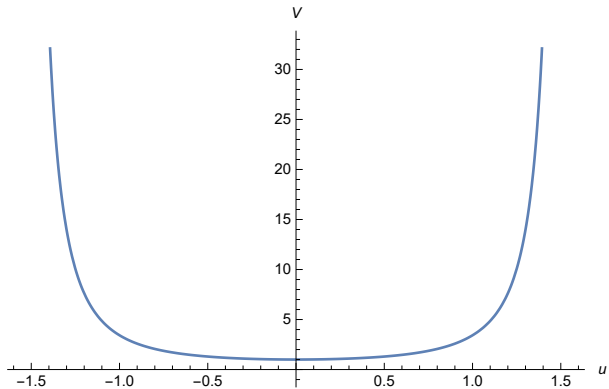}
\caption{The potential $1/\cos^2 u$. The coordinates are such that we are inside one period and the two AdS boundaries are where the potential blows up. The energies of the associated Schr\"{o}dinger problem are below zero.}
\label{fig:potentialAdS2}
\end{figure}

The fluctuation equation for an uncharged scalar on the global metric~\eqref{AdS2metric} of Euclidean signature takes the Schr\"{o}dinger form
\be\label{3e}
- \phi'' + \left(\frac{m^2}{\cos^2 u} \right)\phi = - k^2 \phi
\ee
with $k$ the Fourier modes dual to $\tau$, (they will be discrete if we make $\tau$  compact).
The conformal dimensions are $\Delta_\pm = \half \pm \half\sqrt{1+4m^2}$.
This equation can be solved analytically in terms of Legendre functions.

Before describing the analytic solution, it is useful to describe the properties of the potential depicted in fig.~\ref{fig:potentialAdS2}.
For $m^2>0$, it has a single minimum at $u=0$ (above zero) and blows up near the $AdS_2$ boundaries at $u = \pm \pi/2$ as
\be
V(u) \sim \frac{m^2}{\left( u \pm \frac{\pi}{2} \right)^2} \, .
\ee
The periodicity results in the presence of multiple
``sectors'', which for $m^2>0$ are completely decoupled since the potential is not penetrable. The energies are negative and below the minimum of the potential. Therefore, there are no bound state solutions to this equation and the boundary value problem has a unique solution, for more details see also the discussion in section~\ref{CorrelatorsMeronwormhole}. In case that $m^2 <0$, one has to distinguish two cases, since the potential admits no ground state if $m^2 < m^2_{BF} = - 1/4$ which is in-line with the system being unstable below the BF-bound, where the conformal dimensions $\Delta_\pm$ become complex. When the mass is between $m^2_{BF} < m^2 < 0$ perturbations naively seem to be able to pass between different ``universes'', but one also needs to impose a Hermiticity condition on the Schr\"{o}dinger operator~\eqref{3e}, discussed in Appendix~\ref{Homogeneousstability}. This condition is a zero probability-flux condition
across the boundaries, which is also consistent with the fact that the space ends there (the conformal factor blows up)\footnote{In a more general context the appropriate condition is that the Euclidean fluctuation operator be an elliptic operator~\cite{Witten:2018lgb}. The Hermiticity condition for the scalar Schr\"{o}dinger ODE is a particular case of this more general property to be imposed on solutions.}.
One can therefore maintain stability in this regime and focus on a single period of the potential. According then to our previous discussion there is no normalisable bound state (for which the fluctuation operator is Hermitean) in the spectrum. In the opposite case (positive energies - normalisable bound solutions) one would have the freedom to add any normalisable solution that vanishes at the boundary, hence we would have a family of different solutions with the same boundary values. We therefore also conclude that the boundary value problem has a unique solution (in analogy with the Fefferman - Graham theorem for a single boundary AdS space). Later on we will see that similar conclusions can be drawn for the higher dimensional examples of two boundary wormholes.
\\
\\
After discussing these properties of the problem, the general solution after the transformation
\be
z= \sin u\sp z \in [-1,1]\;,
\ee
can be written as
\be\label{4e}
\phi(z) = (1-z^2)^{\frac{1}{4}}  \, \left( \, C_1  P^\mu_\nu(z) + C_2  Q^\mu_\nu (z) \, \right) \, ,
\ee
with $P^\mu_\nu(z) , \, Q^\mu_\nu(z)$ the associated Legendre functions of the first and second kind and
where \footnote{For these indices the functions are called \emph{conical-Mehler} functions. They also appear as kernels in the Mehler-Fock transform.}
\be\label{41e}
\m= \half \sqrt{1+4 m^2}~~~{\rm and}~~~ \n = - \half + i k \;\;\;.
\ee
From this solution, we can derive the BtB propagator as well as the btB propagators and then the correlators analytically. In particular the un-normalised btB propagator with a source inserted on the boundary located at $z=-1$\footnote{We call the boundary located at $z=-1$ boundary one or the first boundary. Equivalently boundary two or second boundary is the one located at $z=1$.} is
\bea\label{5e}
C_1 K^1(z, k) &=& C_1 (1-z^2)^{\frac{1}{4}} \left(\frac{\pi}{\sin \pi \mu} P^{\mu}_\n(z) - \frac{2}{\cos \pi \mu} Q^\m_\n(z) \right)\, \nn \\
 &=& - C_1 (1-z^2)^{\frac{1}{4}} \frac{2 \pi  \,  \Gamma(\n + \m + 1)}{\sin (2 \pi \mu) \Gamma(\n - \m +1)} P^{-\mu}_\nu (z)
\eea
This propagator has the leading asymptotic behaviour at the first boundary and the subleading at the second i.e. in case of an irrelevant deformation, this solution blows up at the first boundary and asymptotes to zero at the second.

 One may normalise it at some cutoff $\epsilon$, by forming the ratio
 \be
 f_k^1 = {K^1(z, k)\over K^1(\epsilon, k)}\;\;\;.
  \ee
  The second linearly independent solution is
\bea\label{6e}
C_2 K^2(z, k) &=& C_2 (1-z^2)^{\frac{1}{4}}  \left(\frac{\pi}{\sin \pi \nu} P^{\mu}_\n(z) - \frac{2}{\cos \pi \nu} Q^\m_\n(z) \right)\, \nn \\
&=& C_2 \frac{ \pi \sin 2 \pi \mu }{\sin 2 \pi \nu} K^1(-z, k)
\eea
having exactly the opposite behaviour with the first btB propagator (after normalising, we obtain $f^1_k(z) = f^2_k(-z)$). Notice also that the two solutions are related through the exchange $\m \leftrightarrow \n$ in the respective prefactors.
To compute the correlator using the general equation~\eqref{Onshellactionmomentum},  we need to differentiate the btB propagator using the contiguous relation of Legendre functions
\be\label{7e}
(1-z^2)\frac{d P_\n^\m(z)}{d z} = -\n z P_\n^\m (z) + (\n + \m ) P_{\n - 1}^\m (z)\;\;\;,
\ee
with the same formula also holding for $Q_\n^\m$. We then find
\be\label{8e}
\frac{d f^1_k}{d z} \Big|_{z=-1+\epsilon} \,
= \, \left[\frac{-(\n + \half) z}{(1-z^2)} + \frac{(\n - \m )   P^{-\mu}_{\nu-1} (z)}{(1-z^2)   P^{-\mu}_\nu (z)}  \right]_{z=-1+\epsilon} \,
\ee
The structure of this ratio upon expanding for small $\epsilon$ is of the form
\be\label{9e}
\frac{d f^1_k}{d z} \Big|_{z=-1+\epsilon} = \epsilon^{-1} \frac{(A_1+ A_2 \epsilon +... ) + \epsilon^{\m}(B_1 + B_2 \epsilon + ...)}{(C_1+ C_2 \epsilon + ...) + \epsilon^\mu (D_1 + D_2 \epsilon+...)}\, ,
\ee
where the various terms depend on $k, \,m$.
To compute the correlator, one should keep only fractional powers in $\epsilon$, since the integer divergences contain analytic terms in $k$ which can be removed by local counterterms. The result after the expansion is
\be\label{10e}
\frac{d f^1_k}{d z} \Big|_{z=-1+\epsilon} =  \epsilon^{-1+  \mu} \, G_{11}(k) \, + ...  \,\, , \qquad G_{11}(k) = \left( \frac{B_1}{C_1} - \frac{A_1 D_1}{C_1^2} \right) \, ,
\ee
where the term in parenthesis is the renormalised $1-1$ correlator that takes the form\footnote{This is the correlator between two points on the same boundary.}
\bea
G_{11}(k) &=& \mathcal{N}_{11}(m) \frac{ \Gamma(\half + \half \sqrt{1+4m^2} + i k) \Gamma(\half + \half \sqrt{1+4m^2} - i k)}{\Gamma(\half + i k) \Gamma(\half - i k)} \nn \\
 \mathcal{N}_{11}(m)   &=& \frac{  \Gamma \left(1-\frac{1}{2} \sqrt{4 m^2+1}\right) }{2^{-1+\frac{1}{2} \sqrt{4 m^2+1}}   \Gamma \left(\frac{1}{2} \sqrt{4 m^2+1}\right)} \, .
\label{AdS11correlatormomentum}
\eea

\begin{figure*}[!tb]
\begin{center}
\includegraphics[width=0.43\textwidth]{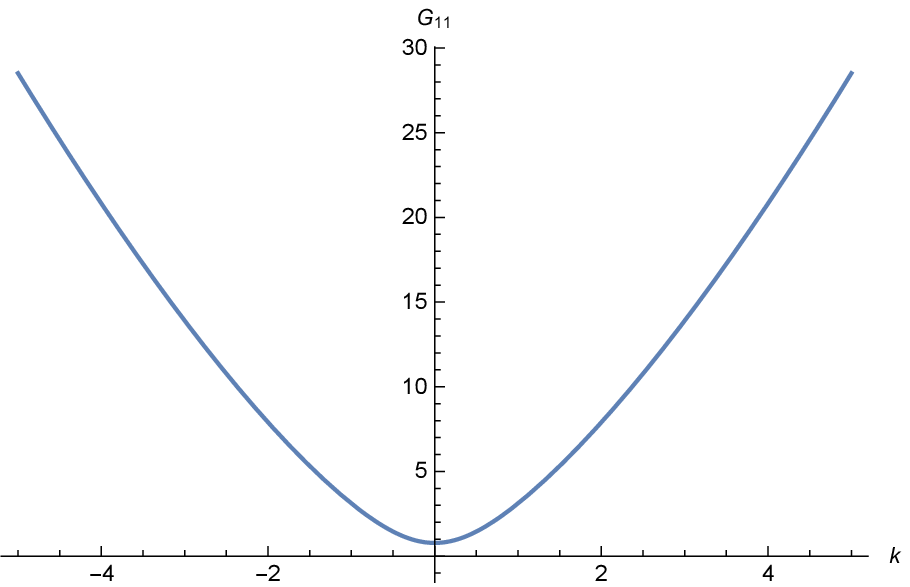}%
\hspace{4mm}
\includegraphics[width=0.43\textwidth]{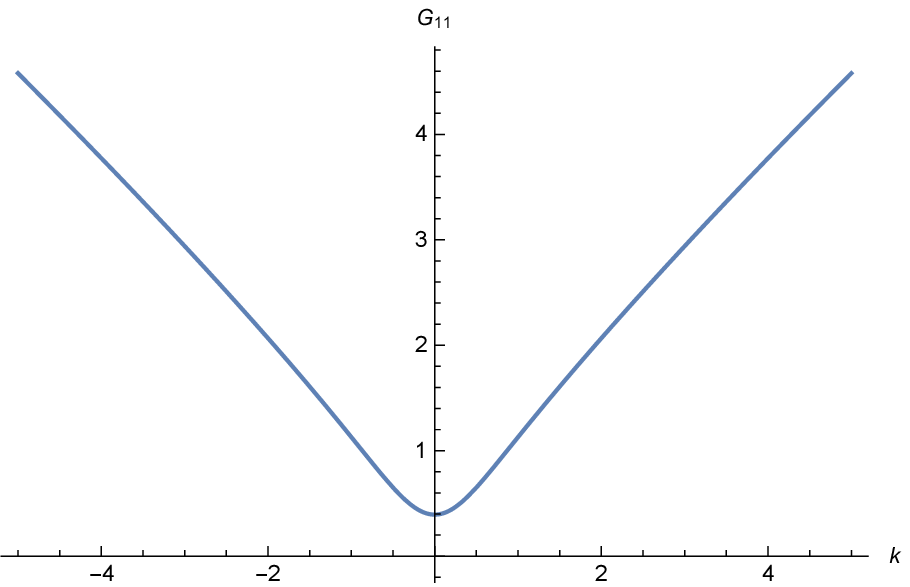}
\end{center}
\caption{Left: The $1-1$ correlator with $m=\half$ as a function of $k$. In the UV it is like the usual $AdS_2$ correlator while in the IR it is gapped with the gap depending on the value of the mass $m$. Right: The same correlator in the case of relevant perturbation ($m^2 = - 0.16$).}
\label{fig:AdS2correlatorir}
\end{figure*}
In figure \ref{fig:AdS2correlatorir} we plot the $1-1$ correlator as a function of the Euclidean momentum $k$. For momenta above the BF bound $m^2_{BF} = -1/4$ the plot is qualitatively similar, the only difference being the change in convexity for large momenta. To be more precise, expanding for low and large momenta, we find that it asymptotes to a constant in the IR and has the expected power law in the UV. It also has a series of poles for $k =\pm i n \pi \pm i \Delta_+ \, , \, n \geq 0$, that have an interpretation of stable excitations in the Lorentzian continuation of the correlator.

\begin{figure*}[!tb]
\begin{center}
\includegraphics[width=0.45\textwidth]{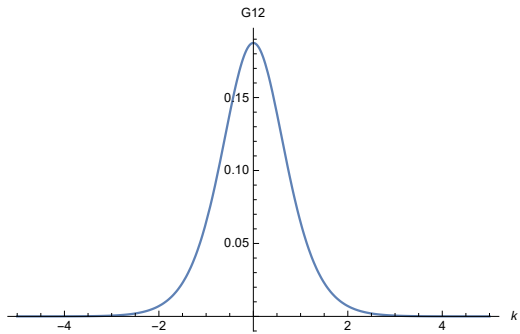}%
\end{center}
\caption{The $1-2$ cross correlator for $m=\half$ as a function of $k$. The correlator is finite and maximises in the IR $k=0$. It remains similar for relevant perturbations as well (the ``bump'' becomes sharper).}
\label{fig:AdS2crosscorrelatorir}
\end{figure*}

One way of acquiring the correlator in position space ($G_{11}^{(\Delta_+)} (\tau)$) is to note that it obeys the recursive relation
\be\label{11e}
G_{11}^{(\Delta_+ +1)} (\tau) = (\Delta_+^2 - \partial_\tau^2) G_{11}^{(\Delta_+ )} (\tau)\, .
\ee
This is to be supplemented with an initial condition that can be found by direct Fourier transform to be
\be\label{12e}
G_{11}^{(1)} (\tau) = - \frac{1}{4 \pi \sinh^2 \left(  \frac{\tau}{2} \right)}
\ee
The final result is then
\be\label{G11space}
G_{11}^{(\Delta_+)} (\tau) =  \frac{ \mathcal{N}_{11}(m)}{\pi \Delta_+ \left(1 + 2 \Delta_+ \right)} \frac{1}{(\sinh \left(  \frac{\tau}{2} \right) )^{2 \Delta_+}}
\ee
This correlator is periodic in real time $t = i \tau$ and exhibits a usual short distance singularity with an exponential decay as $\tau \rightarrow \infty$.
\\
\\
To compute the $1-2$ cross correlator we use the \emph{connection formula} that relates solutions among the two boundaries
\be
P^\m_\n(x)= \cos \left[ (\mu + \nu) \pi \right] P^\m_\n(-x)  - \frac{2}{\pi} Q^\m_\n(-x) \sin \left[ (\n + \m )\pi \right]\eqp
 \ee
After a similar computation, we obtain a correlator that reaches a maximum  in the IR (finite maximum) and vanishes in the UV.  It reads
\bea\label{13e}
G_{12}(k) &=& \mathcal{N}_{12}(m) \Gamma(\half + \half \sqrt{1+4m^2} + i k) \Gamma(\half + \half \sqrt{1+4m^2} - i k) \, , \nn \\
\mathcal{N}_{12}(m) &=& \frac{ 2^{-\frac{1}{2} \sqrt{4 m^2+1}- \frac{5}{2}} \sin \left(\frac{\pi  \sqrt{4 m^2+1}}{2}\right) \left(\sqrt{4 m^2+1}+1\right) \Gamma \left(1-\frac{1}{2} \sqrt{4 m^2+1}\right) }{\pi  \Gamma \left(1+ \frac{1}{2} \sqrt{4 m^2+1}\right)} \nn \\
\eea
where $\mathcal{N}_{12}(m)$ is a mass dependent prefactor. The plot of this function can be seen in figure \ref{fig:AdS2crosscorrelatorir} for $m=1/2$ as a function of momentum $k$. If one stays above the BF bound $m^2_{BF} = - 1/4$ (see section~\ref{AdS2Stability}), the shape of the plot remains qualitatively the same,
looking like a finite bump that becomes progressively sharper as one approaches the BF bound. Exactly at the BF bound the cross correlator is found to vanish due to the prefactor. Such a shape will also persist in the higher dimensional analogues we will study in the next sections.
One can directly Fourier transform this expression to find the position space correlator
\bea\label{G12space}
G_{12}(\tau) &=& \mathcal{N}_{12}(m) \pi 2^{-1 - \sqrt{1+4m^2}} \Gamma(1 + \sqrt{1+4m^2}) \left( \sech \left( \frac{\tau}{2} \right) \right)^{1 + \sqrt{1+4m^2}} \, \nn \\
&=&  \frac{\tilde{\mathcal{N}}_{12}(m)}{(\cosh \left( \frac{\tau}{2} \right))^{2 \Delta_+}}
\eea
This correlator most importantly does not exhibit short distance singularities as $\tau \rightarrow 0$.

On the other hand the correlator~\eqref{G12space} is periodic in real time $t =i \tau$ where it exhibits singularities for $t = i \tau = (2n+1) \pi$, which are multiples of the time a light signal needs to cross between the two boundaries in Lorentzian signature. The Lorentzian periodicity of the correlator comes from the fact that we used the universal cover of the $AdS_2$ hyperboloid. The position space form of the correlators was also discussed in~\cite{Maldacena:2018lmt}, where it was also argued that these Lorentzian signature singularities for the cross-correlator should be regulated by UV effects once one embeds $AdS_2$ in a UV complete description, such as the one of the SYK model.

\subsection{Charged scalar in $AdS_2$}\label{chargedAdS2}

\begin{figure*}[!tb]
\begin{center}
\includegraphics[width=0.43\textwidth]{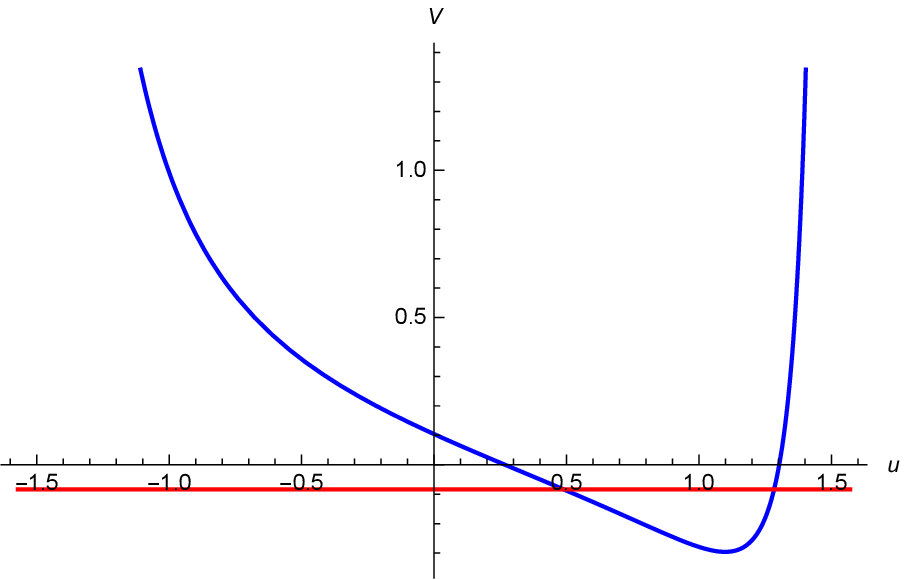}%
\hspace{4mm}
\includegraphics[width=0.43\textwidth]{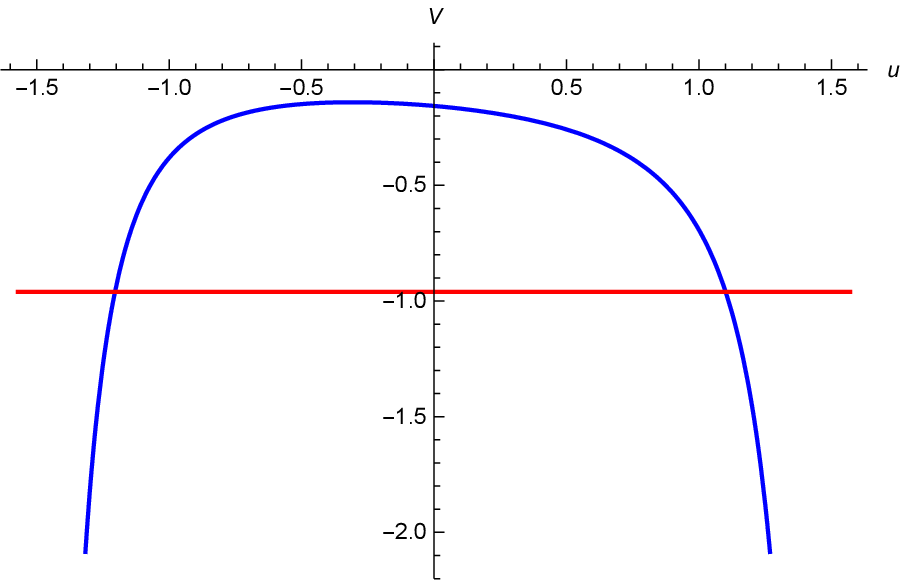}
\end{center}
\caption{Left: The potential for $m^2=-1/16,q^2=1/6,k_r=1/4$. It develops an asymmetric well near the right boundary. The red line is the value of the energy. Right: The potential for $m^2 = -1/6, q^2=1/25, k_r = 1$. It is now asymmetric and unbounded below, the red line again denoting the value of the energy.}
\label{fig:Morse-Rosen}
\end{figure*}

We can treat in a similar analytic fashion, also  the case of a charged scalar in $AdS_2$ for the EMD solution with non trivial gauge field background. The fluctuation equation for minimally coupled scalar with charge $q$ takes the form
\be\label{14e}
-D_{\mu}(\sqrt{g}g^{\mu\nu}D_{\nu}\Phi) + \sqrt{g} m^2 \Phi
\ee
where
\be\label{15e}
D_{\mu}=\partial_{\mu}+i q A_{\mu}\,.
\ee
For our case, the $A_{\mu}=A_{\tau}(r)$ and the metric components are only functions of $r$, where $r$ is either $u$ or $\tilde{u}$ depending on which metric for the $AdS_{2}$ solution ---presented in subsection \ref{AdS2_metr}--- we use. Then the fluctuation equation takes the form:
\be\label{16e}
-\sqrt{g} g^{\tau\tau}\partial_{\tau}^{2}\Phi - 2 i q \sqrt{g} g^{\tau\tau} A_{\tau}\partial_{\tau}\Phi
+q^{2}\sqrt{g} g^{\tau\tau}A_{\tau}^2\Phi -\partial_{r}(\sqrt{g}g^{rr})\partial_{r}\Phi - \sqrt{g} g^{rr}\partial_{r}^{2}\Phi + \sqrt{g}m^2\Phi =0\,.
\ee
For the metric $ds^2 = dr^{2}+\cosh{r}^2d\tau^{2}$ and for $\Phi = e^{i k\tau}\tilde{\phi}(r)$ the equation is:
\be\label{17e}
\partial_{r}^{2}\tilde{\phi}+\tanh r \partial_{r}\tilde{\phi}- \left(\frac{\left(k+qA_{\tau}\right)^{2}}{\cosh^{2} r}+m^{2}\right)\tilde{\phi}=0\,,
\ee
with $A_{\tau}=\mu -\sinh r $. If we substitute the $A_{\tau}$ in the fluctuation equation we obtain,
\be\label{18e}
\partial_{r}^{2}\tilde{\phi}+\tanh r \partial_{r}\tilde{\phi} - \left(\frac{(k+q\mu)^{2}+  q^{2} }{\cosh^{2} r}-2 q (q\mu +k)\frac{\sinh r}{\cosh^{2} r}+m^{2}- q^{2} \right)\tilde{\phi}=0\,,
\ee
This can be turned into a Schr\"odinger form by redefining
\be
\Psi(r) =\sqrt{\tanh r} ~\tilde \phi (r)\eqp
\ee
\be\label{182e}
- \Psi''(r)  + \left(\frac{(k+q\mu)^{2}+  q^{2} + \frac{1}{4} }{\cosh^{2} r}-2 q (q\mu +k)\frac{\sinh r}{\cosh^{2} r} \right) \Psi(r) =\left(q^{2} - m^{2} - \frac{1}{4} \right) \Psi(r) \eqp
\ee
This equation will be useful to test the effect that charge has on the stability of linearised perturbations, see~\ref{AdS2Stability}.
\\
\\
For the metric $ds^2 = (du^{2}+d\tau^{2})/\cos^{2}u$ and for $\Phi = e^{i k\tau}\phi(u)$ the equation is:
\be\label{19e}
\partial_{u}^{2}\phi -\left(\left(k+qA_{\tau}\right)^{2}+\frac{m^{2}}{\cos^{2}u}\right)\phi=0\,,
\ee
with $A_{\tau}=\mu- \tan u $ and $-\frac{\pi}{2}<u<\frac{\pi}{2}$. If we substitute the $A_{\tau}$ in the fluctuation equation we obtain,
\be\label{20e}
\partial_{u}^{2}\phi -\left(\left(k+\mu q\right)^{2}- q^{2} -2 \left(q\mu +k\right) q \tan u+\frac{ q^{2} +m^{2}}{\cos^{2}u}\right)\phi=0 ,,
\ee
or
\be\label{21e}
-\phi''(u) + \left(\frac{q^2 + m^2}{\cos^2 u} - 2 k_r q \tan u \right) \phi(u) = \left(-k_r^2 + q^2 \right) \phi(u) \, ,
\ee
where in eqn.~\eqref{21e} we have used the shifted variable $k_r = k + q \mu$, which is the new natural momentum variable.
The resulting potential for the Schr\"{o}dinger equation is of a generalised Poschl-Teller or Morse-Rosen type~\cite{Morse}. It is symmetric only for zero values of $k_r$, else it becomes asymmetric and develops an infinite well as $k_r \rightarrow \infty$ near one of the two boundaries, see fig.~\ref{fig:Morse-Rosen}. For $m^2>0$ the minimum of the potential is at
\be\label{211e}
V_{min} = m^2 + q^2 - \frac{q^2 k_r^2}{m^2 + q^2} \, ,
\ee
and the allowed energies $E = - k_r^2 + q^2$ are always below $V_{min}$, so that no bound states exist. In case $m^2 < 0$, one needs to take also into account the BF-bound of Appendix~\ref{AdS2Stability}, that is $0 > m^2 \geq m^2_{qBF} = - 1/4 + q^2$, else the background is unstable. This also implies that in this case $q^2 < 1/4$. We then need to further consider the two subcases:
\begin{itemize}

\item In the first case the potential is bounded below $m^2 + q^2 >0$ so that $1/8 < q^2 < 1/4$. This then means that bound states can exist for
\be\label{222e}
E^{bound} = -k^2_{min} + q^2 \geq V_{min} \quad \Rightarrow  \quad  0 \geq m^2 + \frac{m^2 k^2}{m^2 + q^2} \, ,
\ee
which can be satisfied for states with low enough momentum $k$ (the high momentum states have energies below the minimum of the potential). Such a case is plotted in the left side of fig.~\ref{fig:Morse-Rosen}.

\item In the second case the potential is unbounded below $0 > m^2 + q^2 \geq - 1/4 + 2 q^2$. This can only hold for $q^2 < 1/8$. Such a case is plotted on the right side of fig.~\ref{fig:Morse-Rosen} and can in principle support a bound state if the potential is not too ``steep''.

\end{itemize}

This would then mean that the FG theorem fails in such  cases since there exist normalisable bound states that violate the uniqueness of the boundary value problem. Nevertheless in addition one should check whether the fluctuation operator is Hermitean for these states, according to the discussion of Appendix~\ref{Homogeneousstability}. We leave the details of such an analysis for the future.

In order to compute the correlation functions, we can simplify and solve the differential equation if we rescale $\phi = e^{k_r u} (\cos u)^q F(u) $ and then redefine $x= \half (1- i \tan u)$ to find
\be\label{22e}
F''(x) + \frac{m^2 + q^2}{x-x^2} F'(x) + \frac{1+ i k_r -q - 2x + 2 q x}{x-x^2} F(x) = 0
\ee
with the general solution,
\be\label{23e}
F(x) = C_1 \, _2F_1\left(\frac{1}{2} - q- A,+\frac{1}{2} - q + A;i k_r-q+1;x\right)+
\ee
$$
 + C_2 (-x)^{q-i k_r} \, _2F_1\left(\frac{1}{2} - i k_r -A  , \frac{1}{2}- i k_r + A ;-i k_r+q+1;x\right)$$
with
\be
 A = \frac{1}{2}\sqrt{4 m^2+4 q^2+1}
\label{22ee}\ee
\\
\\
After repeating the same steps we followed to compute the uncharged correlator, we find the charged 1-1 correlator
\bea\label{24e}
G^q_{11} &=& N^q_{11} \Big| \Gamma \left(\frac{1}{2} - i k_r + A \right) \Big|^2 \left( \cos \pi q \cos A \cosh \pi k_r - \sin \pi q \sin A \sinh \pi k_r \right) \nn \\
  N^q_{11} &=& \frac{\sec \left(\pi  \left(q-A\right)\right) \Gamma \left(1-A\right) \Gamma
   \left( q+ A+ \frac{1}{2}\right)}{\pi \Gamma \left(2 A\right) \Gamma \left(q-A+\frac{1}{2}\right)} \, ,
\eea
which is a real expression and the charged 1-2 correlator
\bea\label{25e}
G_{12} &=& N^q_{12} \frac{  \Big|\Gamma \left(\frac{1}{2} - i k_r + A \right) \Big|^2 \cosh \pi (k_r - i A)}{\cosh \pi ( k_r - i q) + \cosh \pi (k_r + i q - 2 i A)} \nn \\
  N^q_{12} &=& \frac{  \left( 1+ 2 A \right) \Gamma \left(\half + q + A\right)}{ \Gamma \left(2 A\right) \Gamma \left(1+ 2A\right) \Gamma \left(\half + q - A\right) } \, ,
\eea
again being a real expression in the allowed range of parameters ($A$ a positive real number). As expected, sending $q\to 0$ the expressions above reproduce the correlators found in subsection \ref{AdS2correlators}. The qualitative behaviour of these correlators is quite similar to those of the uncharged scalar even for non-zero $q$ (but with respect to the parameter $k_r = k + \mu q$). In particular, in the UV, the 1-1 and the 1-2 correlators scale as
\be \label{26e}
G_{11}^{UV} \sim k_r^{2 A}   \, , \qquad G_{12}^{UV} \sim k_r^{2 A} e^{- \pi k_r} \, ,
\ee
which is the usual short distance power law behaviour of the 1-1 correlator versus an exponential decaying behaviour for the cross-correlator. The coefficient of the exponential -$\pi$- seems universal and is the same both for charged and uncharged scalar since it does not depend on the parameters $m,q$. In particular it is coming from the part of the connection matrix relating the solutions at the two boundaries that does not depend on $m,q$.
On the other hand in the IR one finds the following expansions
\be \label{27e}
G_{11}^{IR}  \sim a_{11} + b_{11} k_r^2 + ...   \, , \qquad G_{12}^{IR} \sim a_{12} + b_{12} k_r^2 + ... \, ,
\ee
with the precise coefficients now depending on the parameters $q,m$. These coefficients have opposite behaviour (when $b_{11}>0$, $b_{12}<0$) so that the 1-1 IR minimum corresponds to an 1-2 IR maximum.

\subsection{Einstein Yang-Mills Wormhole}\label{CorrelatorsMeronwormhole}

In this section we compute the scalar two-point correlators for the meron wormhole presented in subsection \ref{EinsteinYangMillswormhole}. The Bulk Green's function in the asymptotically flat version of the meron wormhole was analysed in~\cite{Betzios:2017krj}.

\begin{figure}[t]
\vskip 10pt
\centering
\includegraphics[width=80mm]{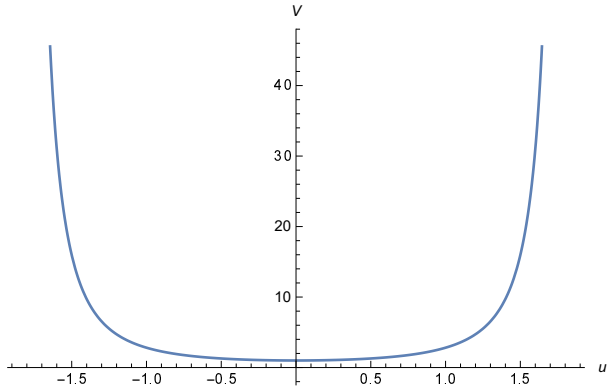}
\caption{The potential $1/\cn^2 u$ for one period of elliptic functions. The coordinates are such that we are inside one period and the two AdS boundaries are where the potential blows up.}
\label{fig:Potential1}
\end{figure}

\begin{figure}[t]
\vskip 10pt
\centering
\includegraphics[width=80mm]{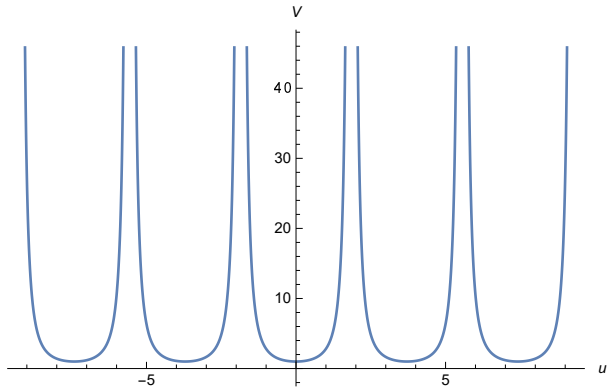}
\caption{The potential $1/ \cn^2 u$, where the periodicity is shown explicitly.}
\label{fig:Potential2}
\end{figure}

One has to use the metric in the elliptic function form~\eqref{ellipticcoords2}, since in these coordinates it is conformal to ESU. The Jacobi elliptic functions along with their properties are defined in Appendix~\ref{Elliptic functions}. We first decompose the modes in $S^3$ harmonics
$\phi(u,\Omega) = \xi_{\ell}(u) \, Y_{\ell m p}(\Omega)$, and use $\partial_\Omega^2 Y_{\ell m p} = - \ell(\ell + 2) Y_{\ell m p}$ with $\ell = 0,1,...$. The fluctuation equation then acquires the form
\be\label{1f}
\left(\frac{d^2}{d u^2} - \frac{\ell(\ell+2)}{2B} \right) \xi(u) + \frac{2 \sn u \dn u}{\cn u} \frac{d \xi(u)}{d u} - \frac{m^2 {k'}^2}{\cn^2 u} \xi(u) =0
\ee
To write the scalar fluctuation equation in Schr\"{o}dinger form we rescale $\xi =  \Psi / \cn u$ and find
\be
-\frac{d^2}{d u^2}  \Psi(u)  + \frac{{k'}^2 \left(m^2  + 2 \right)}{\cn^2 u} \Psi(u) = - \frac{ (\ell+1)^2}{2B}  \Psi(u)
\label{WormE=0}
\ee
This is a similar equation to the Lam\'e equation. The elliptic modulus  and the complementary modulus are
\be
k^2 = {(B+1/2)\over 2B}\sp {k'}^2 = {(B-1/2)\over 2B}\eqp
 \ee
 The parameter $u$ is defined for $u \in [-K , K]$ with $K$ the elliptic period. These endpoints define the two boundaries. The potential is drawn in figures~\ref{fig:Potential1} and~\ref{fig:Potential2} and one can notice that it is periodic.
Then one may ask  if the periodic walls of the  potential can be penetrated quantum mechanically. To examine such possibility, we expand $\cn u$ around $u = K$ which is where the potential blows up. We then use
\be
\cn(u + K) = - {k'} {\sn u \over  \dn u}
\ee
 to find
\be
V(u+K) = \frac{(m^2 + 2)\dn^2 u}{\sn^2 u} \approx  \frac{ m^2+ 2}{u^2}+{\cal O}(u^0)
\label{2f}
\ee
This is precisely the same divergence as in the UV region of AdS in equation~\eqref{AdSGlobalE=0}. Consequently, the space just ends there ---boundary--- and the potential is not penetrable~\footnote{Let us note that in defining the correlators we might need only a part of the global monodromy properties of the ODE since we only restrict to one well of the potential.}.
Another important property is that solutions diverge at one or the other boundary (for $m^2+2 > 0$). The reason is that, due to the negative energies of the Schr\"{o}dinger problem, there are no normalisable solutions to this equation.
This is similar to the single boundary global Euclidean $AdS_4$, (see appendix~\ref{AdSODE}) and the two-boundary global $AdS_2$, see~\ref{AdS2correlators}  where we do not find normalisable solutions to the homogeneous fluctuation equation either\footnote{This is a general feature of the solutions we examine in this paper.}. As previously discussed in the literature, this feature leads to uniqueness of the boundary value problem~\cite{Witten:1998qj}.

In the opposite case (positive energies - normalisable bound solutions) one would have the freedom to add any normalisable solution that vanishes at the boundary, hence we would have a family of different solutions with the same boundary values.

In the limit where  $B=\half$, (the limit where the two sides of the wormhole pinch-off and we have two disconnected $AdS_{4}$ spaces) the equation \eqref{WormE=0} reduces to equation\eqref{AdSGlobalE=0}. In this limit the boundary value problem changes nature, and only one type of correlation function can be defined.

Another observation is that the potential of the meron wormhole can be expressed in terms of an infinite number of $AdS_2$-like potentials\footnote{An equivalent observation holds for the metric. Interestingly, it was argued in \cite{Maldacena:2004rf}  that in wormholes with slices that are finite volume hyperbolic, one would obtain an sum of an infinite number of copies of the AdS$_2$ correlators. This is what is explicitly realized here.}, see~\eqref{AdS2correlators} using the identity
\be
\frac{k'^2}{\cn^2 ( u , k)} = k'^2 - \frac{E(k)}{K(k)} + \left[\frac{\pi}{2K(k)}\right]^2 \sum_{n= - \infty}^\infty \sec^2 \left( \frac{\pi}{2 K(k)} (u + i 2 n K'(k)) \right)
\label{Wormsuperpos}
\ee
A similar decomposition was employed in~\cite{Dunne:1999zc} and it would be interesting to examine if the correlators of the meron wormhole can be computed using superposition of the exact $AdS_2$ solutions.

\subsubsection{Numerical correlators for the Einstein Yang-Mills wormhole}

\begin{figure*}[!tb]
\begin{center}
\includegraphics[width=0.48\textwidth]{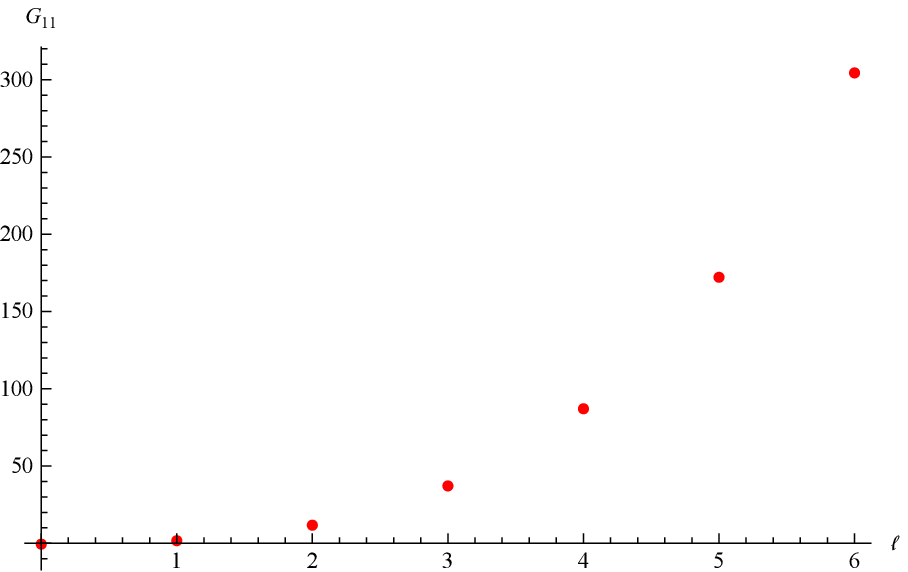}
\hspace{3mm}
\includegraphics[width=0.48\textwidth]{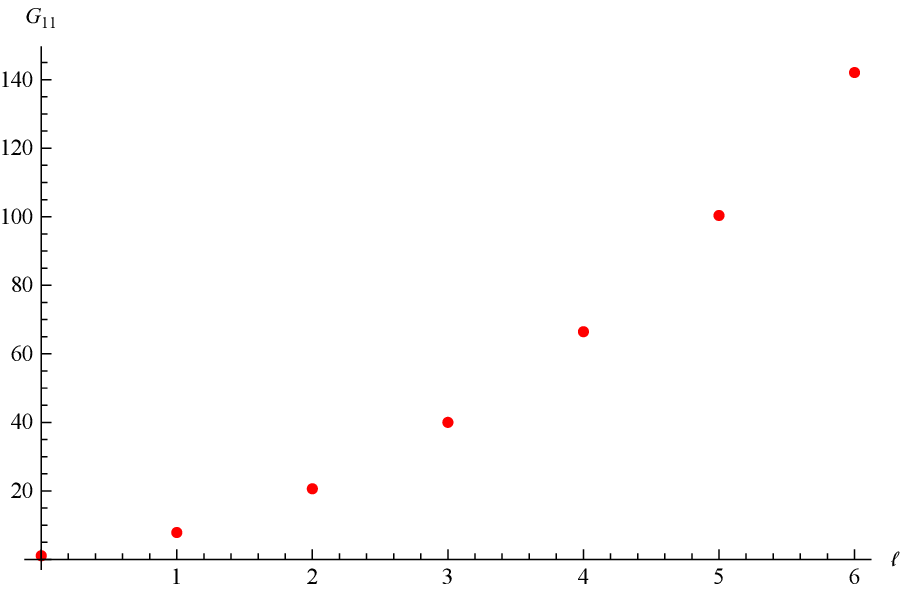}
\end{center}
\caption{Left: The $1-1$ correlator with $m=1, B=1$ as a function of $l$ (irrelevant). Right: The $1-1$ correlator with $m^2 = -1, B=1$ as a function of $l$ (relevant). They resemble the analytic correlator of $AdS_2$ in that they grow in the UV and saturate to a small finite non zero minimum in the IR.}
\label{fig:numericalG11}
\end{figure*}

\begin{figure*}[!tb]
\begin{center}
\includegraphics[width=0.48\textwidth]{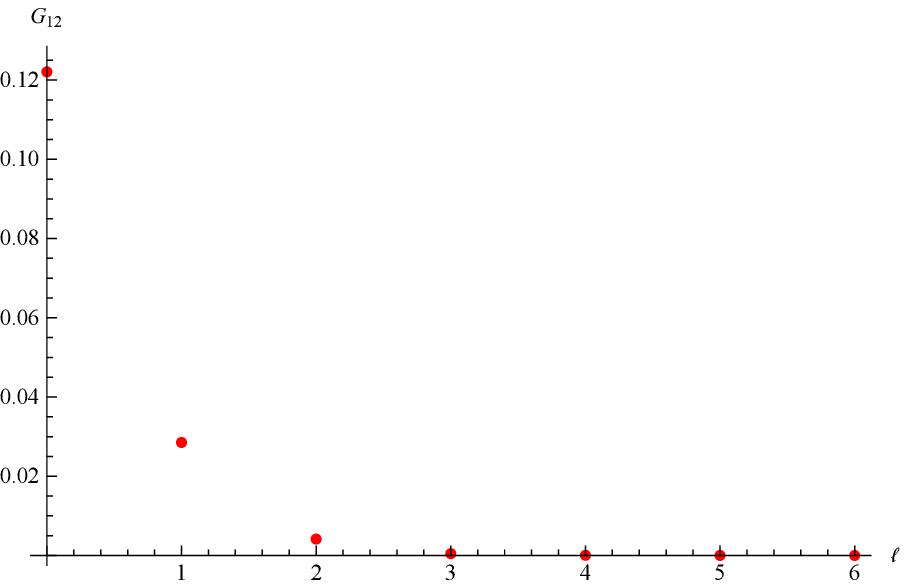}
\hspace{3mm}
\includegraphics[width=0.48\textwidth]{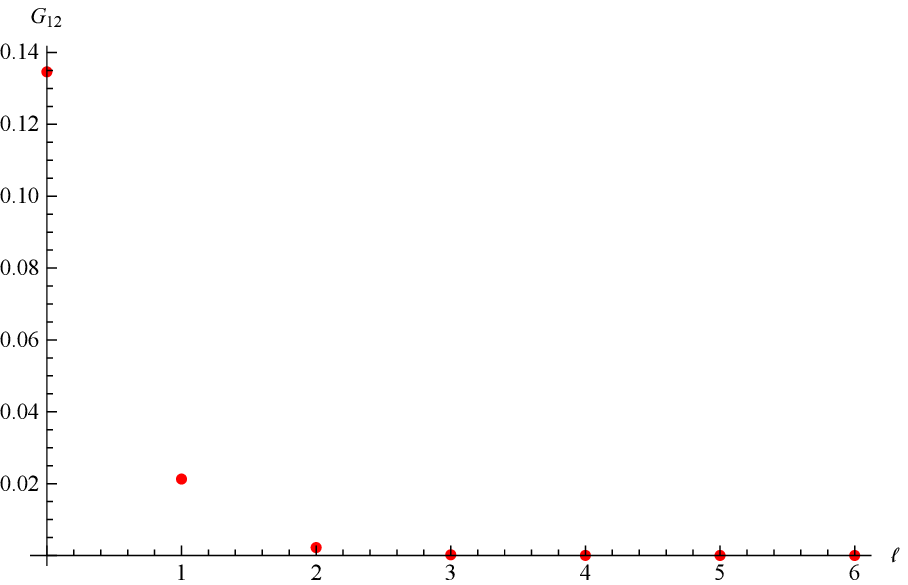}
\end{center}
\caption{Left: The $1-2$ correlator with $m=1, B=1$ as a function of $l$ (irrelevant operator). Right: The relevant case $m^2 = -1$, $B=1$. They drop to zero in the UV with a scale depending on the wormhole size as well as in the conformal dimension of the perturbing operator. In the IR they approach a finite maximum. }
\label{fig:numericalG12}
\end{figure*}

Since the scalar ODE~\eqref{WormE=0} of the Einstein Yang-Mills system is of the Lam\'e type,  it has four regular singular points and the global monodromy properties of the solutions are not known. Consequently,  we resort to numerical techniques to compute the correlators. In particular, we use numerical integration with asymptotic matching to the analytical AdS solutions obtained from expanding~\eqref{WormE=0} near the two boundaries. The results for the behaviour of the 1-1 correlator of an irrelevant operator of scaling dimension $\Delta = 3/2 + \sqrt{9/4 + m^2} $ with
$m=1$ can be seen in the left plot of figure~\ref{fig:numericalG11} and for the 1-2 correlator of two scalar operators of the same conformal dimension in the left plot of figure~\ref{fig:numericalG12}. These plots show the behaviour of these correlators as a function of the $S^3$ mode $\ell$. Similar plots for the case of a relevant operator with $\Delta = 3/2 + \sqrt{5}/2$ are found in the right side plots of figure~\ref{fig:numericalG11} and figure~\ref{fig:numericalG12}.

One notices for the cross - correlator $1-2$ the generic feature that the low $\ell$ IR modes pass through the other side of the wormhole throat and therefore the correlation between the two boundaries grows in the IR (and saturates to a finite value) while it diminishes in the UV. For a correlator on a single boundary such as $1-1$, we find the opposite behaviour, with an unbounded UV growth (resulting to a short distance singularity) and a finite minimum in the IR.
This finite minimum is due to the $S^3$ having finite volume. The cross correlator on the other hand does not exhibit any short distance singularity. These behaviours persist for various values of the throat size $B$, as long as the geometry does not pinch off.
In the case of $AdS_2$ we obtained analytic expressions of such correlators that have the same qualitative behaviour, see~\ref{AdS2correlators}. In the next section we analyse the case of the Dilaton wormhole and we again find a similar behaviour for these correlators.

\subsection{Einstein Dilaton wormhole}\label{ODEforEinsteinDilaton}

In this subsection we study the wormhole solutions of the Einstein Dilaton system presented in the subsection \ref{EinsteiDilatonwormhole}, specifically the three-dimensional solutions with hyperbolic boundary with metric \eqref{elliptichyperbolic}.

The homogeneous fluctuation equation can be most succinctly written in the coordinates~\eqref{elliptichyperbolic}, in order to be put into Schr\"{o}dinger form. We split the modes
\be
\phi  = \xi_{s}(u) F_s (H_2/\Gamma)
 \ee
 with
 \be
 -\Box_{\mathcal{M}_T} F_s = s(1-s) F_s\eqc
  \ee
where the transverse space manifold is $\mathcal{M}_T = \mathbb{H}/\Gamma$ ( $\Gamma$ labeling a Fuchsian discrete $PGL(2,R)$ subgroup) and $F_s$ the specific hyperbolic space quotient eigenfunctions. We therefore need to use the properties of the Laplacian on two-dimensional hyperbolic space and discrete quotients thereof. Some known results are the following (see~\cite{Sarnak} and references within):
\begin{itemize}

\item For $\mathcal{M}_T = \mathbb{H}/\Gamma$ compact, $-\Box_\mathcal{M}$ has discrete spectrum in $[0,\infty)$

\item For $\mathcal{M}_T = \mathbb{H}/\Gamma$ non-compact, $-\Box_\mathcal{M}$ has both discrete spectrum in $[0,\infty)$
and absolutely continuous spectrum in $[1/4, \infty)$.

\item If $\mathcal{M}_T = \mathbb{H}/\Gamma$ has infinite area (2-volume), $-\Box_\mathcal{M}$ has discrete spectrum in $(0,1/4)$ and absolutely continuous spectrum in $[1/4, \infty)$, and therefore, no embedded discrete eigenvalues.

\end{itemize}
We therefore conclude that, in all the cases above,  $ s(1-s) \geq 0$ which can be either discrete and/or continuous,  the later starting always from $1/4$.

\begin{figure*}[!tb]
\begin{center}
\includegraphics[width=0.45\textwidth]{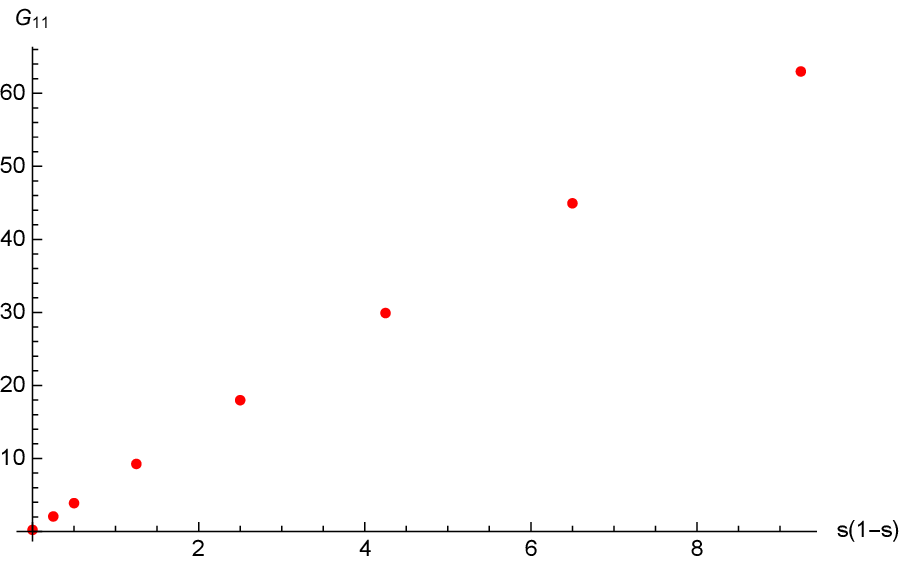}
\hspace{4mm}
\includegraphics[width=0.45\textwidth]{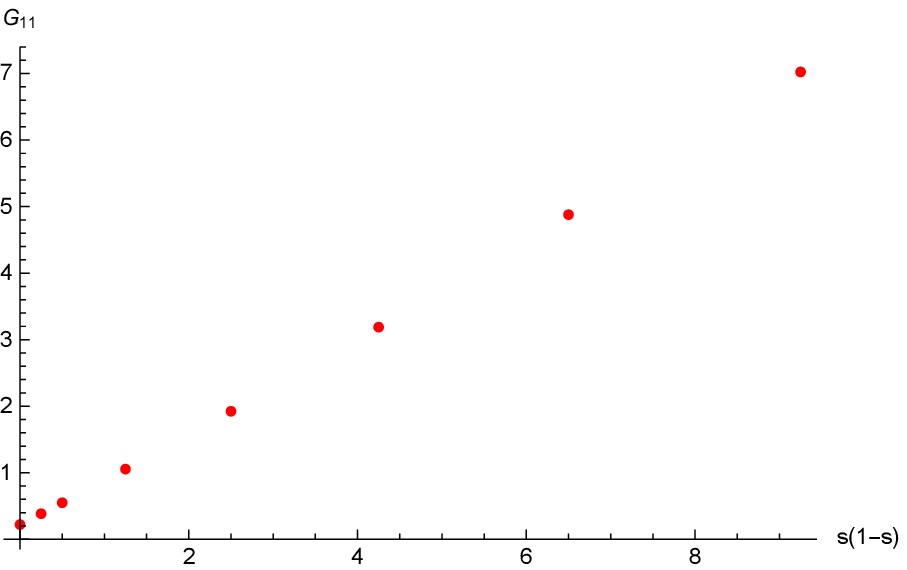}
\end{center}
\caption{Left: The $1-1$ correlator with $m^2 = 1, B=1$ as a function of $s(1-s)$ (irrelevant operator). Right: The relevant case $m^2 = -0.1$, $B=1$. They both resemble the analogous correlators of the meron wormhole and $AdS_2$, in that they acquire a small finite minimum in the IR and blow up in the UV.}
\label{fig:numericalG11H2}
\end{figure*}

\begin{figure*}[!tb]
\begin{center}
\includegraphics[width=0.45\textwidth]{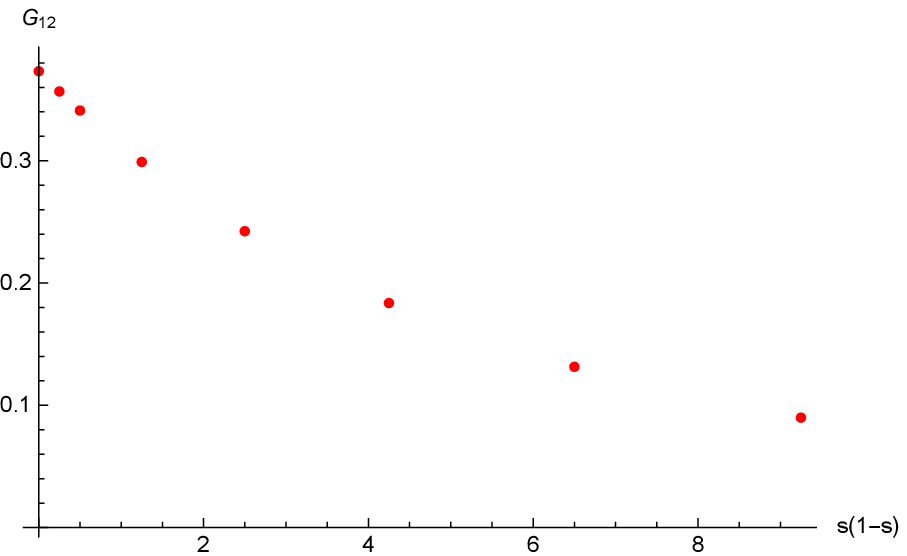}
\hspace{4mm}
\includegraphics[width=0.45\textwidth]{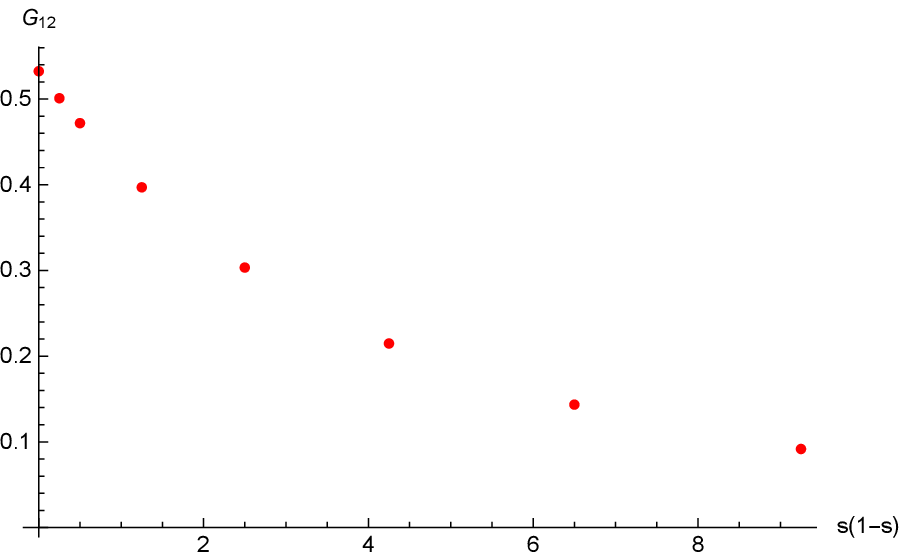}
\end{center}
\caption{Left: The $1-2$ correlator with $m^2 = 1, B=1$ as a function of $s(1-s)$ (irrelevant operator). Right: The relevant case $m^2 = -0.1$, $B=1$. They resemble the correlators of the meron wormhole and $AdS_2$ in that they vanish in the UV and obtain a finite maximum in the IR.}
\label{fig:numericalG12H2}
\end{figure*}

The fluctuation equation now becomes
\be\label{1g}
\xi''(u) - \frac{ s(1-s)}{2B} \xi(u) +\frac{\dn u\sn u}{\cn u } \xi'(u) - m^2\frac{k^{\prime 2}}{\cn^2 u} \xi(u) = 0
\ee
One can turn this into a Schr\"{o}dinger form using
 \be
 \xi(u) = \sqrt{\frac{\cn u}{k^{\prime}}}\Psi(u)
 \ee
\be\label{2g}
- \Psi''(u) +\left(\frac{k^{\prime } (m^2 +1)}{\cn^2 u}   \right) \Psi(u)  = -\frac{1}{2B}(s(1-s) ) \Psi(u)
\ee
We find that the potential is similar to the meron case: it has a minimum in the center and it blows up in the AdS boundaries. The main difference with the meron wormhole is that the transverse space modes are those of  $\mathcal{M}_T = \mathbb{H}/\Gamma$ and not of $S^3$. Similarly to that case we again numerically integrate eq.~\eqref{2g}
to find the correlators plotted in~\ref{fig:numericalG11H2} and~\ref{fig:numericalG12H2}, for both relevant and irrelevant scalar operators having dimension $\Delta = 1+ \sqrt{1+ m^2}$  with $m^2 = -0.1$ and $m^2 = 1$ respectively. The results are again qualitatively similar to the ones of the meron wormhole.
\\
\\
In conclusion, we have computed boundary to boundary scalar correlators between points on the same boundary such as the $1-1$ correlator as well as cross- correlators $1-2$ between points located on two different boundaries. We performed the calculation for the three types of solutions presented in section \ref{sol_asym_reg}. In the $AdS_{2}$ case one can do the computation analytically, while in the cases of three-dimensional hyperbolic wormhole as well as in the case of the four-dimensional meron wormhole, we used numerical techniques.

The common feature in all these cases is that the cross - correlator $1-2$ in momentum space  has a maximum in the IR and asymptotes to zero in the UV.
Moreover, it does not have any short distance singularity in position space, while on the other hand,  the $1-1$ correlator has a non zero minimum in the IR, and diverges in the UV. This result is in accordance with the known qualitative behaviour of single boundary correlators on compact spaces. Based on this results, a generic holographic interpretation of our solutions in terms of a UV-soft coupling between two Euclidean theories will be provided in section~\ref{field theory analogue}.

\section{Wilson lines}\label{Wilsonlines}

\begin{figure}[t]
\vskip 10pt
\centering
\includegraphics[width=90mm]{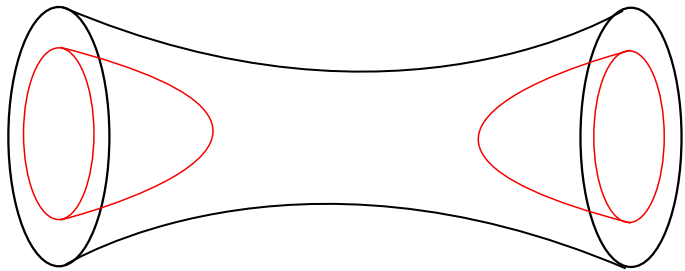}
\caption{A depiction of the factorised correlator $\langle W(C_1)\rangle \langle W(C_2) \rangle$ with loops residing on opposite boundaries. The disk slices are three spheres and each loop is a circle sitting on one of the two asymptotic $S^3$'s. This configuration scales as $O(N^2)$.}
\label{fig:Loop}
\end{figure}

\begin{figure}[t]
\vskip 10pt
\centering
\includegraphics[width=80mm]{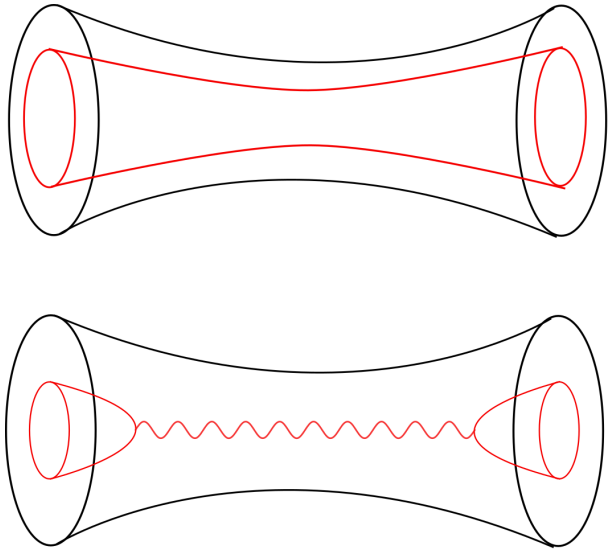}
\caption{The different behaviours of a loop-loop correlator. In both cases the correlator scales as $O(N^0)$. As the world-sheet connecting the loops approaches the string length in the middle of the throat, the single surface configuration ceases to be trustworthy. In the second disconnected case the loops can interact only via exchange of gravitons and other perturbative bulk modes.}
\label{fig:Loopcorrelator}
\end{figure}

In this section we study non-local observables in the wormhole solutions. Specifically, we study expectation values of Wilson loops denoted by $\langle W(C) \rangle$ and loop - loop correlators $\langle W(C_1) W(C_2) \rangle$, with the two loops residing on different boundaries (see the schematic figures~\ref{fig:Loop} and~\ref{fig:Loopcorrelator}). The dual of the loops are fundamental string world-sheets hanging from the boundaries. This offers a natural normalisation for a single loop $\langle W(C) \rangle \sim 1/g_{s} \sim O(N)$, since it corresponds to a world-sheet surface of disk topology $D$ with a single boundary $C = \partial D$.

We shall analyze in detail only the case of the meron wormhole but we expect a similar behavior in the other examples as well.
If we consider a single loop, we find that the dual bulk string minimal surface, cannot pass through the throat. The reason for this is that the meron wormhole background is symmetric and therefore the effective potential that the string is subjected to has a minimum in the middle of the throat~\footnote{This can be evaded if the loop is charged and rotating. In this case the magnetic flux could drive it through the other side of the throat. The world-sheet action in this case can depend on the magnetic field and area alone does not determine the saddle point.  Moreover, the effective potential now is not symmetric due to current/magnetic effects as it is found in the case of a charged scalar in $AdS_2$, see~\ref{chargedAdS2}. In addition, we could study a dual $D_1$-brane that couples directly to the bulk magnetic two-form flux $F$. We shall leave such interesting possibilities for future studies.}.

We first study the behaviour of the expectation value of a  single Wilson loop, located  on one of the two boundaries.  We shall  find that the on-shell action, scales as the area of the loop on the $S^3$, for a large sized loop (close to the equator).

Subsequently, we consider the correlation function of two loops of equal size\footnote{One could also consider loops with different sizes across the two boundaries. To study such a possibility, one must generalize our ansatz{\"e}, and this results in a more complicated system. We will not pursue it further here as we do not expect to extract from it any extra important information.}, the first on one boundary and the other on the opposite boundary.
There is a disconnected contribution to this correlator, as shown in~\ref{fig:Loop}.

There is a contribution to the connected correlator, described by a single world-sheet connecting the two boundary loops and having a cylinder topology $S^1 \times \mathbb{R}$ as shown on the top of ~\ref{fig:Loopcorrelator}. The connected two-point function described by this configuration scales as $O(N^0)$ as the world-sheet has the topology of a cylinder. As we shall find, this semiclassical contribution exists for any size of the boundary loops.

On the other hand, there is another contribution to the connected correlator that arises from the disconnected contribution in~\ref{fig:Loop} by an exchange of a supergraviton between the two string world-sheets.
 The disconnected contribution is of order $O(N^2)$ as it is the product of two disks. An exchange of a supergraviton costs a factor of $1/N^2$, so that the end result is again of order ${\cal O}(N^0)$. For the $N=4$ SYM Wilson loops this contribution was calculated in \cite{BCFM} from holography and in \cite{SZa} from a QFT point of view.
It involves the coupling of the various bulk modes to the world-sheet of the string. Its form is as follows
\be
\langle W_1(C) W_2(C)\rangle=\langle W_1(C)\rangle \langle W_1(C)\rangle \sum_i~\int d{\cal A}_1 \int d{\cal A}_2~f^i_1f^i_2G^i_{12}
\label{equationsofthepropagatorsinthebulk}
\ee
where the disconnected multiplicative factor is given by the semiclassical contribution in figure \ref{fig:Loop}. The sum is over all relevant supergravity modes, $f^i_{1,2}$ are the (local) couplings of the mode i to one or the other string world-sheet in the bottom figure \ref{fig:Loopcorrelator} and $G^i_{12}$ denotes the propagator of the i-th bulk mode.
Finally the two integrals $d{\cal A}_{1,2}$ sum over the position of the two endpoints of the bulk propagator over the two disconnected world-sheets of figure \ref{fig:Loopcorrelator}.
 Unfortunately, we do not know the relevant bulk propagators (that are more complicated than those in AdS), neither the couplings of the appropriate fields to the string world-sheet action  without an explicit embedding in string/M- theory\footnote{With the exception of the meron solution.}.
We will however return to this issue towards the end of this section.

We note that an analogous behaviour for the Wilson loop two-point function  was described by Gross and Ooguri,  \cite{GO,Za}. In that case, one considers a correlator with both loops residing on a single boundary. The transition between a single surface and two disconnected surfaces  is driven by separating the loops in transverse space. Beyond a critical separation the single connected surface collapses and only the exchange of bulk modes remains giving a contribution to the connected correlator.
In our case we take the Wilson loops to be at opposite boundaries but on the same position on the spatial sphere.
Moreover, the connected world-sheet exists for all sizes of the boundary loops.
There is also in our case the contribution given by a supergraviton exchange between the world-sheets of figure \ref{fig:Loop}.
We therefore conclude that the disconnected contribution to the cross Wilson loop connected correlation function is non-zero.

In the following, we focus on non-rotating loops, whose dual dynamics is governed by the dual Nambu-Goto action.
To compute such a loop-loop correlator, one needs to minimise the Nambu-Goto action, together with appropriate boundary conditions that fix the size of the loop at the two asymptotic boundaries. We therefore consider
\be
S_{NG} = \frac{1}{4\pi \alpha'} \int d \sigma d \tau \sqrt{\gamma}\, , \qquad \gamma_{\alpha \beta} = \partial_\alpha X^M \partial_\beta X^N G_{M N}\, ,
\label{NGaction}
\ee
where $G_{M N}$ is the Euclidean target space metric
\be\label{6h}
\frac{ds^2}{L^2} = G_{M N} d X^M d X^N = dr^2 + \left(B \cosh(2 r) - \half \right) d \Omega_3^2 \, ,
\ee
and $\gamma_{\alpha\beta}$ is the induced metric on the world-sheet that can have either a disk $D$, or cylinder $S^1 \times \mathbb{R}$ topology. We re-introduced the AdS length scale $L$, in order to keep track of dimensions, since $B$ is a dimensionless parameter. In these coordinates, $r$ is a dimensionless parameter as well.

In practice one needs to employ an appropriate coordinate system and gauge fixing procedure. In such a case, everything depends on a single world-sheet coordinate,  and  the minimal area problem  translates to ODEs instead of PDEs. We are interested in circular Wilson-loops located on  $S^3$. Therefore, the dual bulk world-sheet must have circular symmetry.

The topology of our target space is $\mathbb{R}\times S^3$ and contains one radial coordinate $X^0 = r$, together with three angular ones parametrising the $S^3$. Since we wish to embed an $S^1$ loop on the three sphere, we must consider different foliations of $S^3$ in terms of $S^2$ and $S^1$. The most efficient way  to parametrize the target space $S^3$,  is using Euler angles (related to the Hopf-fibration). For more details see appendix~\ref{MaurerHopf}. The fibers in this construction are non intersecting great circles of the $S^3$. $t_3$ is the coordinate on the fiber which is a great circle. The unit radius $S^3$ metric takes the form
 \bea\label{8h}
 d \Omega_3^2 &=& {1 \over 4} \biggl( d t_1^2 + d t_2^2 + d t_3^2 + 2 \cos\, t_1 \, d t_2 \rd t_3   \biggr) \, , \nn \\
&\text{with}& \qquad  0\le   t_1 <\pi, \qquad   0\le t_2 < 2\pi, \qquad   -2\pi\le   t_3 <2\pi \, .
 \eea
In order to gauge fix, we need to understand which angle can be used to parametrise the relative size of the loop compared to the three-sphere radius, since the phase transition will depend on this ratio. In order to do this, we embed a three-sphere of radius $R(r)$ in $R^4$ and use the usual angular coordinates
\bea\label{3h}
R^2(r) &=& Y_0^2(r) + Y_1^2(r) + Y_2^2(r) + Y_3^2(r) \, , \nn \\
d\Omega^2_3 &=&  \left(d\psi^2 + \sin^2\left(\psi\right)d\theta^2 + \sin^2\left(\psi\right)\sin^2\left(\theta\right)d\phi^2\right) \, .
\eea
A circular loop located on the $Y_3, \, Y_4$ plane having radius $S(r)$ will be given by
\be\label{4h}
\quad Y_3= S(r) \cos \phi \, , \quad Y_4= S(r) \sin \phi \, , \quad \psi(r),\theta(r) \, .
\ee
Moreover, one can show using the above that
\be\label{5h}
\frac{S(r)}{R(r)} = \sin \psi(r) \sin \theta(r) =  \sin \Omega(r)
\ee
where $\Omega \in [0, \pi]$ is an angle parameter, that dictates the relative size of the loop $S(r)$ to the $S^3$ radius $R(r)$ at each radial slice. We then use the relation between the Euler angles $t_i$ and the usual angles $\psi, \theta, \phi$, found in appendix~\ref{angles connection}, that results in the identification $t_1 = 2 \Omega$. This means that $t_1$ can serve as the parameter that governs the relative size of the Wilson loop on the $S^3$. We can now gauge-fix the world-sheet coordinates as follows:
\be\label{1h}
X^0 \equiv r = \tau\, , \qquad  t_2 = \sigma \, , \qquad t_{1,3} = t_{1,3}(\tau)\, ,
\ee
with $t_i$ the Euler angles of~\eqref{8h}. We can further use a left-over symmetry (for a non rotating loop on the remaining $S^2$), to set $t_3 = const$. The more general case of non-trivial $t_3$ is treated in Appendix~\ref{LoopAppendix}. Then one can write the Nambu-Goto action~\eqref{NGaction} as
\be\label{9h}
S_{NG} = \frac{L^2}{2 \alpha'} \int d r \sqrt{B \cosh (2 r) - \half} \sqrt{1+ \frac{\dot{t}_1^2}{4} \left(B \cosh (2 r) - \half \right)}
\ee
The EOM can be integrated once to give
\be\label{10h}
\dot{t}_1 = \pm \frac{8 C}{\sqrt{B \cosh (2r) - \half} \sqrt{\left( B \cosh (2r) - \half \right)^2 - 16 C^2} }
\ee
with $C$ an arbitrary constant. The simple  solution has $C=0$ and is a constant size loop connecting the two boundaries. This is the only solution with this property. There are other solutions with $C\not=0$ and these will give rise to disconnected surfaces in the bulk.
Therefore there will be a bulk point where the surface will shrink to zero size. We denote this bulk point with $r_m$. At this point $\dot{t}_1(r_m) = \infty$ and $t_1(r_m) = 0$, the first is due to smoothness and the later due to the zero size condition. The first of these conditions fixes
\be\label{11h}
C = \frac{B \cosh 2 r_m - \half}{4}
\ee
Then, the integral of equation \eqref{10h} is
\be\label{12h}
t_1(r) = \pm \int_{r_m}^r d r \frac{2 \left( B \cosh (2r_m) - \half \right) }{\sqrt{B \cosh (2r) - \half} \sqrt{\left( B \cosh (2r) - \half \right)^2 - \left( B \cosh (2r_m) - \half \right)^2 }} \, ,
\ee
where the endpoint is defined such that $t_1(r_m) = 0$. There is still a one parameter freedom. Performing the following integral
\be\label{12hi}
t_1(\infty) = \pm \int_{r_m}^{\infty} d r \frac{2 \left( B \cosh (2r_m) - \half \right) }{\sqrt{B \cosh (2r) - \half} \sqrt{\left( B \cosh (2r) - \half \right)^2 - \left( B \cosh (2r_m) - \half \right)^2 }}= t^*
\ee
we define $t_1(\infty) = t^* $ as the asymptotic value of the angle at the boundary. For large $B$ one finds the simple relation
\be\label{122hi}
t^* \simeq \pm \int_{r_m}^{\infty} d r \frac{2 \cosh (2r_m)  }{\sqrt{B \cosh (2r)} \sqrt{\left( \cosh (2r) \right)^2 - \left(  \cosh (2r_m) \right)^2 }} =
\ee
$$
 =\half \int_{0}^{x_m} d x \frac{1}{\sqrt{B \left(1- x \right) \left( x_m  -x \right) }}
= \frac{1}{\sqrt{B}} \log \frac{ \left( 1 + \sqrt{x_m} \right)}{ \left( 1 - \sqrt{x_m} \right)} = \frac{2}{\sqrt{B}} \log \coth r_m   \, ,
$$
with
\be\label{123hi}
x_m = {1\over ( \cosh 2 r_m)^2} < 1\;.
\ee
\begin{figure*}[!tb]
\begin{center}
\includegraphics[width=0.45\textwidth]{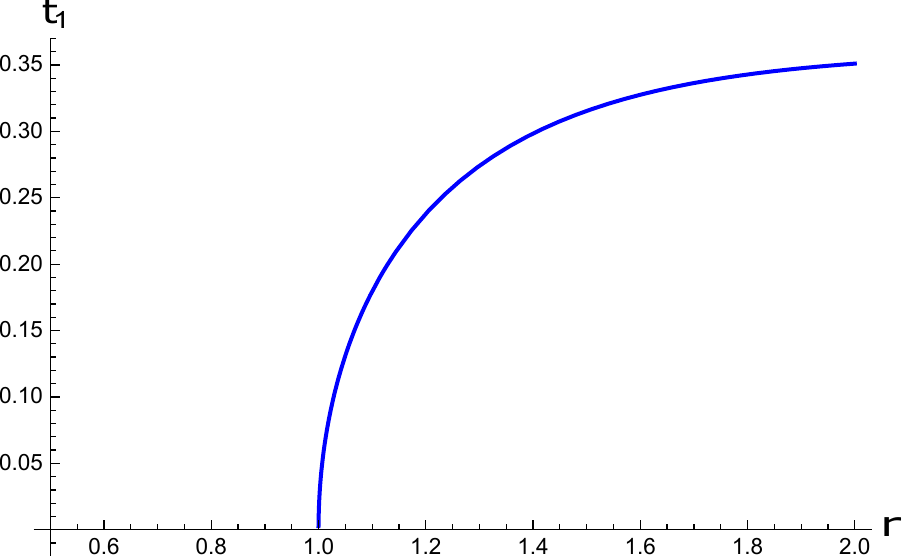}%
\hspace{4mm}
\includegraphics[width=0.45\textwidth]{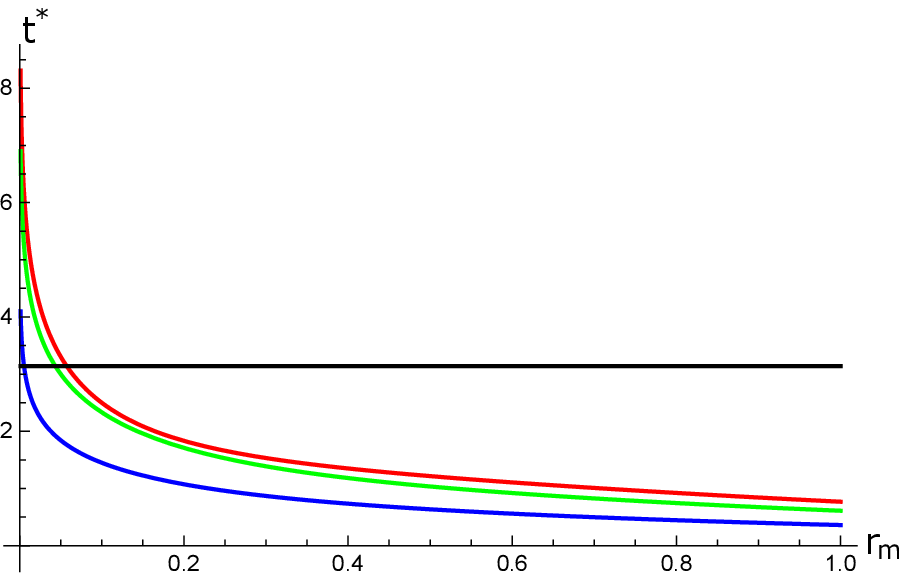}
\end{center}
\caption{Left: The angle $t_1$ as a function of radial position for $B=3$. It vanishes at $r=r_m$ where the loop shrinks to zero size. Right: The asymptotic value of the angle $t^*$ as a function of $r_m$. Large loops on the boundary penetrate further in the bulk (smaller $r_m$). There is a saturation on how far in the bulk the disconnected string can go, where the line crosses $t^* = \pi$, that is a maximal boundary size loop. The lines are for  $B={0.6,1,3}$ from red to blue.}
\label{fig:angle}
\end{figure*}
The on-shell action for the connected solution is (using the results of section~\ref{ESUellipticmetric})
\be\label{13h}
S_{NG}^{c} = \frac{L^2}{\alpha'} \int_{0}^\infty d r \sqrt{B \cosh (2 r) - \half} \, = \frac{\left(B - \half \right) L^2}{\alpha' \sqrt{2B}} \int_{0}^{K(k)}  \frac{d u}{\cn^2 (u, k)} \, .
\ee
This can be integrated to give ($E$ is the elliptic integral of the second kind)
\be\label{15h}
S_{NG}^{c} = \frac{L^2}{\alpha'} \left[ 2 \sqrt{\left(B - \half\right)} \,  E\left(i r, \frac{2B}{B - \half} \right) \right]_0^\infty \, .
\ee
The value of the expression above at  $r=0$ vanishes.
Therefore, the only contribution originates from the UV $r \rightarrow \infty$ limit, and contains divergent and finite subleading pieces that depend on $B$.
The on-shell action for the product of two disconnected solutions is
\bea\label{14h}
S_{NG}^d &=& \frac{L^2}{\alpha'} \int_{r_m}^\infty d r \frac{\left(B \cosh (2 r) - \half \right)^{3/2}}{\sqrt{\left( B \cosh (2r) - \half \right)^2 - \left( B \cosh (2r_m) - \half \right)^2 }}  \, \nn \\
&=& \frac{\left(B - \half\right)L^2}{\alpha' \sqrt{2B}} \int_{u_m}^{K(k)}  \frac{\cn^2 (u_m, k)}{\cn^2 (u, k)} \frac{ d u }{\sqrt{\cn^4 (u_m, k) - \cn^4 (u, k)}} \, .
\eea
This can be only expressed in terms of hyperelliptic integrals. An analytic expression can be given in the limit of large $B$
\bea\label{16h}
S_{NG}^d &\approx& \frac{\sqrt{B} L^2}{\alpha'} \int_{r_m}^\infty d r \frac{\left(\cosh (2 r) \right)^{3/2}}{\sqrt{\left( \cosh (2r)  \right)^2 - \left( \cosh (2r_m) \right)^2 }}  \,  \nn \\
&=& \frac{\sqrt{B} L^2}{4 \alpha'} \int^{x_m}_0 d x \frac{x^{-5/4}}{\sqrt{\left(1 -x \right) \left(1 - x/x_m  \right) }} = -\frac{\sqrt{B} L^2 \sqrt{1-x} \sqrt{1-\frac{x}{{x_m}}}}{\alpha' \sqrt[4]{x}} \bigg|_0^{x_m} + \nn \\
&+& \frac{ 3 \sqrt{B} L^2 x^{7/4}
   F_1\left(\frac{7}{4};\frac{1}{2},\frac{1}{2};\frac{11}{4};x,\frac{x}{{x_m}}\right)}{7 \alpha'
   {x_m}}\bigg|_0^{x_m}  -\frac{\sqrt{B} L^2 x^{3/4} ({x_m}+1)
   F_1\left(\frac{3}{4};\frac{1}{2},\frac{1}{2};\frac{7}{4};x,\frac{x}{{x_m}}\right)}{3 \alpha'
   {x_m}} \bigg|_0^{x_m}  \nn \\
\eea
with $x_m = 1/ \cosh^2 (2 r_m) < 1$, where the specific $F_1$ is Appell's function. The first term exhibits a UV singularity as $x \rightarrow 0$ that is to be subtracted. This UV divergence is the same in the connected solution~\eqref{13h} as well. The other two terms obtain a contribution only from the $x\to x_m$ limit. We then define the renormalised disconnected action as
\bea\label{162h}
S_{NG}^{d-ren}(x_m) &=& \lim_{x \rightarrow 0} \left[ S_{NG}^d  \, + \,  \frac{\sqrt{B} L^2 \sqrt{1-x} \sqrt{1-\frac{x}{{x_m}}}}{\alpha' \sqrt[4]{x}} \bigg|_x^{x_m} \right] \nn \\ &\approx & \frac{\sqrt{B} L^2}{\alpha'}
\frac{\pi ^{3/2} \left(9 {x_m} \,
   _2F_1\left(\frac{1}{2},\frac{7}{4};\frac{9}{4}; {x_m}\right)-5 ({x_m}+1) \,
   _2F_1\left(\frac{1}{2},\frac{3}{4};\frac{5}{4}; {x_m}\right)\right)}{8 \sqrt{2}
   \sqrt[4]{{x_m}} \Gamma \left(\frac{5}{4}\right) \Gamma \left(\frac{9}{4}\right)} \, . \nn \\
\eea

\begin{figure}[t]
\vskip 10pt
\centering
\includegraphics[width=80mm]{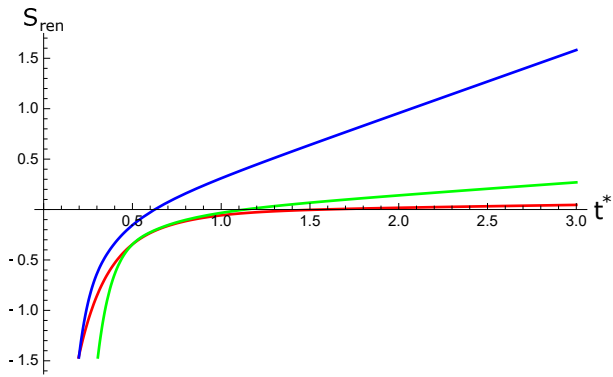}
\caption{The renormalised action $S^{d-ren}$ as a function of $t^*$. The plots are for $B={0.6 , \, 1, \, 3}$ from red to blue. For large loop, the angle $t^*$ approaches $\pi$ and the action is found to scale approximately linearly with $t^*$, while for smaller $t^*$ it grows slower (logarithmically). The approximately linear behaviour is reached faster for large values of $B$ and the slope scales as $\sqrt{B}$.}
\label{fig:loopactionregularised}
\end{figure}

In order to distinguish whether the Wilson loop one-point function exhibits area or perimeter law or some other behaviour, we provide the formula for the length of the loop in terms of $t^*$ using~\eqref{5h}
\be\label{19h}
L^{loop} =  2 \pi R \sin \frac{t^*}{2}
\ee
where $R = L \sqrt{B}$ is the radius of the asymptotic $S^3$ that can be read off from the asymptotic form of the bulk metric~\eqref{6h}.
On the other hand, the area enclosed by the loop on the $S^3$ is given by
\be\label{20h}
A^{loop} = \frac{\pi R^2}{2} \int_0^{t^*} dt_1 = \frac{\pi R^2}{2} t^* \, ,
\ee
which is linear in $t^*$ and quadratic in $R= L \sqrt{B}$. Using~\eqref{12hi} and~\eqref{122hi} we find that $t^* \sim 1/\sqrt{B}$ for large $B$ (decompactification limit). The area then scales as $A^{loop}  \sim L^2 \sqrt{B}$, which matches the overall scaling of the on-shell action~\eqref{16h} in the same limit. On the other hand,  the perimeter scales always linearly with $L$ and can never match the on-shell action behaviour. In addition,  in this decompactification limit, we can compare the two functions~\eqref{162h} and~\eqref{20h} using~\eqref{122hi}. We find that in the limit of $x_m \rightarrow 1$, they both scale and diverge logarithmically as $-\log(x_m-1)$, but differ in a constant term. This is the limit for a relatively large loop that goes deep in the bulk.

In order to further compare the area and the on-shell action as functions of $t^*$ for various values of $B$,  we can resort to numerical techniques. The general result can be seen in fig.~\ref{fig:loopactionregularised}. We find that indeed the Wilson loop one-point function always acquires an asymptotic behaviour linear in $t^*$,  that matches the behaviour of the area~\eqref{20h}. This is for the limit of a large loop close to the equator and is achieved faster for large values of $B$. If one takes the decompactification limit, this is an area law as long as $R\gg \ell_s$, where $\ell_s$ is the fundamental string length.
 Therefore, the bulk theory appears to be dual to a confining boundary theory.
 For small Wilson loops, there is some screening mechanism that results in a slower growth (a logarithmic rate).

\begin{figure*}[!tb]
\begin{center}
\includegraphics[width=0.45\textwidth]{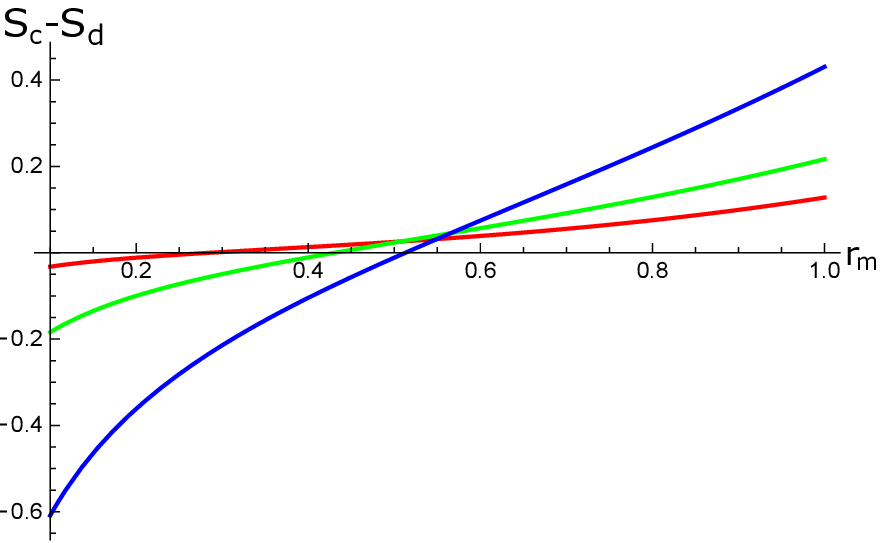}%
\hspace{4mm}
\includegraphics[width=0.45\textwidth]{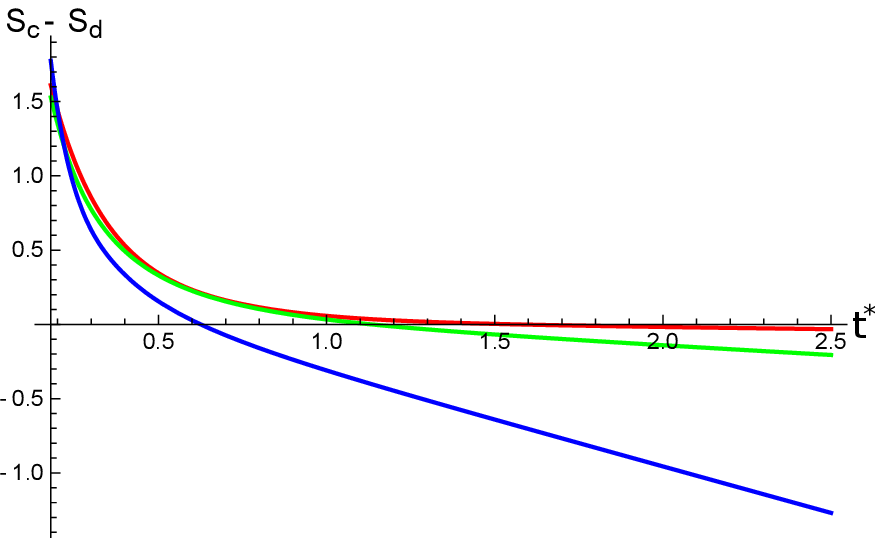}
\end{center}
\caption{The difference of the connected to the disconnected action $S_{conn}-S_{disc}$. The plots are for $B={0.6 , \, 1, \, 3}$ from red to blue. For large $r_m$/small $t^*$ that correspond to a small loop size the disconnected solution dominates, while for small $r_m$/large $t^*$ that correspond to a large boundary loop size the connected dominates. As B approaches $1/2$, the geometry pinches-off and there is only the disconnected solution. This analysis not complete: one has to take into account the contribution of bulk perturbative modes as described in the main text.}
\label{fig:loopactiondifference}
\end{figure*}

Finally, by changing the asymptotic parameter $t^*$ that controls the size of the loops,  we can compare the semiclassical area of the connected contribution to the area of the disconnected contribution.

 The result is in the plot  \ref{fig:loopactiondifference}, where this action difference is shown. This difference is free of UV divergences. We find that the disconnected solution is dominant for small relative asymptotic loop size. As we increase that size, the connected solution has smaller area.
However in order to turn this difference into a comparison between the two connected contributions in figure \ref{fig:Loopcorrelator} we need to have a calculation of the intermediate propagator corrections in (\ref{equationsofthepropagatorsinthebulk}). The sharp phase transition might then smoothen into a cross-over behaviour.  Unfortunately, this is difficult to compute at the moment.

We conclude with a final comment. From the work of~\cite{Maldacena:2004rf} we know that the meron wormhole can be embedded in eleven-dimensional supergravity compactified on $S^7/\mathbb{Z}_k$, whose dual description is in terms of ABJM theory~\cite{Aharony:2008ug}. This theory has a $U(N)_k \times U(N)_{-k}$ gauge symmetry and can be represented as a quiver with two nodes that interact through bi-fundamental matter fields. It would be interesting then to consider the ABJM theory deformed by the presence of a background source for the $R$-symmetry currents that is the same as the one in our bulk solution.
Note that this source makes sense only in Euclidean space, and becomes complex in Minkowski space as described in appendix~\ref{LorentzianHopf}.
One could then try to study Wilson loop correlation functions in such a setup at strong coupling. This would then indicate whether such a state could be given a wormhole interpretation corresponding to the meron wormhole solution.

\section{Multi-trace deformations}\label{Doubletracedeformations}
In this section we study multi-trace deformations \cite{Witten:2001ua}, \cite{Hartman:2006dy} of the state dual to the wormhole background.

Multi-trace deformations of the boundary quantum field theory correspond ---through the holographic dictionary--- to imposing mixed boundary conditions to the dual bulk field. Such deformations can trigger non-trivial RG-flows.

The gravitational solutions that we study in this paper have two boundaries. Therefore, there are more possibilities: one can add deformations to any or both of the boundaries. The simplest one is to deform the theory at one of the two boundaries, for example by adding the term
\be\label{6.a}
g \int_{1} d^d x ~O_1^2(x)\,.
\ee
where $O_{1}$ is an operator of the schematic form $O_{1}=Tr[\Phi^2]$ where $\Phi$ a scalar field such that $O_{1}^{2}=Tr[\Phi^2]^2$ induces a double-trace deformation with coupling $g$ on one of the two boundaries ---in this instance on the first boundary. Equivalently one can perform the same deformation on the second boundary.

A third non-trivial possibility is to add a deformation of the type
\be
g \int d^d x \, O_1(x) O_2(x) \, ,
 \ee
that couples directly the two boundaries.
Assuming a symmetric wormhole case with two boundaries at $r \rightarrow \pm \infty$, such a double-trace deformation can be employed in the bulk description using the cross-boundary conditions

\bea\label{cross_boundary}
\phi(r,x)|_{\partial\mathcal{M}_{1}} &\sim& a_{1}(x) r^{-\Delta} + \beta_1(x) r^{\Delta - d} \, , \quad \beta_2(x) = g \, \alpha_1(x) \nn \\
\phi(r,x)|_{\partial\mathcal{M}_{2}} &\sim&  a_{2}(x)r^{-\Delta} + \beta_2(x) r^{\Delta - d} \, , \quad \beta_1(x) = g \, \alpha_2(x)
\eea
This simply means that the vev of the operator at the one boundary is correlated to the source of the operator at the second boundary and vice-versa.
Therefore one needs both the leading and the subleading behaviour to be normalisable at each boundary (or $\Delta < d$). The generic multi-trace deformation would then correspond to
\bea\label{multi_trace_bulk}
\phi(r,x)|_{\partial\mathcal{M}_{1}} &\sim& a_{1}(x) r^{-\Delta} + \beta_1(x) r^{\Delta -d} \, , \quad \beta_1(x) = f\left( \alpha_1(x),\, \alpha_2(x) \right) \, ,  \nn \\
\phi(r,x)|_{\partial\mathcal{M}_{2}} &\sim&  a_{2}(x)r^{-\Delta} + \beta_2(x) r^{\Delta - d} \, , \quad \beta_2(x) = g \left( \alpha_1(x),\, \alpha_2(x) \right) \, .
\eea

We now study generic double trace deformations ($i,\, j$ label the two boundaries) of the form
\be
S_{int} =\half C_{ij} \int d^d x ~O_i(x) O_j(x)\sp i,j=1,2 \, ,
 \ee
where $C_{ij}$ is a $2 \times 2$ coupling constant matrix whose elements have mass dimension $d-\Delta_{i}-\Delta_{j}$.
These deformations change the two point functions computed on the initial wormhole state. To compute this effect, we perform the following Hubbard-Stratonovich transformation
\be\label{d1}
\exp\left(S_{int}\right)= N_{0}\int \prod_{i=1}^2 \mathcal{D}\zeta_{i} \exp\left[\int d^{d}x\left( - \half \zeta_{i}(x)C^{-1}_{ij}\zeta_{j}(x)+\zeta_{j}(x)\mathcal{O}_{j}(x)\right)\right] \, .
\ee
The deformed generating functional can be written as
\be\label{d2}
e^{- \mathcal{W}(J)} = \mathcal{Z}\left(J\right)=\int \left[\prod_{i=1}^2 \mathcal{D}\zeta_{i}\, Z\left(J_{j}+\zeta_{j}\right)\right]~e^{\int d^{d}x\left( -\half \zeta_{i}(x)C^{-1}_{ij}\zeta_{j}(x)\right)}
\ee
where
\be\label{d3}
Z\left(J\right)= \langle W|e^{\int d^{d}x J_{j}\mathcal{O}_{j}}|W\rangle
\ee
is the generating functional of correlators in the original wormhole state $|W\rangle$.
We work in the large $N$ quadratic approximation where
\be\label{d4}
Z\left(J_{i},J_{j}\right)=e^{\frac{1}{2}\int d^{d}x d^{d}y\,J_{i}(x)G_{ij}(x-y)J_{j}(y)}= e^{\frac{1}{2}\int \frac{d^{d}p}{(2\pi)^{d}}J_{i}(p)G_{ij}(p)J_{j}(-p)}
\ee
with $G_{ij}(x-y)=\langle\mathcal{O}_{i}(x)\mathcal{O}_{j}(y)\rangle_W$ the undeformed two point functions.

By integrating out the auxiliary fields $\zeta_{i}$ we obtain
\bea\label{d5}
\mathcal{W}\left(J \right) &=& - \half \int \frac{d^d p}{(2 \pi)^d} J_{k}\left(G_{ki}\left(C_{ij}^{-1}-G_{ij}\right)^{-1}G_{jl}+G_{kl}\right)J_{l}\nn\\
&=& - \half \int \frac{d^d p}{(2 \pi)^d} J_{i}\left(-C_{ij}+G_{ij}^{-1}\right)^{-1}J_{j}\, .
\eea
To illustrate the effect of the deformation on the correlators we present in a matrix form the deformed correlators $\tilde{G}_{ij}$.
\begin{itemize}

\item When only $C_{11}$ is turned on we find the deformed correlator $\tilde{G}_{ij,11}$
\be\label{GIJ11}
\tilde{G}_{ij,11} =
\left( \begin{matrix}
 \, \frac{G_{11}}{1-C_{11} G_{11}} \, & \, \, \frac{G_{12}}{1-C_{11} G_{11}} \, \\
\, \frac{G_{12}}{1-C_{11} G_{11}} \, & \, \,  G_{22} + \frac{ C_{11} G_{12}^2 }{1-C_{11} G_{11}} \, \\
\end{matrix} \right) \, .
\ee
The case that only $C_{22}$ is turned on is very similar and we omit it.

\item In the case that only $C_{12}$ is turned on the deformed correlators take the following form
\be\label{GIJ12}
\tilde{G}_{ij,12}=\begin{pmatrix} \,
\frac{G_{11}}{(C_{12} G_{12}-1)^2-C_{12}^2 G_{11} G_{22}} \,  &  \, \, \frac{C_{12}(G_{11} G_{22}- G_{12}^2)+G_{12}}{(C_{12} G_{12}-1)^2-C_{12}^2 G_{11} G_{22}} \\ \\
 \, \frac{C_{12}(G_{11} G_{22}- G_{12}^2)+G_{12}}{(C_{12} G_{12}-1)^2-C_{12}^2 G_{11} G_{22}} \, & \, \, \frac{G_{22}}{(C_{12} G_{12}-1)^2-C_{12}^2 G_{11} G_{22}} \\
\end{pmatrix}
\ee

\end{itemize}

We notice above that the correlators can now exhibit poles at finite momentum, for example if $C_{11} G_{11} = 1$. However, a pole in a Euclidean correlator at real values of the momentum indicates an instability. Therefore, allowed deformations should have couplings $C_{ij}$ so that this does not happen\footnote{For explicit examples of such deformations and associated stability criteria see \cite{KN1}.}.  We will assume then that the coefficients $C_{ij}$ are such that no poles can develop at finite Euclidean momentum, so that we ensure the stability of the theory after deformation.

We now present in more detail how of a deformed correlator looks in the UV and in the IR, using representative examples, the rest of the cases admit a similar analysis.

The simplest analysis is for the IR structure. Using the results of section \ref{HomogeneousODES},
\bea\label{GIR}
G_{11}^{IR}& \approx &a_{11}+b_{11} p^{2} + ...\\\nn
G_{22}^{IR}& \approx &a_{22}+b_{22} p^{2} + ...\\\nn
G_{12}^{IR}& \approx &a_{12}-b_{12} p^{2} + ...
\eea
where $a_{ij},\, b_{ij}$ positive (dimensionfull) parameters. Combining these IR expansions with~\eqref{GIJ11},~\eqref{GIJ12} it is straightforward to see that all the deformed correlators retain a similar IR structure $G^{IR} \sim A + B p^2 + ...$, with new renormalised coefficients that depend on the original parameters. We therefore conclude that the relevant double trace deformations drive the theory in a new IR fixed point with similar behaviour to that of the original state, unless there is a change in sign of the quadratic $p^2$ term so that the IR minimum
becomes a maximum and vice-versa. It would be interesting to study further whether such a possibility can actually be realised in an explicit example, or if the deformations are completely innocuous for the IR structure of the theory.

More interesting is the UV analysis. One uses the UV expansions of the undeformed correlators
\bea\label{GUV}
G_{11}^{UV} & \approx & g_{11} ~p^{(2\Delta_{1} - d)} + ...\\\nn
G_{22}^{UV} & \approx & g_{22} ~p^{(2\Delta_{2} - d)} + ...\\\nn
G_{12}^{UV} & \approx & g_{12} ~p^{-\Delta_{12}} + ...
\eea
where $g_{ij},\, a$ are dimensionless constants and $\Delta_{12}>d$ is a positive constant such that the cross-correlator $G_{12}^{UV}$ asymptotes to zero for large momenta and there is no short-distance singularity, based on the results of section \ref{HomogeneousODES}\footnote{Notice that one could also take the cross-correlator to be exponentially suppressed $G_{12}^{UV} \sim p^a e^{- b p}$, the qualitative results remain unaltered.}.

In the case where only $C_{11}\not =0$, we must take the operator $O_1$ to be such that $O_1^2$ is relevant, $2\Delta_1\leq d$.
In that case the short distance structure of $G_{11}$ and $G_{12}$ remains unaltered, but the short distance structure of $G_{22}$ could in principle  change.
Indeed, from \eqref{GIJ11}, if $\Delta_{12}<{d\over 2}-\Delta_2$, then the $G_{12}^2$ term would dominate the $G_{22}$ term. However, the constraint $\Delta_{12}>d$ forbids this case.

We now examine the representative example of the $11$ correlator when only the coupling $C_{12}$ is turned on. In this case we obtain
\be \label{G1112}
\tilde{G}_{11,12}=\frac{G_{11}}{\left(1 - C_{12} G_{12}\right)^2 - C_{12}^{2} G_{11} G_{22}} \, .
\ee
In the case of theories where the spectra at the two boundaries are symmetric, $\Delta_1=\Delta_2$ and the UV structure of the deformed correlator $\tilde{G}_{11,12}$ is similar to $G_{11}$. However, if the dimensions on the two boundaries are different,  and if $\Delta_1+\Delta_2-d>0$, then
\be
\lim_{p\to\infty}\tilde{G}_{11,12}=-{1\over C_{12}G_{22}}
\ee

However, the most radical change happens in the deformed
$\tilde{G}_{12,12}$ correlator
\be \label{G1212}
\tilde{G}_{12,12} =  \frac{C_{12}(G_{11} G_{22}- G_{12}^2)+G_{12}}{(C_{12} G_{12}-1)^2-C_{12}^2 G_{11} G_{22}} \, .
\ee
If we assume the canonical case $\Delta_1+\Delta_2-d<0$, then the denominator is near one, in the UV limit.
However, in the numerator, the term $G_{11}G_{22}$ dominates in the UV.
If
\be
{d\over 2} ~~<~~\Delta_1+\Delta_2~~<~~d
\ee
then the cross correlator now acquires a singularity at short distance.
It is not clear whether  the whole setup allows double-trace deformations
with $C_{12}\not=0$. If it does however,  this seems to be in agreement with
qualitative properties of the putative field theory example discussed in the next section.

\section{A potential field theory analogue\label{field theory analogue}}

A basic property of the theories we study is that the two sectors interact ``mildly" and,  in particular,  mixed correlators do not have short distance singularities when operators for different boundaries collide.
Moreover, their short distance structure is very soft and the Fourier coefficients bounded above.

There are several levels of interaction between two large-N theories. The strongest interaction is mediated by charged/colored field (bifundamentals).
A weaker set of interactions are generated by multi-trace interactions. Of course at low energies, multi-trace interactions can be the avatar of bifundamental interactions at high energies. On the other hand, in special cases,  multi-trace interactions can be UV complete, \cite{KN1}.

In the case of local multi-trace interactions, as argued in \cite{K,ACK}, the  geometric bulk picture of the AdS duals is in terms of asymptotically AdS spaces identified at their respective boundaries. This is a radically different picture of what we have here. Moreover, it is clear that the existence of standard local interactions between two theories, will make cross correlators have short distance singularities as the correlators of the individual theories.

It is natural to expect that we must explore theories whose cross interactions are softer at shorter distances. Moreover, another property of the wormhole solutions we are exploring is that most probably they do not have sensible Minkowski signature continuations. Therefore their putative field theory duals will not exist as regular field theories in Lorentzian signature. Some discussion on two possible types of analytic continuation and the problems one encounters is given in Appendices~\ref{Bang-Crunch} and~\ref{LorentzianHopf}.

In view of the above, in this section we shall try to construct some simple field theory analogues with similar  properties.

We assume we have two Euclidean theories $S_1$  and $ S_2$ and two local operators $O_1(x)\in S_1$ and $O_2(x)\in S_2$ that define the (non-local) interaction between the two theories,
\be
S=S_1+S_2+\lambda \int d^d x~d^d y ~O_1(x)O_2(y)f(x-y)
\label{a1}\ee
where $f(x)$ is a smooth function to be specified shortly.

This can be written in momentum space as
\be
S=S_1+S_2+\lambda\int {d^d p\over (2\pi)^d}O_1(p)O_2(-p)\tilde f(p)
\label{a2}\ee
where we define the Fourier transform as usual,
\be
O_1(x)=\int {d^d p\over (2\pi)^d}e^{ip\cdot x} ~O_1(p)
\label{aa3}\ee
Consider now the cross two-point function
\be
\langle O_1(x)O_2(y)\rangle=\lambda\int d^d z_1 d^d z_2\langle O_1(x)O_1(z_1)\rangle \langle O_2(y)O_2(z_2)\rangle f(z_1-z_2)+{\cal O}(\lambda^2)=
\label{a4}\ee
$$
=\lambda\int {d^d p\over (2\pi)^d}G_{11}(p)G_{22}(-p)\tilde f(p)~e^{ip\cdot (x-y)}+\cdots
$$
where
\be
\langle O_1(p)O_1(q)\rangle\equiv G_{11}(p)(2\pi)^d\delta(p+q)
\sp
\langle O_2(p)O_2(q)\rangle\equiv G_{22}(p)(2\pi)^d\delta(p+q)
\label{a5}\ee
If the UV scaling dimension of $O_{1,2}$ is $\Delta_{1,2}$ then at short distances\footnote{We assume $\Delta_{1,2}$ are not integers.}
\be
 G_{11}(p)\sim p^{2\Delta_1-d}\sp   G_{22}(p)\sim p^{2\Delta_2-d}
\label{a6} \ee
If we want that  \eqref{a4} has no short distance singularity\footnote{Notice that we can also achieve this condition setting $\tilde{f}(p) \sim p^b e^{- c p}$, the qualitative results remain unaltered.}, then we must have that as $p\to\infty$
\be
\tilde{f}(p)\sim p^{-a}\sp \langle O_1(x)O_2(y)\rangle\sim {1\over |x-y|^{2\Delta_1+2\Delta_2-d-a}}\eqp
\label{a7}\ee
We therefore require that
\be
a>2\Delta_1+2\Delta_2-d\geq d-4\eqc
\label{a8}\ee
where in the last inequality in \eqref{a8} we used the unitarity bound on scalar scaling dimensions.
If \eqref{a8} is valid, then  the two-point function in \eqref{a4} is regular at short distances.
Note that in order for the interaction in \eqref{a1} to be relevant we must have $\Delta_1+\Delta_2-d<a$ which is subsumed by \eqref{a8} as expected.

Further, the all order result in $\l$ can be obtained from section 4 of \cite{ECA} by making the coupling non-local and rotating to Euclidean space,
\be
\langle O_1O_2\rangle (p)=\lambda{G_{11}(p)G_{22}(-p)\tilde f(p)\over 1+\l^2 ~G_{11}(p)G_{22}(-p)\tilde f(p)^2}+\cdots
\label{a88}\ee
The ellipsis stands for corrections due to higher-point functions that are subleading at large-N. \eqref{a88} implies that the higher orders in $\l$ induce softer and softer interactions.

We shall now make a free field example of this setup.
Consider two scalars $\tilde\phi_{1,2}$ interacting via a non-local soft interaction
\be
S=\int{d^dp\over (2\pi)^d}\left[\tilde\phi_1(p)(p^2+\tilde m^2)\tilde\phi_1(-p)+\tilde\phi_2(p)(p^2+\tilde m^2)\tilde\phi_2(-p)+2{M^4\over p^2+\tilde \Lambda^2}\tilde\phi_1(p)\tilde\phi_2(-p)\right]
\label{a9}\ee
$$
=\int{d^dq\over (2\pi)^d}\left[\phi_1(q)(q^2+ m^2)\phi_1(-q)+\phi_2(q)(q^2+m^2)\phi_2(-q)+{2\over q^2+ \Lambda^2}\phi_1(q)\phi_2(-q)\right]
$$
where
\be
q={p\over M}\sp m={\tilde m\over M}\sp \Lambda={\tilde \Lambda \over M}\sp \phi_{i}=M^{d+2\over 2}\tilde \phi_{i}
\label{a10}\ee
are all dimensionless.
We may now diagonalize the action above as
\be
S=\int{d^dq\over (2\pi)^d}\left[\phi_+(q)D_+(q)\phi_+(-q)+\phi_-(q)D_-(q)\phi_-(-q)\right]
\label{a11}\ee
with
\be
\phi_{\pm}={\phi_1\pm\phi_2\over \sqrt{2}}\sp D_{\pm}(q)=q^2+m^2\pm{1\over q^2+\Lambda^2}
\label{a12}\ee

We would like now to arrange that the Euclidean theory ($p^2\geq 0$) is well defined and for this we must have that
$D_{\pm}(q)>0$ for all Euclidean momenta.
For this we must demand that
\be
m^2\Lambda^2>1
\label{a13}\ee
so that $D_{\pm}(q)>0$ for all Euclidean momenta.
In this case, the Euclidean propagators $D_{\pm}^{-1}$ are well defined and finite everywhere.

We now obtain
\be
\langle \phi_1\phi_1\rangle(q)\equiv G_{11}(q)=G_{22}(q)\equiv \langle \phi_2\phi_2\rangle(q)={D_{+}^{-1}(q)+D_{-}^{-1}(q)\over 2}=
\label{a14}\ee
$$
={(q^2+m^2)(q^2+\Lambda^2)^2\over (q^2+m^2)^2(q^2+\Lambda^2)^2-1}={1\over q^2+m^2-{1\over (q^2+m^2)(q^2+\Lambda^2)^2}}
$$
Therefore $G_{11}$ at short distances behaves as in the original theory
\be
G_{11}={1\over q^2+m^2+{\cal O}(q^{-6})}
\label{a16}\ee
The same applies to large distances but the effective mass $\hat m^2$ is different
\be
G_{11}={1\over m^2\left(1-{1\over m^4\Lambda^4}\right)+\left(1+{1\over m^4\Lambda^4}+{2\over m^2\Lambda^6}\right){q^2}+{\cal O}(q^{4})}
\label{a25}\ee
\be
\hat m^2={\left(1-{1\over m^4\Lambda^4}\right)\over \left(1+{1\over m^4\Lambda^4}+{2\over m^2\Lambda^6}\right)}m^2>0
\label{a26}\ee
This is a signal of the IR-relevance of the interaction.

On the other hand
\be
\langle \phi_1\phi_2\rangle(q)\equiv G_{12}(q)={D_{+}^{-1}(q)-D_{-}^{-1}(q)\over 2}=-{q^2+\Lambda^2\over  (q^2+m^2)^2(q^2+\Lambda^2)^2-1}
\label{a15}\ee
which is suppressed at short distances
\be
G_{12}\sim {1\over q^6}+{\cal O}(q^{-4})
\label{a17}\ee
This implies that in real space, $G_{12}(x-y)$ asymptotes to a constant as $x\to y$ as long as $d<6$. This is qualitatively similar to what we found in the holographic examples.

The form of the interaction of the two theories in (\ref{a9}) suggests that it can be resolved by integrating in two new fields. This is better visible in the diagonal form (\ref{a11}). We can integrate in new scalar field linearly coupled to $\phi_{\pm}$ respectively, but it is clear that one of them is always a ghost\footnote{In a Euclidean theory this means that the relevant kinetic term is negative.}. Therefore in this coupled non-local theory, it is not possible to resolve the interaction and make it local in a sensible fashion.

\begin{figure*}[!tb]
\begin{center}
\includegraphics[width=0.5\textwidth]{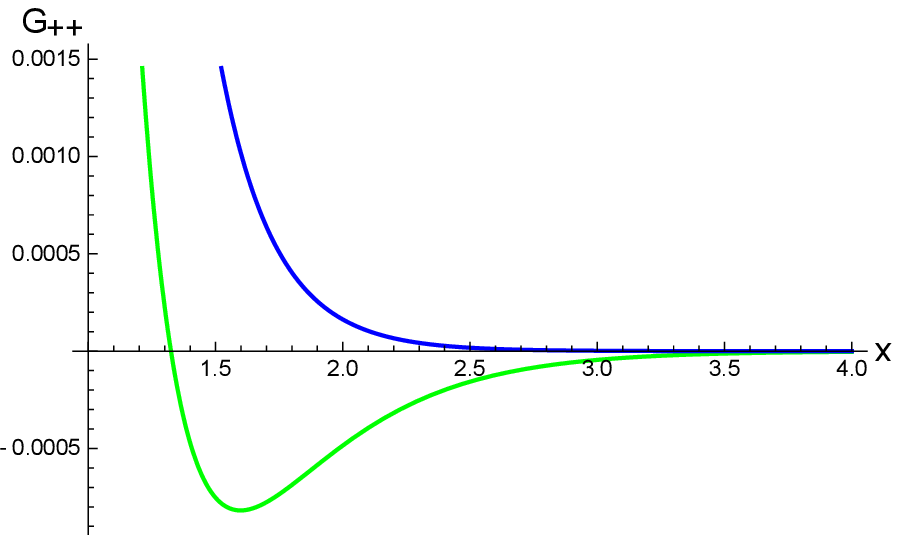}
\hspace{1mm}
\includegraphics[width=0.47\textwidth]{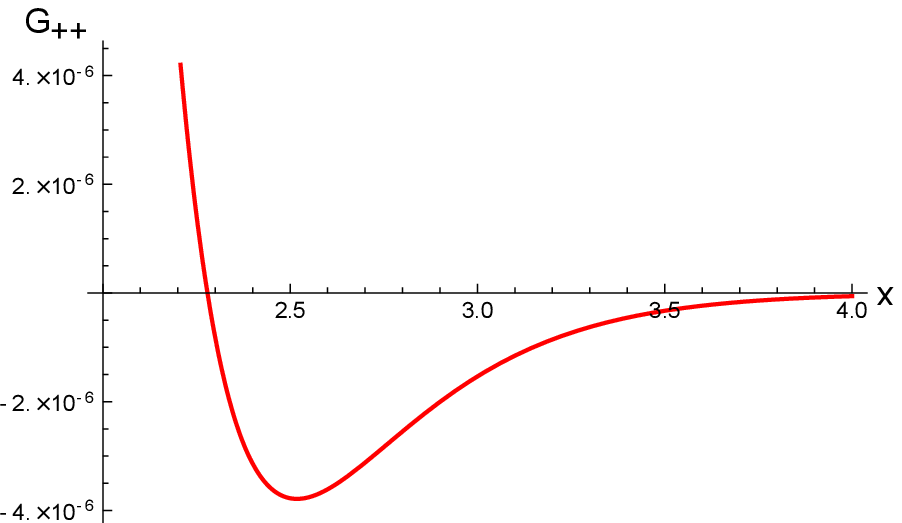}
\end{center}
\caption{The correlator $G_{++}(x)$. Left: The case $|m^2-\Lambda^2|\geq 2$. The blue line represents the case $\Lambda^2 - m^2 \geq 2$ and the green line the case $m^2-\Lambda^2 \geq 2$. Right: The case  $|m^2-\Lambda^2|< 2$. Only the blue line corresponds to a reflection positive correlator for all distances.}
\label{fig:reflectionpos}
\end{figure*}

The fact that the inverse propagators $D_{\pm}$ are positive and non-vanishing in momentum space does not guarantee reflection positivity.
We will now explicitly analyze this property by Fourier transforming the Euclidean propagator. Without loss of generality, we shall present here the $d=3$ case, where the formulae are more compact.
We first write
\be\label{a20}
{1\over D_+}={q^2+\Lambda^2\over (q^2+m^2)(q^2+\Lambda^2)+1}={q^2+\Lambda^2\over (q^2+q_+^2)(q^2+q_-^2)}={A_+\over q^2+q_+^2}+{A_-\over q^2+q_-^2}
\ee
with
\be\label{a21}
A_{\pm}={1\over 2}\left[1\pm{{\Lambda^2-m^2\over 2}\over \sqrt{{(m^2-\Lambda^2)^2\over 4}-1}}\right]\sp q^2_{\pm}={m^2+\Lambda^2\over 2}\mp \sqrt{{(m^2-\Lambda^2)^2\over 4}-1}
\ee
The relevant expression for the two-point function is
\be
G_{++}(x) = \langle \phi_+(x) \phi_+(-x) \rangle =\int {d^3q\over (2\pi)^3}{e^{2iq\cdot x}\over D_+(q)}=
\label{a27}\ee
\be=
  \frac{1}{2|x|} \int_0^\infty \frac{d q}{(2 \pi)^2} q \sin \left( 2 q |x| \right)  \left( \frac{A_+}{q^2 + q_+^2} + \frac{A_-}{q^2 + q_-^2} \right)=
\label{a28}\ee
\be
= \frac{1}{8 \pi |x|}\left( A_+ e^{- 2 q_+ |x|} + A_- e^{- 2 q_- |x|} \right) \, .
\label{a30}
\ee
When $|m^2-\Lambda^2|\geq 2$, then $q_{\pm}\equiv \sqrt{q_{\pm}^2}$ are real and positive. On the other hand $A_{+}>0$ but $A_-<0$ when $\Lambda^2>m^2$.
The mode with the positive coefficient $A_+$ is found to decay slower that the one with the negative and this results into the correlator to be always positive definite as shown in the blue graph of figure~\ref{fig:reflectionpos}.
However in the opposite case,  $m^2-\Lambda^2>2$ the correlator is positive for large and small distances, but negative at intermediate distances. This is the green plot in fig.~\ref{fig:reflectionpos}.
Therefore in this case the theory obeys reflection positivity when $\Lambda^2-m^2\geq 2$.

On the other hand, when  $|m^2-\Lambda^2|< 2$ then both both $A_{\pm}$ and $q^2_{\pm}$ are complex numbers with
\be
(q_{+}^2)^*=q^2_-\sp (A_+)^*=A_-
\label{a29}\ee
In this case the propagator is {\em not} reflection positive as from (\ref{a30}) we obtain
\be
G_{++}(x) = \langle \phi_+(x) \phi_+(-x) \rangle ={R\over |x|}\cos\left[2\sqrt{{1-(m^2-\Lambda^2)^2\over 4}}|x|+\theta\right]~e^{-2\sqrt{m^2\Lambda^2\over 2}|x|}
\label{a34}\ee
where $R,\theta$ can be written in terms of $m,\Lambda$. A plot of this case can be seen on the right-hand side of fig.~\ref{fig:reflectionpos}.

We also calculate the minus two-point function using
\be\label{a22}
{1\over D_-}={q^2+\Lambda^2\over (q^2+m^2)(q^2+\Lambda^2)-1}={q^2+\Lambda^2\over (q^2+\bar q_+^2)(q^2+\bar q_-^2)}={\bar A_+\over q^2+\bar q_+^2}+{\bar A_-\over q^2+\bar q_-^2}
\ee
with
\be\label{a23}
\bar q^2_{\pm}={m^2+\Lambda^2\over 2}\mp \sqrt{{(m^2-\Lambda^2)^2\over 4}+1}
\ee
and
\be\label{a24}
\bar A_{\pm}={1\over 2}\left[1\pm{{\Lambda^2-m^2\over 2}\over \sqrt{{(m^2-\Lambda^2)^2\over 4}+1}}\right]
\ee
 In this case $\bar q^2_{\pm}$ are always real and positive while $\bar A_{\pm}$ are also real and positive. We obtain
\be
G_{--}(x) = \langle \phi_-(x) \phi_-(-x) \rangle =\int {d^3q\over (2\pi)^3}{e^{2iq\cdot x}\over D_-(q)}=
\label{a31}\ee
\be=
  \frac{1}{2 |x|} \int_0^\infty \frac{d q}{(2 \pi)^2} q \sin \left( 2 q |x| \right)  \left( \frac{\bar A_+}{q^2 + \bar q_+^2} + \frac{\bar A_-}{q^2 + \bar q_-^2} \right)=
\label{a32}\ee
\be
= \frac{1}{8 \pi |x|}\left( \bar A_+ e^{- 2 \bar q_+ |x|} + \bar A_- e^{- 2 \bar q_- |x|} \right) \, .
\label{a33}
\ee
This two-point function is manifestly reflection positive for all values of the parameters.

We conclude our Euclidean discussion that when
\be
m^2\Lambda^2>1~~~{\rm  and}~~~  \Lambda^2-m^2\geq 2
\label{a35}\ee
the theory in (\ref{a9}) is reflection positive, quadratic and satisfies all Euclidean QFT axioms including cluster decomposition. Interestingly, in this example, reflection positivity puts a lower bound on the amount of non-locality in the inter-theory coupling, controlled by $\Lambda$.

We now shall attempt to rotate this theory to Minkowski space by the standard
rescription $p^0\to ip^0$ or $x^0\to ix^0$.
This will allow in particular $p^2$ to be positive, negative or zero.

We shall find now the  poles of the propagators of \eqref{a11} or equivalently the zeros of the functions $D_{\pm}(q)$. This will be useful in order to understand if the analytically continued theory to Lorentzian signature is sensible.

We start with $D_+$,
\be\label{a18}
D_{+}(q)=0~~~\to~~~ (q^2+m^2)(q^2+\Lambda^2)+1=0
\ee
with two solutions, $- q_{\pm}^2$ with $q_{\pm}^2$ given in (\ref{a21}).

If $|m^2-\Lambda^2|>2$ both $q^2_\pm$ are real and positive and therefore they correspond to positive mass squared solutions.
If $|m^2-\Lambda^2|<2$ the solutions are complex, and therefore the analogue of the mass square is complex. This is the case where the Euclidean theory violates reflection positivity.

From (\ref{a20}) we observe that the propagator has two poles, that correspond to positive masses squared (when $|m^2-\Lambda^2|>2$), but the respective residues have opposite signs and therefore one of them is a ghost.

On the other hand, in the minus sector the spectrum is healthy as can be checked from (\ref{a22}).

We conclude that the free theory in question, exists only as a healthy theory in Euclidean space, when (\ref{a35}) is valid, where it exhibits the phenomena we have seen in the holographic model.
It is not clear however, if analogous non-trivial interacting theories are possible in Lorentzian signature.
Moreover, here we encounter an apparent puzzle: this theory seems to violate the Osterwalder-Schrader theorem, \cite{Osterwalder:1973dx} but we do not understand why.

It is clear that this theory is a special case of a large class of possible theories that can be constructed by generalizing the ingredients used here.
One clear generalization is to use a more general function $f(p^2)$ to couple the two operators.
One such parametrization is
\be
f(p^2)=\sum_{n=1}^{\infty}{c_n\over p^2+\Lambda_n^2}
\ee
We can also add interactions in the the two individual theories, or interaction vertices that include composite operators with more than two elementary fields.
It is not a priori clear what are the properties of such theories and whether they are sensible at least as Euclidean theories, but we expect that a subset of them will be. It is not also clear whether a subset of them will have a healthy Minkowski continuation but this is an interesting question worthy of further study.

\section*{Acknowledgements}\label{ACKNOWL}
\addcontentsline{toc}{section}{Acknowledgements}

We thank Costas Bachas, Matteo Baggioli, Oscar~J.C.~Dias, Juan Maldacena, Kyriakos Papadodimas, Kostas Skenderis and Konstantin Zarembo for useful discussions.

\noindent This work is supported in part by the Advanced ERC grant SM-GRAV, No 669288. O.P. acknowledges support by the STFC Ernest Rutherford grants ST/K005391/1 and ST/M004147/1 at the initial stages of this project when she was a member of the University of Southampton.

\newpage
\appendix
\renewcommand{\theequation}{\thesection.\arabic{equation}}
\addcontentsline{toc}{section}{Appendix\label{app}}
\section*{Appendices}

\section{Global AdS ODE}\label{AdSODE}

As a warmup and to fix the notation, contrasting with a well known case, we study here the fluctuation equation for $AdS_4$ using the global metric in the form~\eqref{AdSESU}. One then finds the differential equation
\be
\xi''(u) - 2 \coth(u) \xi'(u) - \left(\ell(\ell+2) + \frac{m^2}{\sinh^2 (u)}  \right) \xi(u) = 0
\ee
we again redefine the wave-function as $\xi(u) = \sinh u \Psi(u)$ to find
\be
- \Psi''(u) + \frac{2+m^2}{\sinh^2 u} \Psi(u) = - (\ell+1)^2 \Psi(u)
\label{AdSGlobalE=0}
\ee
which  has a convex potential bounded from below but negative energies leading to non-normalizable solutions. The boundary is at $u=0$ and the IR at $u = - \infty$.
Expanding the potential near the UV $u=0$ one finds

\be
V_{u}\sim {s(s+1)\over u^2}\, , \qquad s= -\half \pm \half \sqrt{9 + 4 m^2}
\label{AdSasympot}
\ee
where we parametrized the coefficient by the real number $s$. If we choose the branch  $s\geq -{1\over 2}$,
the two linearly independent solutions are
\be
\Psi_{\pm}\sim u^{{1\over 2}\pm \left(s+{1\over 2}\right)}
\ee
$\Psi_{+}$ is normalizable when $s> -1$ (always), while $\Psi_-$ is normalisable when $s<0$ and non-normalisable when $s>0$. The bulk field $\xi(u)$ has the two asymptotic behaviours
\be
\xi_\pm(u) \sim u^{{3\over 2}\pm \left(s+{1\over 2}\right)}
\ee
from which one finds the canonical conformal dimensions. The relation is $\Delta_\pm = {3\over 2}\pm \left(s+{1\over 2}\right)$.
The conformal dimensions in AdS are then
\be
\Delta_\pm = \frac{3}{2} \pm \frac{1}{2} \sqrt{9+ 4 m^2}
\ee
In terms of these if $\Delta_\pm >1$ the corresponding solution is normalisable ($\Delta_+$ is always in this range), while if $\Delta_- <1$ it is not normalisable.
In addition solving the equation in the IR $u \rightarrow - \infty$ one finds the solutions
\be
\Psi^{IR}_{\pm}\sim e^{\pm (\ell+1) u}
\ee
Since one of the two solutions is divergent in the IR, one can also impose IR regularity to remove it.

\section{Fluctuations and stability}\label{fluctuationstability}

In order to further describe the properties of Euclidean wormholes, one can first analyse perturbations around such backgrounds. The question of perturbative stability of Euclidean wormholes is a long time unsettled question of great importance, since this would indicate their physical role in the semi-classical Euclidean path integral. If they are local minima, one should include such saddles in a similar fashion to other instantons which affect the vacuum structure of the theory. Nevertheless even if they are found to be unstable, they could potentially still play an interesting role perturbatively if their lifetime is parametrically large and possibly non-perturbatively in some similar fashion to the role of unstable saddles in resurgence theory~\cite{Basar:2013eka}.

To test their stability, one should check the spectrum of normalisable modes and whether there exist modes with negative Euclidean energy. Possible instabilities have been discussed in the literature, but the results are not conclusive since generically the graviton fluctuations mix with those of other fields (such as the non-abelian gauge field in the case of merons) and one needs to study the fully coupled system of perturbations and clarify which modes are physical and which can be cancelled by gauge symmetry ghosts in order to derive unambiguous results.
\\
\\
In the rest, we shall examine in detail probe scalar modes and therefore the spectrum and the properties of
\be
(-\Box+m^2) \, \phi \, = \, E \, \phi \, .
\label{Probescalar}
\ee
A normalizable mode of this equation
with $E<0$ will tend to destabilize the background. If the system is stable, normalisable modes with $E>0$ can be used to construct the propagator in the bulk of the space-time. All the equations we shall study in the rest, will reduce to ODE's due to the radial dependence of the background fields. It will be convenient then to bring the fluctuation equation into a Schr\"{o}dinger form with $E$ the energy of the Schr\"{o}dinger problem, so that stability can be readily tested.
In particular for a metric of the form
\be\label{F2}
d s^2 = e^{2 \Omega} \left(du^2 + ds_{d}^2 \right)\, , \qquad ds_{d}^2 = g_{i j}(u) dx^i dx^j \, ,
\ee
where $g_{ij}(u)= e^{2 f(u)} \tilde{g}_{ij}(x)$  one finds the scalar fluctuation equation
\be\label{Genericfluctuationeqn1}
\left( \partial_u^2 + e^{- 2 f}  \Box_{\tilde{g}} \right) \phi +\left[ (d-1) \Omega' + d  f' \right]\phi' - m^2 e^{2 \Omega} \phi = 0
\ee
To remove the first derivative of $\phi$ and bring the equation into Schr\"{o}dinger form, we perform the following transformation $\phi = e^{-\beta}\Psi$ where:
\be\label{F3}
2 \beta' = (d-1) \Omega' + d f'
\ee
and after an integration
\be\label{F4}
\beta = \frac{d-1}{2} \Omega +\frac{d}{2} f
\ee
then the differential equation takes its final form
\be\label{F5}
-\Psi'' + \left(m^2 e^{2 \Omega} + \beta'' +  {\beta '}^{2}-  e^{- 2 f} \Box_{\tilde{g}} \right) \Psi = 0
\ee
in terms of the metric functions $\Omega(u), f(u)$.
\\
\\
Let us now use as a warmup example the case of pure AdS. We expand the modes in $S^3$ harmonics
$\phi=\chi_{\ell}(r)Y_{\ell m p}(\Omega_3)$ with $-\Box_\Omega^2 Y_{\ell m p} =  \ell(\ell + 2) Y_{\ell m p}$. Then,
the radial fluctuation ODE becomes
\bea\label{F6}
\chi_{\ell}''+3 \coth r \chi_{\ell}'+\left( E-m^2-\frac{\ell(\ell+2)}{\sinh^2 r} \right) \chi_{\ell}=0
\eea
One can then rescale $\chi(r) = \sinh(r)^{-3/2} \psi(r)$ so that
\be\label{AdSglobalcoords}
-\psi_{\ell}'' + \left[  \frac{\ell(\ell+2)}{\sinh^2 r} + \frac{3}{4}  \frac{1}{\sinh^2 r}  \right] \psi_{\ell} = \left(E-m^2 - \frac{9}{4} \right)\psi_{\ell}
\ee
This potential has a divergence (blows up) at $r=0$, see fig.~\ref{fig:Potentials}. This is a manifestation that pure AdS ends at $r=0$. The spectrum in these coordinates is continuous and bounded below. The constant shift $9/4$ corresponds to the fact that AdS can support negative $m^2$ as long as one stays above the BF bound $m^2_{BF}=-9/4$.
\\
\\
After this warmup, we will now perform a preliminary stability analysis of our solutions and make a comparison with results obtained previously in the literature for similar solutions (mainly in~\cite{Maldacena:2004rf} and~\cite{Hertog:2018kbz}).

\subsection{Einstein Yang-Mills system}

\begin{figure*}[!tb]
\begin{center}
\includegraphics[width=0.45\textwidth]{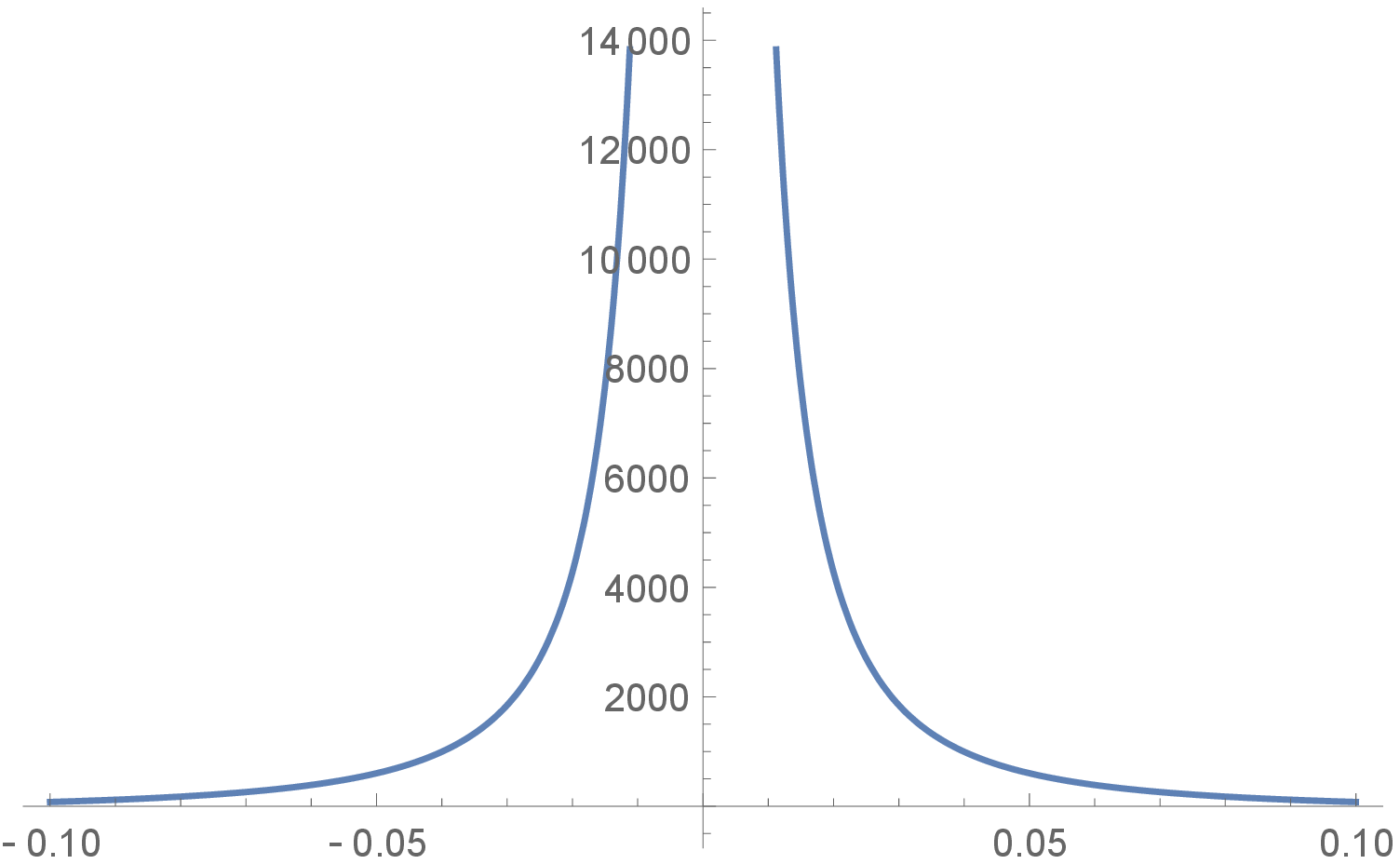}%
\hspace{4mm}
\includegraphics[width=0.43\textwidth]{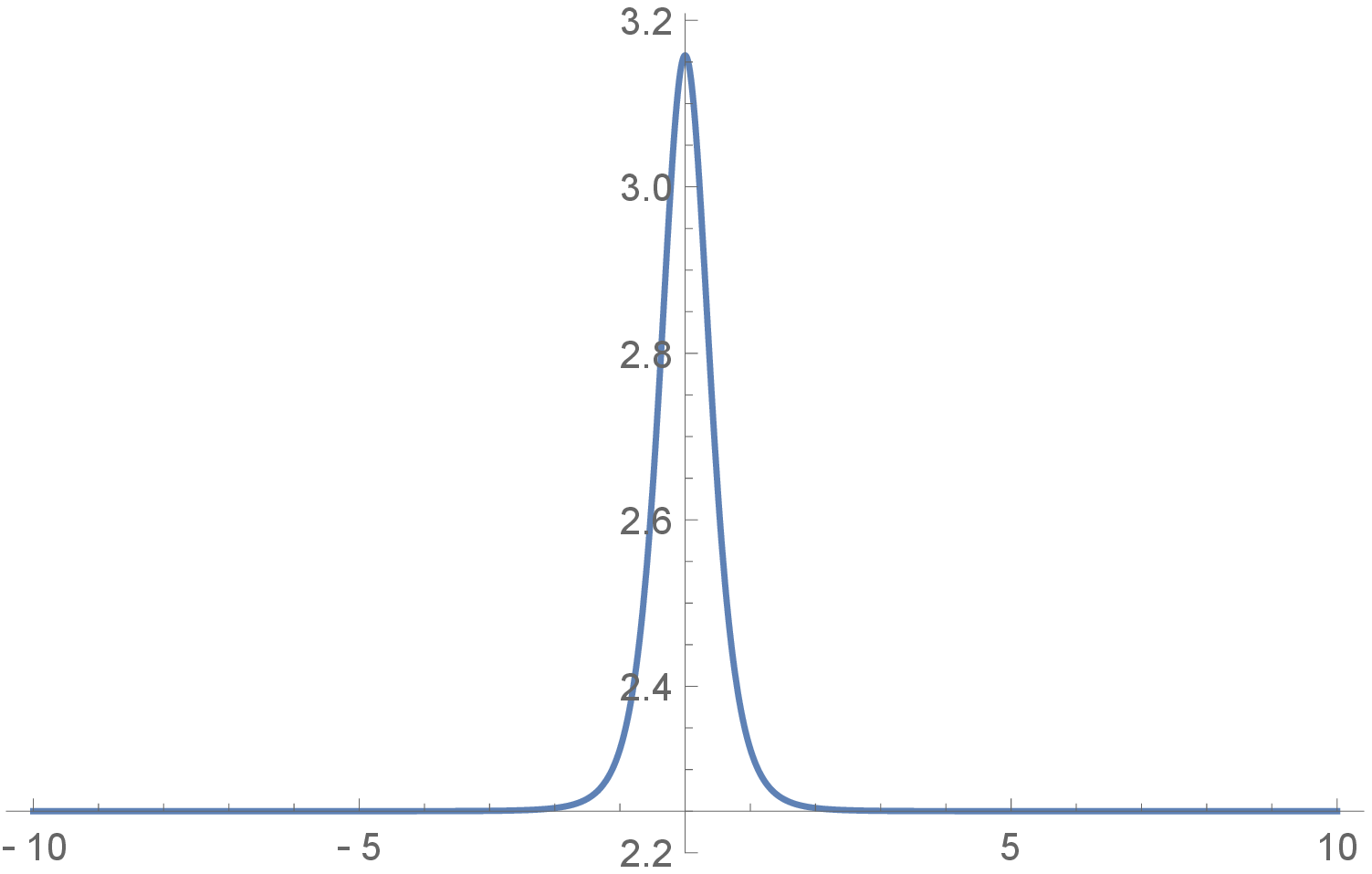}
\end{center}
\caption{The potentials for AdS (impenetrable) and for the meron wormhole.}
\label{fig:Potentials}
\end{figure*}

In the case of the meron wormhole~\ref{EinsteinYangMillswormhole} we expand the modes in $S^3$ harmonics
$\phi=\chi_{\ell}(r)Y_{\ell m p}(\Omega_3)$ with $-\Box_\Omega^2 Y_{\ell m p} =  \ell(\ell + 2) Y_{\ell m p}$. Then,
the radial fluctuation ODE becomes
\bea\label{F7}
\chi_{\ell}''+3 w'\chi_{\ell}'+(E-m^2-\ell(\ell+2)e^{-2w})\chi_{\ell}=0 \nn \\
e^{2 w(r)} = B \cosh (2 r) - \half \, , \qquad w' = \frac{B \sinh 2 r}{B \cosh (2 r) - \half}
\eea
In order to bring the fluctuation equation into a Schr\"{o}dinger form
we rescale $\chi(r) = \left(B \cosh 2r - \half \right)^{-3/4} \Psi(r)$, and find the differential equation
\be\label{F8}
-\Psi_{\ell}'' + \left[   \frac{\ell(\ell+2)}{B \cosh 2 r - \half} + \frac{3}{4} \frac{B \cosh 2 r + B^2 - 3/4}{(B \cosh 2 r - \half)^2}  \right] \phi_{\ell} = \left( E-m^2 - \frac{9}{4} \right) \Psi_{\ell}
\ee
The potential now looks like a bump with finite width and height for any allowed value  of $B \geq \half$, see fig.~\ref{fig:Potentials}. The spectrum is continuous and bounded below. Fluctuations now can penetrate and pass through the other side so the two sides ``talk to each other'' and one finds both a reflection and transmission amplitude. The constant shift $9/4$ corresponds again to the fact that this wormhole solution can support negative $m^2$ as long as one stays above the BF bound $-9/4$. Notice that for $B\rightarrow \half$ the equation reduces to the one of AdS and the potential becomes singular at $r=0$. For $B \rightarrow \infty$ one finds that the angular momentum modes decouple since the three sphere acquires infinite size. A preliminary analysis of the gauge field modes~\cite{Maldacena:2004rf}, indicated that this solution could have instabilities for some regime of parameters. These instabilities would be related to instabilities that inflict the pure meron solution, but the results are not conclusive, since on this wormhole background the gauge field modes are coupled to metric perturbations and one has to split properly the fluctuations into spin-2, spin-1 and spin-0 components as discussed in~\cite{Betzios:2017krj}.

\subsection{Einstein Dilaton theory}

\begin{figure*}[!tb]
\begin{center}
\includegraphics[width=0.45\textwidth]{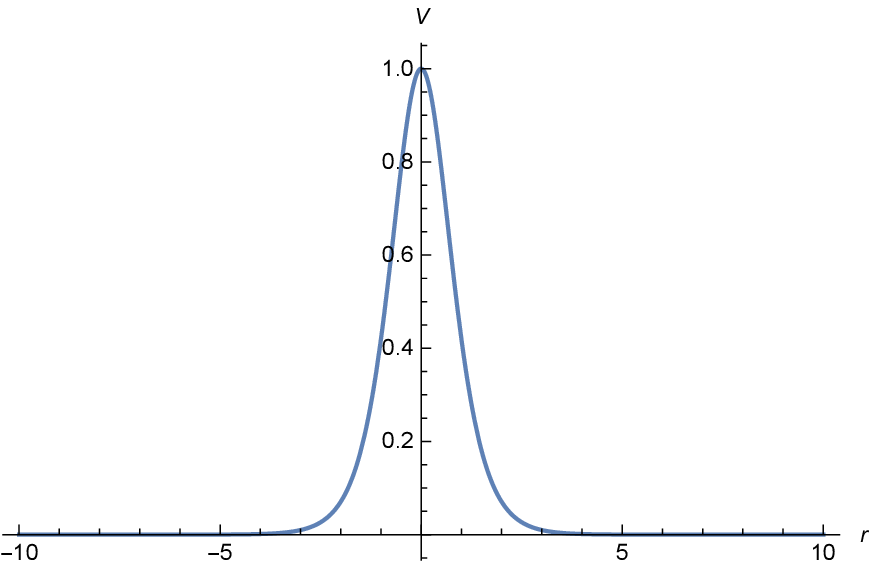}%
\hspace{4mm}
\includegraphics[width=0.43\textwidth]{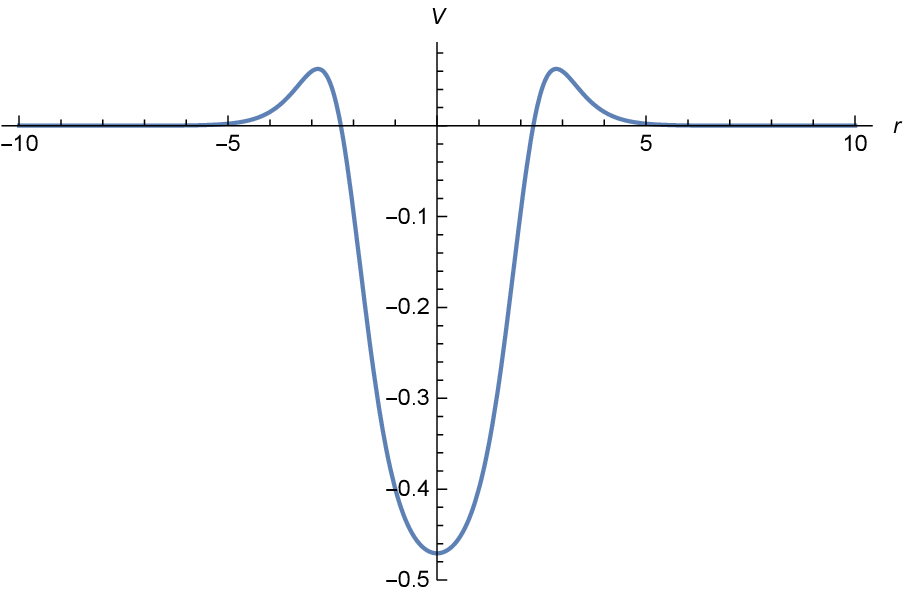}
\end{center}
\caption{On the left: the potential for $s(1-s)=1/4\, , B=1/2$. On the right: the potential for $s(1-s)=1/4, B=1/100$. There exist negative energy bound states of the fluctuation equation in case $B<1/2$.}
\label{fig:Potentials2}
\end{figure*}

We shall now study the probe scalar equation~\eqref{Probescalar} on the symmetric background of section~ \ref{Hyperbolicslicing} splitting
$\phi  = \chi_{s}(\tr) F_s (H_2)$ with $-\Box_{\mathcal{M}} F_s = s(1-s) F_s$, with $\mathcal{M} = \mathbb{H}/\Gamma$ ( $\Gamma$ labeling a Fuchsian discrete $PGL(2,R)$ subgroup) and $ F_s$ the specific hyperbolic space quotient eigenfunctions. We therefore need to use the properties of the Laplacian on two-dimensional hyperbolic space and discrete quotients thereof. We use the following properties:
\begin{itemize}

\item For $\mathcal{M} = \mathbb{H}/\Gamma$ compact, $-\Box_\mathcal{M}$ has discrete spectrum in $[0,\infty)$

\item For $\mathcal{M} = \mathbb{H}/\Gamma$ non-compact, $-\Box_\mathcal{M}$ has both discrete spectrum in $[0,\infty)$
and absolutely continuous spectrum in $[1/4, \infty)$.

\item If $\mathcal{M} = \mathbb{H}/\Gamma$ has infinite area (2-volume), $-\Box_\mathcal{M}$ has discrete spectrum in $(0,1/4)$ and absolutely continuous spectrum in $[1/4, \infty)$, and therefore, no embedded discrete eigenvalues in the continuum.

\end{itemize}
Using this we find that in all cases $s(1-s) \geq 0$.
The equation for the radial modes takes the form
\bea\label{F9}
\chi_{s}''+2 A'\chi_{\ell}'+(E-m^2-s(1-s)e^{-2A})\chi_{\ell}=0 \nn \\
 e^{2 A} = \left(B  \cosh (2 r) + \frac{1}{2} \right) \, , \qquad A' = 2B \sinh (2 r)
\eea
One can write this equation in Schr\"{o}dinger form by rescaling $\Psi = e^A \chi$ to obtain
\be\label{F10}
- \Psi_s'' +\left( \frac{s(1-s)}{ \left(B  \cosh (2 r) + \frac{1}{2} \right) } + \frac{B^2 - \frac{1}{4}}{\left( B  \cosh (2 r) +  \half \right)^2} \right) \Psi_s = (E-m^2-1) \Psi_s \, .
\ee
Since the parameter of the wormhole throat size $B$ obeys now just $B>0$,
this potential has a quite interesting behaviour: it could either have a single maximum but there is also the possibility for it to exhibit two maxima and one minimum depending on the values of $s,B$.
These two cases are depicted in~\ref{fig:Potentials2}. The physical meaning is that we can have both continuous (scattering) and bound states. The first type of states was also encountered in the meron wormhole. In this case fluctuations can penetrate and pass through the other side. Therefore, the two sides ``communicate'' and one finds both a reflection and a transmission amplitude.

On the other hand, the normalisable bound states correspond to states that are localised near the throat and since they can have energies below zero, they  lead to a perturbative instability for some regime of parameters. In order to remove the instability, we need to impose $B>1/2$, in line to what was found in the meron case, although in the latter case this condition was coming from the fact that the space pinches off, while now it is a stability criterion. The BF bound then is again given by the $AdS_3$ condition $m^2_{BF} = -1$. On the other hand, if we stay in the region $B<1/2$, even if the bound states have $E-m^2-1<0$, we find that for large positive $m^2$ these bound states will also have positive $E$, and therefore there can exist localised states that are still stable. To quantify this, we note that for $B<1/2$, the lowest energy supported states are for
\be\label{F10B}
E_{min}-m^2-1 \, = \, V_{min} \, = \, \frac{s(1-s)}{B+\half} - \frac{\frac{1}{4}- B^2}{\left(B+\half \right)^2}\, ,
\ee
so that iff
\be\label{F10C}
E_{min} = m^2 + 1 + \frac{s(1-s)}{B+\half} - \frac{\frac{1}{4}- B^2}{\left(B+\half \right)^2} > 0 \,
\ee
then the operator has both scattering and bound states but they are stable. This condition is most strict for $s=0$, so that iff
\be\label{F10D}
 m^2 >   \frac{\frac{1}{4}- B^2}{\left(B+\half \right)^2} - 1  \equiv m^2_{WBF} \, ,
\ee
then all states are stable. $m^2_{WBF}$ then can be thought of as an analogue of the BF bound for this wormhole solution.

In contrast to this result, Maldacena-Maoz~\cite{Maldacena:2004rf} found that quotients of pure AdS written in hyperbolic coordinates suffer from perturbative instabilities for probe scalars, not present in our solution for $B > 1/2$. We notice that we can obtain the pure AdS quotients for $B= \half$, which is the limit when the potential starts to form a well and negative normalisable states can appear. Nevertheless one can extend the stability for $B<1/2$, provided that $m^2 > m^2_{WBF}$, but one has to understand the role of the bound normalisable modes. It would be also interesting to complete our stability analysis using the tools of~\cite{Hertog:2018kbz}, since in this more recent work another similar wormhole supported by an \emph{axion} field was found to have an infinite number of unstable modes. Nevertheless there is a significant difference of our solution compared to these single axion wormholes, since we have a positive definite kinetic term for our scalar field, while the afforementioned axionic solution comes from an action with the wrong sign for the axion kinetic term~\footnote{This is also in line with the statement in~\cite{Hertog:2018kbz} that dilaton solutions do not suffer from these instabilities found for axions.}.

\subsection{$AdS_2$}\label{AdS2Stability}

\begin{figure}[t]
\vskip 10pt
\centering
\includegraphics[width=80mm]{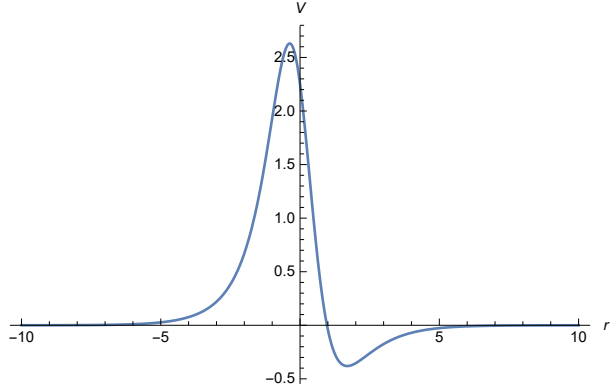}
\caption{The potential for $q=1,k=1$, it can develop a well.}
\label{fig:potentialchargedads2}
\end{figure}

In the case of global $AdS_2$~\eqref{AdS2metric}, one needs to perform a coordinate transformation in order to bring the metric into a form that leads to the appropriate Schr\"{o}dinger problem. This is the $1-1$ transformation
\be\label{F11}
r = \log \frac{\cos \left(\frac{u}{2} \right) + \sin \left(\frac{u}{2} \right)}{\cos \left(\frac{u}{2} \right) - \sin \left(\frac{u}{2} \right) } \, , \quad u \in [-\frac{\pi}{2}, \frac{\pi}{2}], \, \, r \in (-\infty, \infty)
\ee
that takes the metric into the form (the two boundaries are at $r = \pm \infty$)
\bea\label{F12}
ds^2_{AdS_2} \, = \, d r^2 + \cosh^2 r \, d \tau^2
\eea
The Schr\"{o}dinger ODE is now
\be\label{F13}
-\Psi_{\ell}'' + \left[   \frac{k^2 + \frac{1}{4}}{ \cosh^2 r }  \right] \Psi_{\ell} = \left( E-m^2 - \frac{1}{4} \right) \Psi_{\ell}
\ee
The potential  looks again like a bump with finite width and height like in fig.~\ref{fig:Potentials}. The spectrum is continuous and bounded below. The BF bound is now $m^2_{BF}= -1/4$, which is again the one expected for $AdS_2$.
\\
\\
The charged case is more interesting. We reproduce the non-homogeneous version of eqn.~\eqref{182e} upon shifting to the parameter $k_r = k + q \mu$
\be\label{F14}
- \Psi''(r)  + \left(\frac{k_r^{2}+  q^{2} + \frac{1}{4} }{\cosh^{2} r}-2 q k_r\frac{\sinh r}{\cosh^{2} r} \right) \Psi(r) =\left(E + q^{2} - m^{2} - \frac{1}{4} \right) \Psi(r) \, .
\ee
The potential has a similar behaviour to the one of~\eqref{F13} for very small or very large $q$. For intermediate values of $q$ it develops a well, which can be seen in fig.~\ref{fig:potentialchargedads2}.

This results to the presence of bound states inside the well. Nevertheless in order to clarify whether these bound states result in an instability or not, one can use that the extremum values of the potential are
\be\label{F15}
 V(r_\pm^*) =  \frac{1}{8}\left(1+ 4 k_r^2 + 4 q^2 \pm \sqrt{16 k_r^4 + (1+ 4 q^2)^2 + k_r^2(8+ 96 q^2)} \right) \, ,
\ee
where the plus sign gives the maximum and the minus the minimum. So, in order to forbid the existence of bound states with negative energy the condition is
\be\label{F16}
 E_{min} = m^2 + \frac{1}{4}  + \frac{1}{8}\left(1+ 4 k_r^2 - 4 q^2 - \sqrt{16 k_r^4 + (1+ 4 q^2)^2 + k_r^2(8+ 96 q^2)} \right)  > 0 \, .
\ee
Therefore is this condition holds, none of the bound or scattering states lead to an instability. This condition is most strict for $k_r=0$, so that iff
\be\label{F17}
  m^2  >  q^2 - \frac{1}{4} \equiv m^2_{qBF}  \, ,
\ee
all the states are stable. This is again the bulk analogue of the dual Euclidean field theory operators having real conformal dimensions.
\\
\\
We conclude that simple probe scalar perturbations do not lead to an instability for the solutions studied in this paper, as long as one stays above the appropriate BF-bound, which is the bulk analogue of the dual theory conformal dimensions remaining real. We expect this dual field theory criterion to extend to other modes and correlation functions as well. Nevertheless our analysis is far from complete- one should study all possible gauge invariant combinations of perturbations around our specific wormhole backgrounds for a complete stability analysis that can settle the physical interpretation of such solutions in the semi-classical Euclidean path integral.

\section{On the BF bound from Homogeneous ODE}\label{Homogeneousstability}

In the previous section~\ref{fluctuationstability}, we studied the stability of the solutions, using the non-homogeneous fluctuation equation, and derived the associated BF bound for the solutions. It would be interesting to derive the BF-bound in an alternate fashion, using the homogeneous ODE. This is also important in order to clarify the discussion provided in section~\ref{AdS2correlators}. In particular it is not enough to only distinguish the existence or not of normalisable states with negative eigenvalues, but one needs to also study whether the fluctuation operator is Hermitean on such states. In a more general context the appropriate condition is that the Euclidean fluctuation operator be an elliptic operator~\cite{Witten:2018lgb}. The Hermiticity condition for the Schr\"{o}dinger ODE we study below is a particular case of this more general property.
The conditions are the following:
\begin{itemize}
\item If the Schr\"{o}dinger operator is Hermitean on a set of eigenfunctions, and if these wavefunctions are normalisable i.e $\int du |\Psi(u)|^2 < \infty$, then this corresponds to a configuration in the bulk having finite Euclidean action.

\item In case such configurations are related to a negative eigenvalue of the operator and there is an infinite number of them, such states would lead to an instability of the system, or in other words the spectrum is not bounded from below and the system will prefer to decay.

\end{itemize}

We focus on a Schr\"{o}dinger operator defined on a domain $\mathcal{D}$ and the corresponding Hermiticity condition on the boundary of the domain $\partial \mathcal{D}$
\be\label{Schroendinger}
- \Psi_E''(u) + V(u) \Psi_E(u) = E \Psi_E(u)\, , \qquad \left[\Psi_{E_1}^*(u) \Psi_{E_2}'(u) - \Psi_{E_2}(u) {\Psi_{E_1}^*}'(u) \right]\vert_{\partial \mathcal{D}} = 0 \, ,
\ee
and check whether there exist negative energy bound states of this operator that satisfy Hermiticity. Notice that the Hermiticity condition when $E_1 = E_2$ is equivalent to the vanishing of the Wronskian (or of the ``probability flux density'' passing through the boundary).
\\
\\
In case of pure $AdS_{d+1}$ (global or Poincare patch) the potential is
\be\label{V1}
V_{GAdS}(u) = \frac{m^2 + \frac{d^2 - 1}{4} }{\sinh^2(u)} \, , \quad V_{PAdS}(u) = \frac{m^2 + \frac{d^2 - 1}{4} }{u^2}
\ee
In this case one finds that in the range $- \frac{d^2}{4} < m^2 < -\frac{d^2 - 1}{4}$, even though the potential changes sign, nevertheless the combination of IR regularity together with the Hermiticity condition in the UV ($u=0$), give
\be\label{V2}
\left(\frac{E_1}{E_2}\right)^{\half \sqrt{4 m^2 + d^2}} + \left(\frac{E_2}{E_1}\right)^{\half \sqrt{4 m^2 + d^2}} = 0
\ee
This equation has no solution and therefore does not allow the presence of any negative energy bound state. The system is therefore stable. Below the value $m^2_{BF} = - d^2/4$, the negative potential is ``too steep'' and the wave-functions become complex and start to oscillate near the boundary. It can then be shown that there exist a discrete spectrum of negative energy normalisable modes, which leads to an instability of the vacuum.
\\
\\
In the cases of the wormholes treated in this paper, or of global $AdS_2$, the boundary value problem changes dramatically and one would therefore have to check for states satisfying the Hermiticity condition using both boundaries.
The potentials are ($d=3$ for the meron wormhole)
\be\label{V3}
V_{GAdS_2}(u) = \frac{m^2}{\cos^2(u)}\, , \quad V_{meron}(u) = \frac{m^2 + 2}{\cn^2(u,k)}
\ee
As long as the potential is convex with $V(u), \, V''(u)>0$ there is no bound state. The question is what happens when the potential switches sign. In the case of $AdS_2$ we find the exact solutions as a linear combination of Legendre functions $P_\nu^\mu(u), Q_\nu^\mu(u)$ with $\mu =\half \sqrt{1+ 4 m^2} , \, \nu = - \half + i k $ with $E = - k^2$, together with the connection formulae relating the two sides of the wormhole
\be\label{V5}
\begin{pmatrix}
P_\nu^\mu(-u) \\ Q_\nu^\mu(-u)
\end{pmatrix} =
\begin{pmatrix}
\cos \pi (\nu+\mu) & - \sin \pi (\nu+\mu)  \\
  - \sin \pi (\nu+\mu)  & - \cos \pi (\nu+\mu)
\end{pmatrix}
\begin{pmatrix}
P_\nu^\mu(u) \\ Q_\nu^\mu(u)
\end{pmatrix}
\ee
It is then a tedious but straightforward exercise using the properties of Legendre functions and asymptotic expansions to show that there exist no normalisable states for which the operator is Hermitean for $\mu \in (0, \half)$.
An easier method to check this is using~\eqref{Schroendinger} for $E_1 = E_2 = - k^2$, in conjunction with the Wronskian of the associated Legendre functions on the real axis
\be\label{V4}
\mathcal{W}\left(P_\nu^\mu , Q_\nu^\mu \right)(u) = \frac{1}{1-u^2} \frac{\Gamma(\nu + \mu + 1)}{\Gamma(\nu - \mu + 1)}\, ,
\ee
which cannot vanish if $\mu$ is a positive real number and $\nu = - \half + i k $. This result is easy to understand since the wave-functions do not change nature as long as $\mu$ is a positive real number and this covers the whole range
$m^2 > -1/4 $. Once $m^2 < - 1/4$, the parameter $\mu$ becomes complex. Then one finds that $P_\nu^\mu(u), P_\nu^\mu(-u)$ or $P_\nu^\mu(u), P_\nu^{-\mu}(u)$ form the correct complex pair of orthogonal states than can be used as a basis for the wave-functions~\cite{Zwillinger}, which are again found to oscillate rapidly near the two boundaries. This of course is in line with the bound derived in section~\ref{AdS2Stability}, where the analysis is much simpler, so we omit the rest of the examples that are treated there.

\section{Maurer-Cartan forms - Hopf Fibration}\label{MaurerHopf}

We denote a group $G$ and the associated Lie algebra $\mathcal{G}$, whose basis elements/generators $T_a$ satisfy
\be\label{m1}
[T_a, T_b]=f_{abc} T_{c}.
\ee
If $g\in G$ is a generic group element of $G$, one defines the  Maurer-Cartan forms $\omega_a$ through
\be\label{m2}
g^{-1}  d g = \sum_{a} T_{a} \omega_a,
\ee
which satisfy
\be\label{m3}
d \omega_a+{1\over 2} f_{abc } ~\omega_b\wedge \omega_c=0.
\ee
We now specialize to $SU(2)$. A basis for the Lie algebra is given by the Pauli matrices
\be\label{m4}
T_a={ i\over 2} \sigma_a.
\ee
The structure constants are then
\be\label{m5}
 f_{abc}=-\epsilon_{abc}.
 \ee
Any element of $SU(2)$ can be written in the form
\be\label{m6}
 g= \left( \begin{matrix} \alpha&  \beta \\
 -\bar \beta &   \bar \alpha \end{matrix} \right), \qquad |\alpha|^2 + |\beta|^2=1.
\ee
We parametrize this element as
\be\label{m7}
|\alpha|=\cos \, {t_1\over 2}, \qquad |\beta|=    \sin \, {t_1\over 2}, \qquad {\rm Arg}\, \alpha={t_2 + t_3\over 2} , \qquad
 {\rm Arg}\, \beta={t_2 - t_3 + \pi \over 2},
 \ee
where $t_i$ are the Euler angles having the range
\be\label{m8}
0\le   t_1 <\pi, \qquad   0\le t_2 < 2\pi, \qquad   -2\pi\le   t_3 <2\pi.
\ee
The general element of $SU(2)$ is explicitly given by
\be\label{m9}
g(t_1, t_2. t_3)
 =\left( \begin{matrix}  \re^{i t_2 /2} &    0 \\
   0  &    \re^{- i t_2 /2}  \end{matrix}
  \right)\left( \begin{matrix} \cos(t_1/2)  &  i \sin(t_1/2)  \\
    i \sin(t_1/2)   &   \cos(t_1/2)  \end{matrix}
   \right)\left( \begin{matrix}  \re^{ i t_3/2} &  0 \\
      0  &    \re^{ - i t_3/2}  \end{matrix}  \right).
    \ee
Analytically continuing $t_1 = i t$ we find the parametrization of $SU(1,1)$. Considering all $t_1,t_2,t_3$ complex with the restriction
\be\label{m10}
0\le   \mathit{Re}[t_1] <\pi, \qquad   0\le\mathit{Re}[t_2]< 2\pi, \qquad   -2\pi\le  \mathit{Re}[t_3] <2\pi.
\ee
we parametrise the generic $SL(2,C)$ element. One can further restrict this to subgroups as above.
\\
\\
We then obtain for the $SU(2)$ element
\be\label{m11}
 \Omega=g^{-1} d g= {\ri\over 2}      \left( \begin{matrix} d t_3 + \cos\, t_1 d t_2&
\re^{- i t_3} ( d t_1 +  i d t_2  \sin\, t_1) \\ \re^{ i t_3} ( d t_1 -   i d t_2  \sin\, t_1)  & - d t_3 -\cos\, t_1 d t_2
 \end{matrix}
   \right).
 \ee
with the explicit Maurer-Cartan forms
 \be\label{m12}
 \ba
 \omega_1=&  \cos\, t_3  d t_1  +  \sin\, t_3 \sin\,t_1 d t_2 ,\\
  \omega_2=&\sin\, t_3  d t_1  - \cos\, t_3 \sin\, t_1  d t_2,\\
 \omega_3=&    \cos\, t_1 d t_2 + d t_3 \, .
 \ea
 \ee
The metric on $SU(2)= S^3$ is induced from the metric on $\IC^2$ in terms of the complex $\alpha, \beta$ parameters of~\eqref{m6}
 \be\label{m13}
 d s^2= \biggl( d |\alpha|^2 + |\alpha|^2 d{\rm Arg}\alpha^2 + d|\beta|^2 + |\beta|^2 d{\rm Arg}\beta^2   \biggr),
 \ee
One can then describe this in terms of euler angles $t_i$ as
 \be
 d s^2= {1 \over 4} \biggl( d t_1^2 + d t_2^2 + d t_3^2 + 2 \cos\, t_1 \, d t_2 d t_3   \biggr) \, = {1 \over 4} \biggl( d t_1^2 +\sin^2\, t_1 d t_2^2 + \left( d t_3 +  \cos\, t_1 \, d t_2 \right)^2   \biggr),
 \label{Hopfmetric}
 \ee
with the inverse metric
 \be\label{m14}
 G^{-1}={4 } \left( \begin{matrix}   1& 0& 0\\ 0& \csc^2 \, t_1& -\cot\, t_1 \, \csc\, t_1 \\0 &  -\cot\, t_1 \, \csc\, t_1&  \csc^2 \, t_1
 \end{matrix}
  \right)
 \ee
and volume element
\be\label{m15}
({\rm det}\, G)^{1/2}= { \sin\, t_1\over 8}.
\ee
These coordinates are used to describe the so-called Hopf fibration of $S^3$ with $S^1$ fibers (an $S^1$ bundle over $S^2$) denoted by: $S^{1}\hookrightarrow S^{3}{\xrightarrow {\ p\,}} \, S^{2} $. The fiber coordinate is $t_3$ while the base $S^2$ is parametrised by $t_1\, , t_2$. The fibers are non intersecting great circles of the $S^3$.

\subsection{Connection with usual angles on $S^3$}\label{angles connection}

One can connect the Euler angle parametrization with the usual in terms of $\psi, \theta, \phi$ that we also use in the main text. We have
\bea\label{m16}
x_1 &=& \cos \psi = \cos {t_1\over 2} \cos {t_2 + t_3\over 2}   \nn \\
x_2 &=&  \sin \psi \cos \theta = \cos {t_1\over 2} \sin {t_2 + t_3\over 2} \nn  \\
x_3 &=&  \sin \psi \sin \theta \cos \phi =  \sin {t_1\over 2}  \cos  {t_2 - t_3 + \pi \over 2} \nn \\
x_4 &=& \sin \psi \sin \theta \sin \phi = \sin {t_1\over 2} \sin {t_2 - t_3 + \pi \over 2}
\eea
From these equations one can immediately find
\be\label{m17}
\sin  {t_1\over 2} = \sin \psi \sin \theta \, , \qquad \phi = {t_2 - t_3 + \pi \over 2}
\ee
which makes clear that $t_1$ can serve as a parameter for the size of a loop that is described in eqn.~\eqref{5h}.

\subsection{Analytic continuation to Lorentzian}\label{LorentzianHopf}

It would be very interesting if we could analytically continue a combination of the Euler angles so that the $S^3$ metric would become that of three dimensional De-Sitter space ($dS_3$). This would provide an interesting extension of the meron wormhole solution~\eqref{EinsteinYangMillswormhole} to a Lorentzian signature cosmology that resembles De-Sitter at late times.

The generic analytic continuation to complex Euler angles results in the group $SL(2,C)$, as we mentioned above. In order to describe the metric of global $dS_3$, we need to analytically continue the usual angle coordinate $\psi$, see~\eqref{angles connection}, into $\psi = i t + \pi/2$. In terms of Euler angles this means $t_1 \rightarrow i t_1$ and $t_2 + t_3 \rightarrow i (t_2 + t_3)$ with their difference remaining real. The generic problem with such continuations is that the gauge field becomes \emph{complex}, and it is not clear whether there exists a configuration with a real gauge field that can produce such a metric~\footnote{Or a complex gauge transformation that turns the gauge field back to a real section.}.

\section{Analytic continuation to a Bang-Crunch Universe}\label{Bang-Crunch}

In this Appendix we discuss the possibility of analytically continuing the radial direction to obtain a geometry resembling a Bang-Crunch universe. This was discussed in~~\cite{Maldacena:2004rf} and in~\cite{Betzios:2017krj}. It is not clear how one can interpret such a continuation from a dual field theory point of view, since the radial direction is naturally identified with the RG scale of the dual theory and we should essentially try to deduce an emergent time
from an otherwise Euclidean theory~\footnote{This is similar in flavor but not exactly the same as what is needed for $dS/CFT$ proposals, since the two boundaries in time are now the two singularities.}. As an example we can analytically continue the radial direction $\tilde{u} =  i u $ of the meron-wormhole metric~\eqref{ellipticcoords2} to obtain the geometry
\be\label{1ff}
ds^2_{BC} = \left( B- \half \right) \cn^2 (\tilde{u}, k')\left(- d \tilde{u}^2 + \frac{d \Omega_3^2}{2 B} \right) \, ,
\ee
where $k'$  is the complementary modulus. Now $\tilde{u}$ plays the role of a \emph{time} coordinate. The lifetime of this universe can be described in terms of spatial $S^3$'s that start at zero size, expand and then contract to a Crunch. The universe has zero size at $\tilde{u} = (2m+1)K $ and maximal finite size at $\tilde{u} = 2 m K$, $m$ being an integer.

The fluctuation equation~\eqref{WormE=0} is now written as
\be\label{2ff}
-\frac{d^2 \Psi(\tilde{u})}{d \tu^2}   + \left(m^2+ 2\right) {k'}^2 \sn^2 (\tu, k') \Psi(\tu) = \left( \frac{(\ell+1)^2}{2B} + \left(m^2+ 2\right) {k'}^2 \right) \Psi(\tu) \, ,
\ee
This is precisely the form of Lam\'e differential equation. This form of the equation can be helpful in studying the properties of perturbative modes in the Bang-Crunch universe.  In particular the analogue of~\eqref{Wormsuperpos} and the spectrum of Lam\'e ODE, show that there exist time dependent field configurations localised in a period of time whose masses exhibit a band structure. Using the global monodromy of the ODE (that unfortunately is not known), one would have been able to compute the relation between early and late time modes on such a cosmological background and therefore understand effects such as particle creation in such a finite lifetime universe. This was achieved for the simpler metric studied in~\cite{Betzios:2017krj}. Nevertheless such geometries contain spacelike singularities where the spacetime as a whole contracts to a point and one would need to understand in more detail under which conditions they can be resolved or understood in string theory. A two dimensional microscopic toy model that discusses this in the context of $c=1$ Liouville theory is~\cite{Betzios:2016lne}.

\section{Wilson loop}\label{LoopAppendix}

Here we derive the equations of motion for the Wilson loop of section~\ref{Wilsonlines} in the case where both $t_1 , \, t_3$ are functions of the radial distance $r$. The action then is
\be\label{9h1}
S_{NG} = \frac{L^2}{2 \alpha'} \int d r \sqrt{B \cosh (2 r) - \half} \sqrt{1+ \frac{\dot{t}_1^2 + \sin^2 t_1 \dot{t}_3^2  }{4} \left(B \cosh (2 r) - \half \right)} \, .
\ee
This action depends on $t_1$ explicitly so it is not easy to find an integral of motion. The simplest equation that can be integrated is the one for $t_3$ which gives
\be\label{10h1}
\dot{t}_3 = \pm \frac{C_3 \sqrt{4+ \dot{t}_1^2(B \cosh (2 r) - \half) } \csc t_1}{\sqrt{B \cosh (2r) - \half} \sqrt{\sin^2 t_1 - C_3^2} } \, .
\ee
The equation of motion for $t_{1}$ is:
\be \label{eomt1I}
\frac{\left(B \cosh (2 r) - \half\right)^{2} \sin(2 t_{1})\dot{t}_{3}}{8 \mathcal{L}}=\frac{d}{dr}\left(\frac{\dot{t}_{1}\left(B \cosh (2 r) - \half\right)^{2}}{4 \mathcal{L}}\right) \, ,
\ee
where $\mathcal{L}$ is the Lagrangian in equation \eqref{9h1}.
If we substitute in \eqref{eomt1I} the equation \eqref{10h1} then we find
\be \label{eqot1III}
\frac{C_{3}^2 \cos (t_{1}) \sqrt{2 B \cosh (2 r)-1} }{8 \sin ^3(t_{1}) } \sqrt{\frac{-2 B  \dot{t}_{1}^2 \cosh (2 r)+ \dot{t}_{1}^2-8}{C_{3}^2 \csc ^2(t_{1})-1}}=\frac{d}{dr}\left(\frac{\dot{t}_{1} (2 B \cosh (2 r)-1)^{3/2}}{8 \sqrt{\frac{-2 B \dot{t}_{1}^2 \cosh (2 r)+\dot{t}_{1}^2-8}{C_{3}^2 \csc ^2(t_{1})-1}}}\right) \, .
\ee
If we plug \eqref{10h1} into the action we get
\be\label{S_on_t3}
S_{NG} = \frac{L^2}{2 \alpha'} \int d r \sqrt{B \cosh (2 r) - \half} \sin t_1 \sqrt{\frac{4+ \dot{t}_1^2 (B \cosh (2 r) - \half)  }{\sin^2 t_1 - C_3^2} }
\ee
which when setting $C_3=0$ ($t_3$ is constant), goes back to the simplified action~\eqref{9h}.

If one varies \eqref{S_on_t3} then the equation of motion for $t_{1}$ is:
\be\label{eomt1II}
\frac{C_{3}^2 \cot (t_{1}) \sqrt{2 B \cosh (2 r)-1} \sqrt{\frac{-2 B \dot{t}_{1}^2 \cosh (2 r)+\dot{t}_{1}^2-8}{C_{3}^2 \csc ^2(t_1)-1}}}{8 \text{C3}^2+4 \cos (2 t_{1})-4}=\frac{d}{dr}\left(\frac{-\dot{t}_{1} (2 B \cosh (2 r)-1)^{3/2}}{8 \left(C_{3}^2 \csc ^2(t_1)-1\right) \sqrt{\frac{-2 B \dot{t}_{1}^2 \cosh (2 r)+\dot{t}_{1}^2-8}{C_{3}^2 \csc ^2(t_{1})-1}}}\right)
\ee
One can explicitly show that equations \eqref{eqot1III} and \eqref{eomt1II} are equivalent.

\section{Singular solutions of Einstein-Dilaton equations}\label{scalarpathologies}

An explicit way to describe the pathologies encountered searching for wormhole solutions in Einstein-Dilaton theory~\eqref{1} with spherical $S^d$ slices, is to assume that
\be
V=-{d(d-1)\over \alpha^2}
\label{7}\ee
The solution of (\ref{6}) and (\ref{5}) is
\be \phi'= \sqrt{2d(d-1)}{C\over \alpha} ~e^{-d A}
\label{8}\ee
and
\be
A'=\pm {1\over \alpha}\sqrt{1+{\alpha^2\over R^2}e^{-2A}+{C^2}e^{-2dA}}
\label{9}\ee

$\bullet$ We first take $d=2$.
Then the solution is
\be
\left({\alpha^2\over R^2}+2 (e^{2A}+\sqrt{C^2+{\alpha^2\over R^2}e^{2A}+e^{4A}})\right)=e^{2{r-r_0\over \alpha}}
\label{10}\ee
which give the following polynomial equation for the scale factor
\be
4e^{2A}=Z+{\tilde C\over Z}-2{\alpha^2\over R^2}\sp Z\equiv e^{2{r-r_0\over \alpha}}\sp \tilde C\equiv {\alpha^4\over R^4}-4C^2
\label{11}\ee
The scale factor vanishes at
\be
Z={\alpha^2\over R^2}\pm 2|C|
\label{11a}\ee

When $r\to\infty$ then the metric  to leading order is
\be
ds^2=dr^2+{1\over 4}e^{2{r-r_0\over \alpha}}R^2~d\Omega_2=dr^2+e^{2{r\over \alpha}}R_{\rm uv}^2~d\Omega_2\sp R_{\rm uv}={R\over 2}e^{-{r_0\over \alpha}}
\label{12}\ee
Therefore $R$ and $r_0$ are not independent "physical parameters". The UV value of the two-sphere curvature is determined by a combination of them.

The solution for $\phi$ is
\be
\phi=\phi_0-{\sqrt{2d(d-1)}}{\rm arctanh}{Z-{\alpha^2\over R^2}\over 2C}
=\phi_0+{\sqrt{2d(d-1)}\over 2}\log{ Z-2C-{\alpha^2\over R^2}\over Z+2C-{\alpha^2\over r^2}}
\label{13}\ee
In all cases, $\phi(r\to\infty)=\phi_0$ and diverges to plus or minus infinity at the point where the scale factor vanishes.
In this regime there is a singularity because $(\p\phi)^2\to \infty$.

\section{Conventions for Einstein Yang Mills}\label{EYMconventions}

Our conventions for the EYM action~\eqref{1b} are:
\be\label{2bb}
 R  =  R_{\mu\nu}g^{\mu\nu}\sp
 R_{\mu\nu}  =  R^{\lambda}_{\mu\lambda\nu}\sp
 R^{\lambda}_{\mu\nu\rho}  =  {\partial}_{\nu} {\Gamma}^{\lambda}_{\mu\rho}-{\partial}_{\rho} {\Gamma}^{\lambda}_{\mu\nu}+{\Gamma}^{\lambda}_{\rho\kappa}{\Gamma}^{\kappa}_{\mu\nu}-{\Gamma}^{\lambda}_{\nu\kappa}{\Gamma}^{\kappa}_{\mu\rho}
\ee
\be
 F^{a}_{\mu\nu}  = {\partial}_{\mu} A^{a}_{\nu}-{\partial}_{\nu} A^{a}_{\mu} + {\epsilon}^{abc}A^{b}_{\mu}A^{c}_{\nu}\sp
 D_{\mu}^{ab}=\delta^{ab}\nabla_{\mu}+  \epsilon^{acb}A_{\mu}^{c}
\ee
The equations of motion for the action \eqref{1b} are
\be\label{3bb}
R_{\mu\nu}-\frac{1}{2}g_{\mu\nu}R ~ = ~ 8\pi G\left(T_{\mu\nu}-\Lambda g_{\mu\nu}\right)\ee
\be
T_{\mu\nu} ~ = ~ \left[ F^{a}_{\mu\rho} {F^{a}_\nu}^\rho - \frac{1}{4} g_{\mu\nu} \left(F^{a}_{\mu\nu}\right)^2\right]\sp
D^{\mu} F^{a}_{\mu\nu} ~ = ~ 0 \, .
\ee
For the metric ansatz defined by
\be\label{4bb}
ds^2 = dr^2 + e^{2 w(r)} d \Omega_3^2 = d r^2 + \frac{f^2(r)}{4} \omega^a \omega^a \, ,
\ee
and for the  gauge field configuration
\be\label{5bb}
A^a = \frac{h(r)}{2} \omega^a
\ee
with $\omega^a$ the Maurer-Cartan forms defined in~\eqref{MaurerHopf}, the Yang-Mills equations can be expressed in differential forms as~\cite{Hosoya:1989zn}
\be
D^* F^a = \epsilon^{a b c} d r \wedge \omega^b \wedge \omega^c \frac{(h^2 - 2 h)(h-1)}{f} = 0 \, , \quad \Rightarrow h = 1,2
\ee
The non trivial configuration that is not a pure gauge is the meron for which $h=1$. The Einstein equations then reduce to an ODE
\be\label{6bb}
\left(\frac{d f}{d r}\right)^2 = 1 - \frac{r_0^2}{f^2} - H^2 f^2 \, ,
\ee
where $r_0^2 = 4 \pi G_N/g_{YM}^2$ and $H^2= 8 \pi G \Lambda /3\, $. The solutions for the various cases of $\Lambda$ are presented in~\ref{EinsteinYangMillswormhole}. More general ``nested wormhole" solutions can be found in~\cite{Rey:1989th}.

\section{Elliptic functions}\label{Elliptic functions}

\begin{figure}[t]
\vskip 10pt
\centering
\includegraphics[width=176mm]{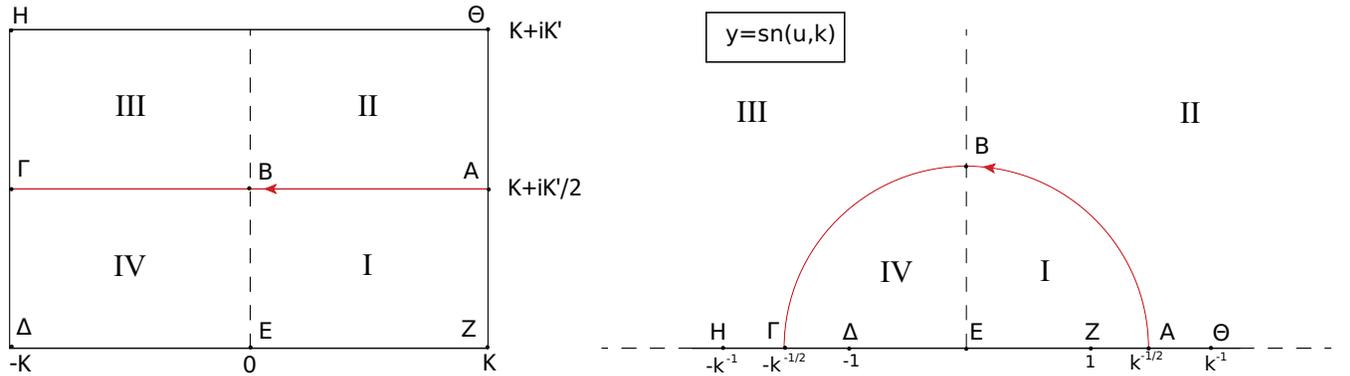}
\caption{The mapping of the rectangle to the upper-half plane via $y= \sn (u, k)$, with a matching of corresponding points. The branch cuts are between $H \Delta$ and $Z \Theta$. Both pictures correspond to the $UHP_1$ quadrant of fig.~\ref{fig:Branchcuts}. }
\label{fig:UHP}
\end{figure}

\begin{figure}[t]
\vskip 10pt
\centering
\includegraphics[width=170mm]{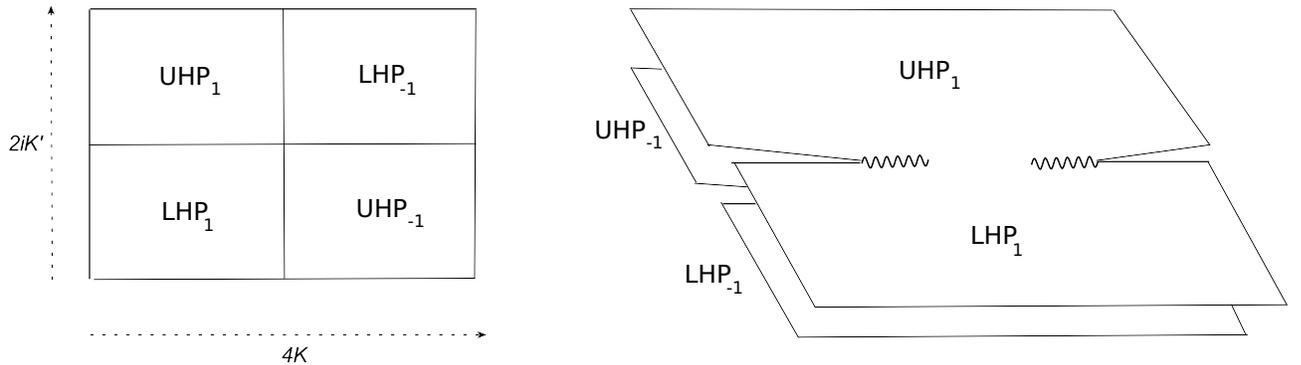}
\caption{The geometry in the complex $y$ plane is of a two-sheeted Riemann surface. The elliptic substitution makes clear that this surface is a torus.}
\label{fig:Branchcuts}
\end{figure}

We collect here some formulae on elliptic functions used in the main text. More details can be found in~\cite{Zwillinger}
\\
\\
We shall define as $0 \leq k, k' \leq 1$ the elliptic modulus/ complementary modulus. They satisfy $k^2 + k'^2 = 1$.
Once these are given, one can compute the elliptic integrals
\bea\label{A1}
F(\phi, k) &=& \int_0^\phi \frac{d \theta}{\sqrt{1- k^2 \sin^2 \theta}}\, , \quad E(\phi, k) = \int_0^\phi d \theta \sqrt{1- k^2 \sin^2 \theta} \, , \nn \\
K(k) &=& F(\pi/2, k) \, , \nn \\
E(k) &=& E(\pi/2, k) \, , \nn \\
K'(k) &=& K(k') \, , \nn \\
E'(k) &=& E(k') \,
\eea
We shall also denote the doubly periodic Jacobian elliptic functions as $\sn (u, k), \cn(u,k) , \dn(u, k)$ etc. They are periodic in appropriate multiples of $K, i K'$, see~\cite{Zwillinger}. They satisfy
\be\label{A2}
\sn^2(u,k) + \cn^2(u,k) = k^2\sn^2(u,k) + \dn^2(u,k) = 1
\ee
The identities
\be\label{A3}
\cn(i u, k') = \frac{1}{\cn (u , k)}\, , \qquad \cn(u + K, k) = - {k'} \frac{\sn (u, k)}{ \dn (u,k)}
\ee
are also useful.
In figs.~\ref{fig:UHP} and~\ref{fig:Branchcuts} we depict the properties of the mapping $y = \sn(u,k)$.

\newpage
\addcontentsline{toc}{section}{References}
 
\end{document}